\documentclass[12pt]{article}

\usepackage{verbatim,color,amssymb}
\usepackage{amsmath}					
\usepackage{amsthm}					
\usepackage{blkarray}
\usepackage{algorithm}
\usepackage{algpseudocode}
\usepackage{natbib}
\usepackage{multirow}
\usepackage{setspace}
\usepackage[mathscr]{euscript}
\usepackage{fancyhdr}
\usepackage{enumitem}
\usepackage{graphicx}
\usepackage{geometry}
\usepackage{lineno}
\usepackage[compact]{titlesec}
\usepackage{listings}

\usepackage{subfig}
\usepackage{caption}
\usepackage{float}
\usepackage{setspace}
\usepackage{tikz}
\usetikzlibrary{arrows}
\usepackage{multirow}
\usepackage{lineno}

\newtheorem{Thm}{\underline{\bf Theorem}}
\newtheorem{Assmp}{{\bf Assumption}}

\newtheorem*{Proof*}{Proof}

\newtheorem{Lem}{\underline{\bf Lemma}}

\def\rR{\mathbb{R}}
\def\eE{\mathbb{E}}

\def\B{{\cal B}}
\def\C{{\cal C}}

\def\F{{\cal F}}

\def\N{{\cal N}}
\def\calP{{\cal P}}
\def\S{{\cal S}}
\def\T{{\cal T}}

\def\Y{{\cal Y}}

\def\wh{\widehat}
\def\wt{\widetilde}

\def\log{\hbox{log}}

\def\var{\hbox{var}}
\def\cov{\hbox{cov}}

\def\Bern{\hbox{Bernoulli}}
\def\Beta{\hbox{Beta}}

\def\Dir{\hbox{Dir}}
\def\DP{\hbox{DP}}

\def\Ga{\hbox{Ga}}
\def\SB{\hbox{SB}}

\def\IG{\hbox{Inv-Ga}}

\def\MVN{\hbox{MVN}}

\def\Normal{\hbox{Normal}}
\def\Poi{\hbox{Poi}}
\def\SB{\hbox{SB}}

\def\Unif{\hbox{Unif}}
\def\Mult{\hbox{Mult}}

\def\HOHMM{\hbox{HOHMM}}

\def\ANNALS{{\it Annals of Statistics}}

\def\P_25_ICML{{\it Proceedings of the 25th international conference on Machine learning}}

\def\refhg{\hangindent=20pt\hangafter=1}
\def\refmark{\par\vskip 2mm\noindent\refhg}

\def\refhg{\hangindent=20pt\hangafter=1}
\def\refmark{\par\vskip 2mm\noindent\refhg}

\def\bse{\begin{eqnarray*}}
\def\ese{\end{eqnarray*}}
\def\be{\begin{eqnarray}}
\def\ee{\end{eqnarray}}
\def\bq{\begin{equation}}
\def\eq{\end{equation}}

\def\wh{\widehat}

\def\trans{^{\rm T}}

\def\th{^{th}}

\def\bc{{\mathbf c}}

\def\b1e{{\mathbf e}}
\def\b1f{{\mathbf f}}

\def\bG{{\mathbf G}}

\def\bk{{\mathbf k}}
\def\bm{{\mathbf m}}
\def\bM{{\mathbf M}}
\def\bn{{\mathbf n}}
\def\bp{{\mathbf p}}
\def\bP{{\mathbf P}}
\def\br{{\mathbf r}}
\def\bs{{\mathbf s}}

\def\bU{{\mathbf U}}

\def\bw{{\mathbf w}}

\def\by{{\mathbf y}}

\def\bz{{\mathbf z}}

\newcommand{\etal}{\emph{et al.~}}
\newcommand{\etam}{\mbox{\boldmath $\eta$}}

\newcommand{\bpi}{\mbox{\boldmath $\pi$}}

\newcommand{\btheta}{\mbox{\boldmath $\theta$}}

\newcommand{\bbeta}{\mbox{\boldmath $\beta$}}

\newcommand{\bzeta}{\mbox{\boldmath $\zeta$}}
\newcommand{\bsigma}{\mbox{\boldmath $\sigma$}}
\newcommand{\bSigma}{\mbox{\boldmath $\Sigma$}}

\newcommand{\blambda}{\mbox{\boldmath $\lambda$}}

\newcommand{\bpsi}{\mbox{\boldmath $\psi$}}

\newcommand{\btau}{\mbox{\boldmath $\tau$}}

\newcommand{\abs}[1]{\left\vert#1\right\vert}
\newcommand{\norm}[1]{\left\Vert#1\right\Vert}

\renewcommand\footnoterule{\kern-3pt \hrule \textwidth 2in \kern 2.6pt}

\newcommand\Algphase[1]{%
\vspace*{-.5\baselineskip}\Statex\hspace*{\dimexpr-\algorithmicindent-2pt\relax}\rule{\textwidth}{0.4pt}%
\Statex\hspace*{-\algorithmicindent}\textbf{#1}%
\vspace*{-.5\baselineskip}\Statex\hspace*{\dimexpr-\algorithmicindent-2pt\relax}\rule{\textwidth}{0.4pt}%
}

\def\boxit#1{\vbox{\hrule\hbox{\vrule\kern6pt \vbox{\kern6pt \textcolor{blue}{#1}\kern6pt}\kern6pt\vrule}\hrule}}

\def\authorfootnote#1{{\let\thefootnote\relax\footnotetext{#1}}}

\pgfarrowsdeclare{biggertip}{biggertip}{%
  \setlength{\arrowsize}{1pt}  
  \addtolength{\arrowsize}{.5\pgflinewidth}  
  \pgfarrowsrightextend{0}  
  \pgfarrowsleftextend{-5\arrowsize}  
}{%
  \setlength{\arrowsize}{1pt}  
  \addtolength{\arrowsize}{.5\pgflinewidth}  
  \pgfpathmoveto{\pgfpoint{-5\arrowsize}{4\arrowsize}}  
  \pgfpathlineto{\pgfpointorigin}  
  \pgfpathlineto{\pgfpoint{-5\arrowsize}{-4\arrowsize}}  
  \pgfusepathqstroke  
}  

\allowdisplaybreaks

\begin{document}
\thispagestyle{empty}
\baselineskip=28pt

\begin{center}
{\LARGE{\bf Bayesian Higher Order Hidden\\ Markov Models}}
\end{center}
\baselineskip=12pt

\vskip 10mm
\begin{center}
Abhra Sarkar\\
Department of Statistics and Data Sciences,\\ University of Texas at Austin,\\ 2317 Speedway D9800, Austin, TX 78712-1823, USA\\
abhra.sarkar@utexas.edu\\ 
 and \\
David B. Dunson\\
Department of Statistical Science,\\ Duke University,\\ Box 90251, Durham NC 27708-0251\\
dunson@duke.edu\\
\end{center}

\vskip 10mm
\begin{center}
{\Large{\bf Abstract}} 
\end{center}
\baselineskip=12pt
We consider the problem of flexible modeling of higher order hidden Markov models 
when the number of latent states and the nature of the serial dependence, including the true order, are unknown. 
We propose flexible Bayesian methods based on tensor factorization techniques 
that can characterize any transition probability with a specified maximal order, 
allowing automated selection of the important lags and capturing higher order interactions among the lags. 
Theoretical results provide insights into identifiability of the emission distributions 
and consistency of the posterior. 
We design Markov chain Monte Carlo algorithms for posterior computation. 
In simulation experiments, the method vastly outperforms competitors not just in higher order settings, 
but, remarkably, also in first order cases.  
Practical utility is illustrated using real world applications.

\vskip 8mm
\baselineskip=12pt
\noindent\underline{\bf Some Key Words}: Bayesian nonparametrics, Conditional tensor factorization, Higher order hidden Markov models, Sequential data, Time series. 

\par\medskip\noindent
\underline{\bf Short/Running Title}: Higher Order Hidden Markov Models

\par\medskip\noindent
\underline{\bf Correpsonding Author}: Abhra Sarkar (abhra.sarkar@utexas.edu)

\clearpage\pagebreak\newpage
\pagenumbering{arabic}
\newlength{\gnat}
\setlength{\gnat}{16pt}
\baselineskip=\gnat

\newpage
\section{Introduction}
Hidden Markov models (HMMs) have been tremendously successful in statistical analyses of sequentially generated data 
\citep{fruhwirth2006finite,mcdonald_zuchhini:1997, cappe2009inference}
in diverse application areas like proteomics \citep{Bae_etal:2005, Lennox_etal:2010}, genomics \citep{guha2008bayesian, Yau_etal:2011, titsias2016statistical},
animal movement \citep{langrock2015nonparametric, quick2017hidden}, 
speech recognition \citep{Rabiner:1989, fox2011sticky}, 
and economics and finance \citep{Hamilton:1990, Albert_Chib_JBES:1993}.

The basic HMM consists of two processes: a \emph{hidden} process $\{c_{t}\}$, which evolves according to a first order Markov chain with discrete state space, 
and a potentially multivariate \emph{observed} process $\{\by_{t}\}$ observed sequentially over a set of discrete time points $t=1,2,\dots,T$.
Specifically, an HMM makes the following set of conditional independence assumptions to model the hidden and the observed processes
\be
p(c_{t} \mid \bc_{1:(t-1)})  &=& p(c_{t}\mid c_{t-1}),	\\
p(\by_{t} \mid \by_{1:(t-1)},\bc_{1:(t-1)}) &=& p(\by_{t} \mid c_{t}).
\ee
The distributions $p(c_{t} \mid c_{t-1})$ and $p(\by_{t} \mid c_{t})$ are often referred to as the \emph{transition distribution} and the \emph{emission distribution}, respectively. 

A challenging problem of the HMM framework is the determination of the cardinality of the state space. 
This is often unknown in practice  
and is determined using model selection approaches \citep{sclove1983time, leroux1992maximum, wang1999markov} or reversible jump type model space exploration techniques \citep{robert2000bayesian}. 
\cite{teh_etal:2006} developed a Bayesian nonparametric approach to HMMs based on the hierarchical Dirichlet process (HDP) 
that defines a prior distribution on transition matrices over a countably infinite number of states. 
The number of latent states for any given dataset can be inferred from its posterior, 
allowing for uncertainty in the analysis and also the possibility that additional states may be required when more data points become available, 
precluding the necessity to decide \emph{a priori} the size of the state space.

One serious limitation of the HDP-HMM in particular and the basic HMM framework in general is the restrictive assumption of first order Markovian dynamics of the latent sequence $\{c_{t}\}$. 
The focus of this article is on higher order HMMs (HOHMMs) that allow $\{c_{t}\}$ to depend on its more distant past.  
An HOHMM of maximal order $q$ thus makes the following set of conditional independence assumptions
\be
p(c_{t} \mid \bc_{1:(t-1)})  &=& p(c_{t}\mid \bc_{(t-q):(t-1)}),	\\
p(\by_{t} \mid \by_{1:(t-1)},\bc_{1:(t-1)}) &=& p(\by_{t} \mid c_{t}).
\ee

We distinguish between an HOHMM of \emph{maximal order $q$} and an HOHMM of \emph{full order $q$}. 
An HOHMM is said to be of maximal order $q$ if conditional on the values of $c_{t-1},\dots,c_{t-q}$, the distribution of $c_{t}$ is independent of its more distant past, 
but the lags actually important in determining the distribution of $c_{t}$ may be an arbitrary subset of $\{c_{t-1},\dots,c_{t-q}\}$. 
In contrast, if the distribution of $c_{t}$ actually varies with the values at all the previous $q$ times points, we call the HOHMM to be of full order $q$. 
The case $q=0$ corresponds to serial independence of the observation sequence $\{\by_{t}\}$. 
Also, we say that an HOHMM of maximal order $q$ has \emph{true} maximal order $q$, if the set of important predictors of $c_{t}$ includes $c_{t-q}$.

While the HOHMM framework relaxes the restrictive first order assumption of the basic HMM, it also brings in a daunting dimensionality challenge. 
Consider, for instance, an HOHMM with $C$ states and maximal order $q$.  
The transition distributions  are now indexed by the $C^{q}$ different possible values of the lags $\bc_{(t-q):(t-1)}$ (rather than just $c_{t-1}$), 
and involve a total number of $(C-1)C^{q}$ parameters, which increases exponentially with the order 
and becomes too large to be estimated efficiently with datasets of the sizes typically encountered in practice. 
The issue is further complicated by the fact that we do not directly observe the values of the latent sequence $\{c_{t}\}$ 
but only their noisy manifestations $\{\by_{t}\}$.

Any HOHMM can be reformulated as a first order HMM by moving either in blocks of $q$ time steps or, more conventionally, in single time steps but with a special $q$-tuple initial latent state 
\citep{mcdonald_zuchhini:1997, cappe2009inference}. 
While convenient for theoretical treatment of HOHMMs, 
for modeling purposes such formulations are not very useful since they require working with large $C^{q} \times C^{q}$ dimensional transition probability matrices with $C^{q}(C^{q}-C)$ structural zeros. 
Associated computational machineries also quickly become practically ineffective even for moderately small values of $C$ and $q$. 
Parsimonious characterization of the transition dynamics in higher order settings is thus extremely important.  
It is also important to obtain an interpretable structure, with unnecessary lags eliminated.  

These daunting challenges to higher order generalizations 
have forced researchers to focus on first order HMMs. 
\cite{thede1999second} used a second order HMM for parts of speech tagging, 
estimating the transition probabilities by weighted mixtures of empirical proportions of subsequences of maximal length three.   
\cite{seifert2012parsimonious} developed an HOHMM with known finite state space and Normal emission densities for modeling array comparative genomic hybridization (aCGH) data. 
Transition dynamics of maximal order $q$ were modeled using state context trees of maximal depth $q$ that divide the set of all possible state combination histories into disjoint sets of equivalent state contexts.
Tree-based strategies employ strict top-down search for important lags 
and hence are not suitable for scenarios when distant lags may be more important than recent ones \citep{jaaskinen_etal:2014, sarkar_dunson:2016}. 

In this article, we develop a novel Bayesian nonparametric approach to HOHMMs 
that can parsimoniously characterize the transition dynamics of any HOHMM with a specified maximal order, 
allows flexibility in modeling the emission distributions, 
admits generalizations to countably infinite state spaces, precluding the necessity to predetermine the number of states,
and allows automated selection of the important lags, 
determining the true order and nature of the serial dependence, 
removing the necessity to decide \emph{a priori} the exact order of the transition dynamics. 

We begin by structuring the transition probabilities $p(c_{t}\mid \bc_{(t-q):(t-1)})$ as a high dimensional conditional probability tensor. 
Adapting the conditional tensor factorization approach of \cite{yang_dunson:2015} to the HOHMM setting, 
we parameterize the probabilities $p(c_{t}\mid \bc_{(t-q):(t-1)})$ as mixtures of `core' probability kernels 
with mixture weights depending on the state combinations of the lags. 
Such a parameterization explicitly identifies the set of important lags 
and implicitly captures complex higher order interactions among the important lags, borrowing strength across the states of the HOHMM 
by sharing the core kernels in a `soft' probabilistic manner.  
The elimination of the redundant lags and the implicit modeling of the interactions among the important ones 
can lead to a significant two fold reduction in the effective number parameters required to flexibly characterize the transition dynamics of the HOHMM.  
We assign sparsity inducing priors that favor such lower dimensional representations of the transition probability tensor. 

We assign a hierarchical Dirichlet prior on the core probability kernels, 
encouraging the model to shrink further towards lower dimensional structures by borrowing strength across these components as well. 
This also facilitates a generalization to countably infinite state space HOHMMs that allow uncertainty in the number of states. 
The HDP-HMM of \cite{teh_etal:2006} corresponds to a special case when the kernel sharing feature is turned off and the order is restricted to one.  

We develop a two-stage Markov chain Monte Carlo (MCMC) algorithm for learning the parameters of the model. 
The first stage selects the important lags implementing a coarser `hard' sharing approximation using a stochastic search variable selection (SSVS) approach \citep{george_mcculloch:1997}. 
The second stage keeps the set of important lags fixed and implements the finer soft kernel sharing feature, 
building on existing computational machineries for the HDP-HMM.

{HOSVD-type factorizations have previously been employed in \cite{sarkar_dunson:2016} to model the transition dynamics of observable state sequences in a higher order Markov chain framework. 
The framework of HOHMM, however, brings in significant additional challenges. 
Unlike an observable Markov process, the states $c_{t}$ are now latent, only their noisy manifestations $\by_{t}$ are available. 
The size of the state space is often unknown and has to be inferred from these noisy data points. 
These issues make infinite state space models particularly relevant in the HOHMM context. 
The emission distributions $p(\by_{t} \mid c_{t})$ have to be additionally modeled 
which brings in identifiability issues and significant computational challenges.
}

The rest of the article is organized as follows. 
Section \ref{sec: models} details the proposed tensor factorization based HOHMMs and their properties. 
Section \ref{sec: posterior computation} describes Markov chain Monte Carlo (MCMC) algorithms for drawing samples from the posterior. 
Section \ref{sec: simulation experiments} presents the results of simulation experiments comparing our method with existing approaches. 
Section \ref{sec: applications} presents some real world applications. 
Section \ref{sec: discussion} contains concluding remarks.

\section{Higher Order Hidden Markov Model}  \label{sec: models}

\subsection{Modeling the Transition Probabilities}  \label{sec: transition}
We build on the idea of higher order singular value decomposition (HOSVD) tensor factorization to develop a nonparametric approach for modeling the transition dynamics of a finite memory HOHMM.
HOSVD \citep{tucker:1966, de_lathauwer_etal:2000} factorizes a $C_{1} \times \dots \times C_{p}$ dimensional $p$-way tensor $\bM=\{m_{x_{1},\dots,x_{p}}\}$ as 
\bse
m_{x_{1},\dots,x_{p}} = \sum_{h_{1}=1}^{k_{1}}\cdots\sum_{h_{p}=1}^{k_{p}}g_{h_{1},\dots,h_{p}} \prod_{j=1}^{p} u_{h_{j}x_{j}}^{(j)}, 
\ese
where the \emph{core tensor} $\bG=\{g_{h_{1},\dots,h_{p}}\}$ captures the interactions between different components and $\bU^{(j)}=\{u_{h_{j}x_{j}}^{(j)}\}$ are component specific weights. 
In our HOHMM setting, the hidden sequence $\{c_{t}\}$ with state space $\{1,\dots,C\}$ has finite memory of true maximal order $q$. 
Given $c_{t-q},\ldots,c_{t-1}$, the distribution of $c_{t}$ is independent of all latent states prior to $t-q$. 
The variables that are important in predicting $c_{t}$ comprise a subset of $\{c_{t-q},\ldots,c_{t-1}\}$, possibly proper but including $c_{t-q}$.

\begin{figure}[h!]
\centering
\begin{center}
\resizebox{12cm}{5cm}{%
\begin{tikzpicture}
	[
		grid/.style={very thin,gray},
		axis/.style={->,blue,thick},
		cube/.style={opacity=.75,very thick,fill=blue!20}]


	\draw[cube] (0,0,0) -- (0,4,0) -- (4,4,0) -- (4,0,0) -- cycle;
	
	\draw[cube] (0,0,0) -- (0,4,0) -- (0,4,4) -- (0,0,4) -- cycle;

	\draw[cube] (0,0,0) -- (4,0,0) -- (4,0,4) -- (0,0,4) -- cycle;

	\draw[cube] (4,0,0) -- (4,4,0) -- (4,4,4) -- (4,0,4) -- cycle;

	\draw[cube] (0,0,4) -- (0,4,4) -- (4,4,4) -- (4,0,4) -- cycle;

	\draw[cube] (0,4,0) -- (4,4,0) -- (4,4,4) -- (0,4,4) -- cycle;

\node[] at (7,2,2) {\Huge$=$};

	\draw[cube] (10,0,0) -- (10,0,2) -- (12,0,2) -- (12,0,0) -- cycle;
	
	\draw[cube] (10,0,0) -- (12,0,0) -- (12,2,0) -- (10,2,0) -- cycle;

	\draw[cube] (10,0,0) -- (10,0,2) -- (10,2,2) -- (10,2,0) -- cycle;

	\draw[cube] (12,0,0) -- (12,0,2) -- (12,2,2) -- (12,2,0) -- cycle;

	\draw[cube] (10,2,2) -- (10,0,2) -- (12,0,2) -- (12,2,2) -- cycle;

	\draw[cube] (10,2,0) -- (10,2,2) -- (12,2,2) -- (12,2,0) -- cycle;

	\draw[cube,fill=red!10] (13,1,0) -- (13,1,2) -- (17,1,2) -- (17,1,0) -- cycle;		

	\draw[cube,fill=red!10] (10,3,1) -- (12,3,1) -- (12,7,1) -- (10,7,1) -- cycle;	
	
	\draw[cube,fill=red!10] (10,1,3) -- (10,1,7) -- (12,1,7) -- (12,1,3) -- cycle;

\node[] at (1.5,1.5,2) {\Huge$\bP$};
\node[] at (11.5,1.5,3) {\Huge$\blambda$};
\node[] at (18.7,1.7,3) {\Huge$\bpi^{(1)}$};
\node[] at (13.5,5.5,3) {\Huge$\bpi^{(2)}$};
\node[] at (8.5,0.5,3) {\Huge$\bpi^{(3)}$};

\end{tikzpicture}
}
\end{center}
\caption{Pictorial representation of the factorization of a transition probability tensor $\bP$ characterizing a Markov chain of maximal order 3 with core tensor $\blambda$ and mode matrices $\bpi^{(j)}, j=1,2,3$.}
\label{fig: HOSVD2}
\end{figure}
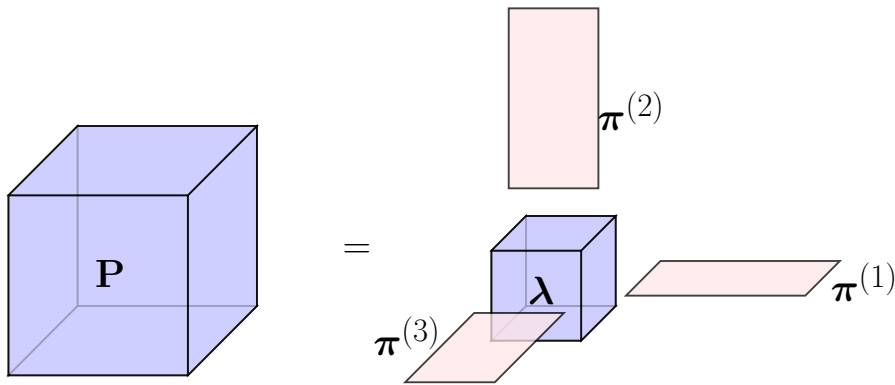

We structure the transition probabilities $p(c_{t} \mid c_{t-q},\ldots,c_{t-1})$ as a $C \times C \times \dots \times C$ dimensional $(q+1)$-way tensor 
and consider the following HOSVD-type factorization  
\be
&&\hspace{-1cm} p(c_{t}\mid c_{t-j},j=1,\dots,q)  = \sum_{h_{1}=1}^{k_{1}}\cdots\sum_{h_{q}=1}^{k_{q}} \lambda_{h_{1},\dots,h_{q}}(c_{t})\prod_{j=1}^{q}\pi_{h_{j}}^{(j)}(c_{t-j}). \label{eq: TFM1}
\ee
See Figure \ref{fig: HOSVD2}. 
Here $1 \leq k_{j} \leq C$ for all $j$ and the parameters $\lambda_{h_{1},\dots,h_{q}}(c_{t})$ and $\pi_{h_{j}}^{(j)}(c_{t-j})$ are all non-negative and satisfy the constraints 
(a) $\sum_{c_{t}=1}^{C}  \lambda_{h_{1},\dots,h_{q}}(c_{t}) =1,~~ \text{for each combination}~(h_{1},\dots,h_{q}),  \label{eq: TFM2}$ and 
(b) $\sum_{h_{j}=1}^{k_{j}} \pi_{h_{j}}^{(j)}(c_{t-j}) = 1, ~~ \text{for each pair }~(j,c_{t-j}). \label{eq: TFM3}$
If follows as a direct consequence of Theorem 1 in \cite{yang_dunson:2015}, a general result on conditional probability tensors, 
that any transition probability tensor can be represented as (\ref{eq: TFM1}) with the parameters satisfying the constraints (a) and (b).

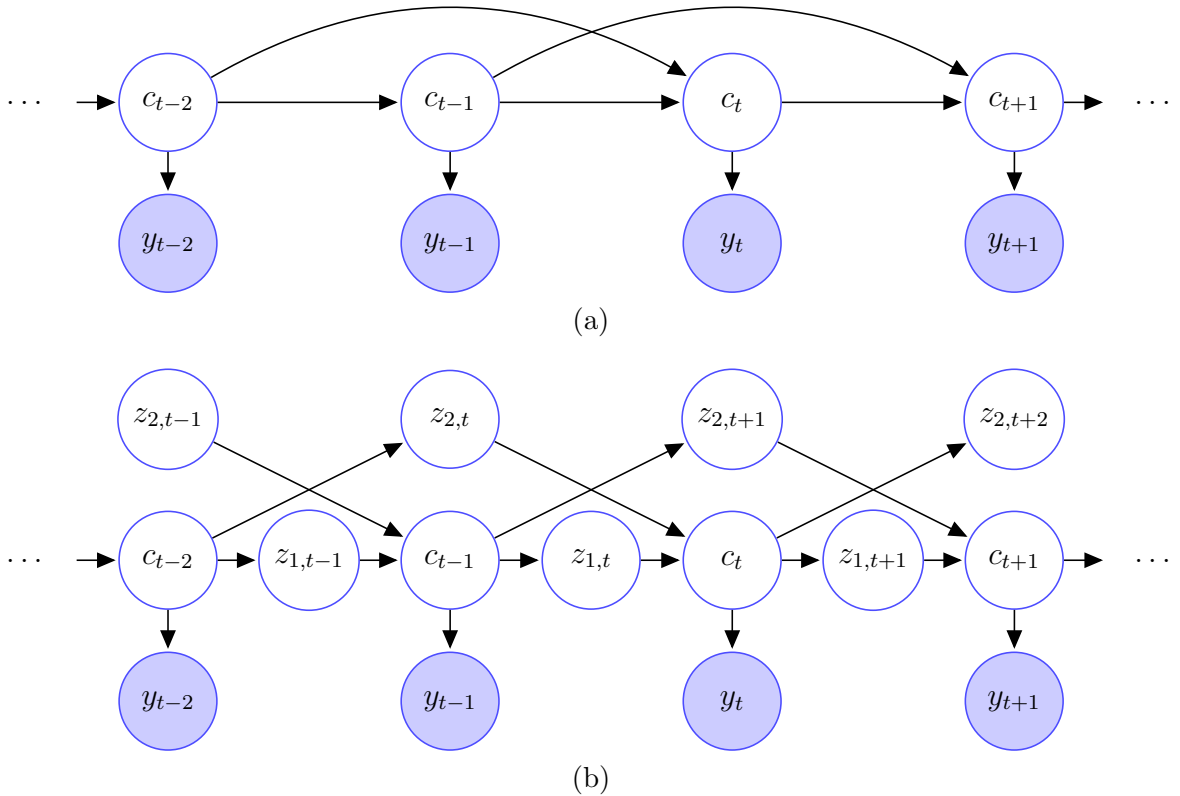
\begin{figure}[h!]
\subfloat[]
{
\centering
\begin{tikzpicture}[scale=1.5,->,>=triangle 45,shorten >=1pt,auto,node distance=2.8cm,semithick]

  \node[minimum size=1.3cm] (c_t-3) at (-0.75,0) {$\dots$};
  \node[style={draw=blue!70,circle}, minimum size=1.3cm] (c_t-2) at (0.5,0) {$c_{t-2}$};
  \node[style={draw=blue!70,circle}, minimum size=1.3cm] (c_t-1) at (3,0) {$c_{t-1}$};
  \node[style={draw=blue!70,circle}, minimum size=1.3cm] (c_t) at (5.5,0) {$c_{t}$};
  \node[style={draw=blue!70,circle}, minimum size=1.3cm] (c_t+1) at (8,0) {$c_{t+1}$};
  \node[minimum size=1.3cm] (c_t+2) at (9.25,0) {$\dots$};
  
  \node[style={draw=blue!70,circle,fill=blue!20}, minimum size=1.3cm] (y_t-2) at (0.5,-1.25) {$y_{t-2}$};
  \node[style={draw=blue!70,circle,fill=blue!20}, minimum size=1.3cm] (y_t-1) at (3,-1.25) {$y_{t-1}$};
  \node[style={draw=blue!70,circle,fill=blue!20}, minimum size=1.3cm] (y_t) at (5.5,-1.25) {$y_{t}$};
  \node[style={draw=blue!70,circle,fill=blue!20}, minimum size=1.3cm] (y_t+1) at (8,-1.25) {$y_{t+1}$};

  \path (c_t-3) edge (c_t-2);
  \path (c_t-2) edge (c_t-1);
  \path (c_t-1) edge (c_t);
  \path (c_t) edge (c_t+1);
  \path (c_t+1) edge (c_t+2);

  \path (c_t-2) edge[bend left] (c_t);
  \path (c_t-1) edge[bend left] (c_t+1);

  \path (c_t-2) edge (y_t-2);
  \path (c_t-1) edge (y_t-1);
  \path (c_t) edge (y_t);
  \path (c_t+1) edge (y_t+1);

\end{tikzpicture}
}

\subfloat[]
{
\begin{tikzpicture}[scale=1.5,->,>=triangle 45,shorten >=1pt,auto,node distance=2.8cm,semithick]
\normalsize

  \node[style={draw=blue!70,circle}, minimum size=1.3cm] (c_t-2) at (0.5,0) {$c_{t-2}$};
  \node[style={draw=blue!70,circle}, minimum size=1.3cm] (c_t-1) at (3,0) {$c_{t-1}$};
  \node[style={draw=blue!70,circle}, minimum size=1.3cm] (c_t) at (5.5,0) {$c_{t}$};
  \node[style={draw=blue!70,circle}, minimum size=1.3cm] (c_t+1) at (8,0) {$c_{t+1}$};
  
  \node[style={draw=blue!70,circle,fill=blue!20}, minimum size=1.3cm] (y_t-2) at (0.5,-1.25) {$y_{t-2}$};
  \node[style={draw=blue!70,circle,fill=blue!20}, minimum size=1.3cm] (y_t-1) at (3,-1.25) {$y_{t-1}$};
  \node[style={draw=blue!70,circle,fill=blue!20}, minimum size=1.3cm] (y_t) at (5.5,-1.25) {$y_{t}$};
  \node[style={draw=blue!70,circle,fill=blue!20}, minimum size=1.3cm] (y_t+1) at (8,-1.25) {$y_{t+1}$};

  \node[minimum size=1.3cm] (z_1_t-2) at (-0.75,0) {$\dots$};
  \node[style={draw=blue!70,circle}, minimum size=1.3cm] (z_2_t-1) at (0.5,1.25) {$z_{2,t-1}$};
  \node[style={draw=blue!70,circle}, minimum size=1.3cm] (z_1_t-1) at (1.75,0) {$z_{1,t-1}$};
  \node[style={draw=blue!70,circle}, minimum size=1.3cm] (z_2_t) at (3,1.25) {$z_{2,t}$};
  \node[style={draw=blue!70,circle}, minimum size=1.3cm] (z_1_t) at (4.25,0) {$z_{1,t}$};
  \node[style={draw=blue!70,circle}, minimum size=1.3cm] (z_2_t+1) at (5.5,1.25) {$z_{2,t+1}$};
  \node[style={draw=blue!70,circle}, minimum size=1.3cm] (z_1_t+1) at (6.75,0) {$z_{1,t+1}$};
  \node[style={draw=blue!70,circle}, minimum size=1.3cm] (z_2_t+2) at (8,1.25) {$z_{2,t+2}$};
  \node[minimum size=1.3cm] (z_1_t+2) at (9.25,0) {$\dots$};

 \path (z_1_t-2) edge (c_t-2);
 \path (z_2_t-1) edge (c_t-1);
 \path (z_1_t-1) edge (c_t-1);
  \path (z_2_t) edge (c_t);
  \path (z_1_t) edge (c_t);
  \path (z_2_t+1) edge (c_t+1);
  \path (z_1_t+1) edge (c_t+1);

  \path (c_t-2) edge (z_1_t-1);
  \path (c_t-1) edge (z_1_t);
  \path (c_t) edge (z_1_t+1);
  \path (c_t+1) edge (z_1_t+2);
  
  \path (c_t-2) edge (z_2_t);
  \path (c_t-1) edge (z_2_t+1);
  \path (c_t) edge (z_2_t+2);

  \path (c_t-2) edge (y_t-2);
  \path (c_t-1) edge (y_t-1);
  \path (c_t) edge (y_t);
  \path (c_t+1) edge (y_t+1);

\normalsize
\end{tikzpicture}
}
\caption{Graphical model depicting the dependence structure of a second order hidden Markov model (a) without and (b) with second level latent variables. 
Shaded and unshaded nodes represent observed and latent variables, respectively.}
\label{fig: graph 2}
\end{figure}

Introducing latent allocation variables $z_{j,t}$ for $j=1,\dots,q$ and $t=q+1,\dots,T$, the latent variables $\{c_{t}\}$ are conditionally independent 
and the factorization can be equivalently represented through the following hierarchical formulation
\be
(c_{t}\mid z_{j,t}=h_{j},j=1,\dots,q)   	&\sim&   \Mult(\{1,\dots,C\},\lambda_{h_{1},\dots,h_{q}}(1),\dots,\lambda_{h_{1},\dots,h_{q}}(C)), \label{eq: interaction}    \\
(z_{j,t}\mid c_{t-j})   	&\sim&   \Mult(\{1,\dots,k_{j}\},\pi_{1}^{(j)}(c_{t-j}),\dots \pi_{k_{j}}^{(j)}(c_{t-j})).  \label{eq: soft clustering}
\ee
See Figure \ref{fig: graph 2}. 
Equation (\ref{eq: soft clustering}) reveals the soft sharing property of the model 
that enables it to borrow strength across the different states of $c_{t-j}$ 
by allowing the $z_{j,t}$'s associated with a particular state of $c_{t-j}$ to be allocated 
to different latent populations, which are shared across all $C$ states of $c_{t-j}$. 
In contrast, a hard sharing model would allocate each $z_{t,j}$ to a single latent population. 
Equation (\ref{eq: interaction}) shows how such soft assignment enables the model to capture complex interactions among the lags in an implicit and parsimonious manner by allowing the latent populations indexed by $(h_{1},\dots,h_{q})$ to be shared among the various state combinations of the lags.

When $k_{j}=1$, $\pi_{1}^{(j)}(c_{t-j})=1$ and $P(c_{t} \mid c_{t-q},\ldots,c_{t-1})$ does not vary with $c_{t-j}$. 
The variable $k_{j}$ thus determines the inclusion of the $j\th$ lag $c_{t-j}$ in the model. 
The variable $k_{j}$ also determines the number of latent classes for the $j\th$ lag $c_{t-j}$. 
The number of parameters in such a factorization is given by $(C-1)\prod_{j=1}^{q}k_{j} + C\sum_{j=1}^{q}(k_{j}-1)$, which will be much smaller than the number of parameters $(C-1)C^{q}$ required to specify a full Markov model of the same maximal order, if $\prod_{j=1}^{q}k_{j} \ll C^{q}$.

As a first proposal, we may assign independent Dirichlet priors on $\blambda_{h_{1},\dots,h_{q}}$ as 
\be
&&\hspace{-1cm} \blambda_{h_{1},\dots,h_{q}}    \sim   \Dir(\alpha,\dots,\alpha),~\hbox{independently for each}~(h_{1},\dots,h_{q}).  \label{eq: DP1}
\ee
The estimation of $\prod_{j=1}^{q}k_{j}$ independent parameters may still be a daunting task in practical applications. 
Even in first order HMMs, single layer independent priors on the transition probability vectors 
have far inferior performance to hierarchical priors that allow information to be better shared between different state-dependent transition distributions.  
See, for example, Figure 10 in \cite{fox2011sticky}.  


Towards a more parsimonious representation of the transition probability tensor, 
we assign a conditionally independent hierarchical Dirichlet prior on $\blambda_{h_{1},\dots,h_{q}}=\{\lambda_{h_{1},\dots,h_{q}}(1),\dots,\lambda_{h_{1},\dots,h_{q}}(C)\}$. 
Specifically, we let 
\be
&&\hspace{-1cm} \blambda_{h_{1},\dots,h_{q}}    \sim   \Dir\{\alpha\lambda_{0}(1),\dots,\alpha\lambda_{0}(C)\},~\hbox{independently for each}~(h_{1},\dots,h_{q}),  \label{eq: hDP1} \\
&&\hspace{-1cm} \blambda_{0}=\{\lambda_{0}(1),\dots,\lambda_{0}(C)\} \sim \Dir(\alpha_{0}/C,\dots,\alpha_{0}/C).  \label{eq: hDP2}
\ee
The `kernels' $\blambda_{h_{1},\dots,h_{q}}$ are associated with the mixture weights in a hierarchical DP. 

The dimension of $\bpi_{k_{j}}^{(j)}(c_{t-j})=\{\pi_{1}^{(j)}(c_{t-j}),\dots,\pi_{k_{j}}^{(j)}(c_{t-j})\}$, unlike the $\blambda_{h_{1},\dots,h_{q}}$'s, varies only linearly with $k_{j}$. 
We assign independent priors on the $\bpi_{k_{j}}^{(j)}(c_{t-j})$'s as 
\be
&& \bpi_{k_{j}}^{(j)}(c_{t-j}) = \{\pi_{1}^{(j)}(c_{t-j}),\dots,\pi_{k_{j}}^{(j)}(c_{t-j})\} \sim \Dir(\gamma_{j},\dots,\gamma_{j}).  \label{eq: prior on pi}
\ee

While the dimension of the core tensor varies with $k_{j}$'s, 
all $\blambda_{h_{1},\dots,h_{q}}$ share the same support $\{1,\dots,C\}$. 
This allows us to avoid conditioning on the $k_{j}$'s while specifying the hierarchical prior on $\blambda_{h_{1},\dots,h_{q}}$. 
The probability vectors $\bpi_{k_{j}}^{(j)}(c_{t-j})$, on the other hand, are supported on $\{1,\dots,k_{j}\}$ for each pair $(j,c_{t-j})$. 
Therefore, unlike $\blambda_{h_{1},\dots,h_{q}}$, conditioning on $k_{j}$, which we have kept implicit in (\ref{eq: prior on pi}), can not be avoided. 

Finally, we assign the following independent priors on $k_{j}$'s 
\be
p_{0,j}(k) &\propto& \exp(-\varphi jk),~~~j=1,\dots,q,~~~k=k_{j,\min},\dots,C,     \label{eq: prior on k_{j}'s}
\ee
where $\varphi>0$, $k_{j,\min}=1$ for $j=1,\dots,q-1$ and $k_{q,\min}=2$.  
The prior $p_{0,j}$ assigns increasing probabilities to smaller values of $k_{j}$ as the lag $j$ becomes more distant, 
reflecting the natural belief that increasing lags have diminishing influence on the distribution of $c_{t}$. 
The larger the value of $\varphi$, the faster is the decay of $p_{0,j}(k)$ with increase in $j$ and $k$, favoring sparser lower order models. 
The restriction $k_{q}>1$ implies that the $q\th$ lag is important so that the true maximal order is $q$. 


\subsection{Modeling the Emission Distributions}  \label{sec: emission}
The generic form of the emission distribution that we consider in this article is  
\bse
p(\by_{t} \mid c_{t}, \bbeta, \bpsi) = f(\by_{t} \mid \bbeta_{c_{t}}, \bpsi).  
\ese
Here $\bbeta=\{\bbeta_{c}: c=1,\dots,C\}$ denotes parameters indexed by the latent process $\{c_{t}\}$, 
whereas $\bpsi$ collects global parameters that do not evolve with time but remain constant and may sometimes be kept implicit.

In the first order HMM literature, parametric choices for the emission distribution are common. 
\cite{leroux:1992} provided sufficient conditions for identifiability in such models. 
There has been some recent interest in flexible models for the emission distributions \citep{Yau_etal:2011, langrock2015nonparametric} that relax restrictive parametric assumptions, 
which can be shown to satisfy sufficient conditions for identifiability \citep{gassiat_etal:2015, alexandrovich2016nonparametric}.
The following lemma establishes such guarantees in higher order settings. 

\begin{Lem} \label{lem: identifiability}
Let $P$ be the transition probability tensor and $\b1f$ be the emission distributions of an HOHMM 
with known state space $\C$ and known true maximal order $q$. 
Let the first order representation of the underlying Markov chain 
be ergodic and stationary with transition probability matrix $\wt{P}$ and stationary and initial distribution $\{\pi(c_{1},\dots,c_{q}): c_{j}\in\C, j=1,\dots,q\}$. 
Let $\wt{P}$ be of full rank and the emission distributions $\b1f=\{f_{c}: c\in\C\}$ be all distinct. 
Then $P$ and $\b1f$ are nonparametrically identifiable from the distribution of $T=(2C^{q}+1)q$ consecutive observations $\by_{1:T}$ up to label swapping of the states. 
\end{Lem}

The proof, deferred to section \ref{appendix: proof of identifiability} in the Supplementary Materials, 
utilizes a similar result on first order HMMs from \cite{alexandrovich2016nonparametric}, noting that by moving in blocks of size $q$ as 
\vspace{-2ex}
\bse
\begin{array}{c c c c c c c}
(c_{1},\dots,c_{q}) & \to & (c_{q+1},\dots,c_{2q}) & \to & (c_{2q+1},\dots,c_{3q}) & \to & \cdots \\
\downarrow &  & \downarrow & & \downarrow &  &  \\
(\by_{1},\dots,\by_{q}) & & (\by_{q+1},\dots,\by_{2q}) & &  (\by_{2q+1},\dots,\by_{3q}) & & \cdots
\end{array} 
\ese 
\vspace{-3ex}\\
an HOHMM of maximal order $q$ with state space $\C=\{1,\dots,C\}$ and transition probability tensor $P$ 
can be represented as a first order HMM with 
expanded state space $\C^{q}$, 
stationary and initial distribution $\pi(c_{1},\dots,c_{q})$, 
emission distributions $\{f_{c_{1}} f_{c_{2}} \cdots f_{c_{q}}: c_{j}\in\C, j=1,\dots,q\}$, 
and $q$-step transition probability matrix $\wt{P}^{q}$, 
where the single-step transition probabilities are given by 
\bse
&&\hspace{-1cm} \wt{P}\{(j_{t-q},\dots,j_{t-1}), (i_{t-q+1},\dots,i_{t})\} \\
&& = p\{(c_{t-q+1}=i_{t-q+1},\dots,c_{t}=i_{t}) \mid (c_{t-q}=j_{t-q},\dots,c_{t-1}=j_{t-1})\} \\
&& = 
\left\{\begin{array}{ll}
P(c_{t}=i_{t}\mid c_{t-q}=j_{t-q},\dots,c_{t-1}=j_{t-1}),  & \text{if}~i_{t-\ell}=j_{t-\ell}~\text{for}~\ell=1,\dots,(q-1),\\
0,	& \text{otherwise}.
\end{array}\right.
\ese
\vspace{-2ex}

Lemma \ref{lem: identifiability} assumes nonsingularity of $\wt{P}$. 
This does not limit its applicability to HOHMMs of full orders but also accommodates lag gaps. 
In this case, the transition probability matrix $\wt{P}$ will have multiple rows sharing the same nonzero elements 
but they will appear in different columns so that $\wt{P}$ could still have full rank. 
Consider, for example, a binary Markov chain of maximal order $2$ with a lag gap at $t-1$ 
so that $P(c_{t}\mid c_{t-2},c_{t-1})=P(c_{t}\mid c_{t-2})$ and $\wt{P}$ is given by 
\vspace{-2ex}
\bse
\wt{P}=
\begin{blockarray}{ccccc}
 		& (1,1) 		& (1,2) 		& (2,1) 		& (2,2)		 \\
\begin{block}{c[cccc]}
    (1,1)~ 	& P(1\mid 1) 	& P(2\mid 1) 	& 0 			& 0 			\\
    (1,2)~ 	& 0 			& 0 			& P(1\mid 1)  	& P(2\mid 1) 	\\
    (2,1)~ 	& P(1\mid 2) 	& P(2\mid 2) 	& 0 			& 0 			\\
    (2,2)~ 	& 0 			& 0 			& P(1\mid 2)  	& P(2\mid 2)	\\
\end{block}
\end{blockarray}~.
\ese
\vspace{-3ex}\\
The implication of the restriction $k_{q}>1$ in (\ref{eq: prior on k_{j}'s}) in ensuring nonsingularity of $\wt{P}$ is now clear. 
%

A result on identifiability of HOHMMs in parametric settings can be derived along the lines of \cite{leroux:1992} 
where only ergodicity of $\wt{P}$ suffices. 
For such choices, an unrestricted independent prior on the $k_{j}$'s would suffice and
the restriction $k_{q}>1$ may be dropped. 
Treating $q$ to be an upper bound on the maximal order, 
the proposed model can then select the important lags itself, 
including zeroth order cases 
which can be viewed as HOHMMs with $P(c_{t}\mid c_{t-q},\dots,c_{t-1})=\pi(c_{t})$.  
In applications of HOHMMs, however, some form of serial dependency would generally be expected 
and we do not pursue the zeroth order cases any further. 
Practical strategies that allow the assumption of known true maximal order to be relaxed are discussed in Section \ref{sec: unknown order and state space}.

In this article, we consider the following families of emission distributions - (a) Normal, (b) Poisson and (c) translated mixture of Normals. 
For Gaussian emission distributions $f(y\mid c_{t}=c) = \Normal(y \mid \mu_{c},\sigma_{c}^{2})$.
We assign conjugate $\Normal(\mu_{0},\sigma_{0}^2) \times \IG(a_{0},b_{0})$ priors on $(\mu_{c},\sigma_{c}^2)$. 
For Poisson emission distributions $f(y\mid c_{t}=c) = \Poi(y \mid \mu_{c})$, 
we assign conjugate $\Ga(a_{0},b_{0})$ prior on $\mu_{c}$, 
with the hyper-parameters chosen such that $E(\mu_{c})=a_{0}b_{0}=\overline{\by}, \var(\mu_{c})=a_{0}b_{0}^{2}=2\var(\by)$.  
Finally, translated mixture Normal emission distributions are constructed  as 
\bse
\textstyle f(y\mid c_{t}=c) = \sum_{s=1}^{S}\pi_{s}  \Normal(y \mid \mu_{c}+\eta_{s},\sigma_{s}^{2}), ~~ \text{subject to}~\sum_{s=1}^{S} \pi_{s}\eta_{s}=0.
\ese
Introducing additional latent variables $s_{t}\in\{1,\dots,S\}$ for each $t$, the model can be rewritten hierarchically as 
\bse
f(y\mid c_{t}=c, s_{t}=s) = \Normal(y \mid \mu_{c}+\eta_{s},\sigma_{s}^{2}),~~~~~p(s_{t}=s) = \pi_{s}.
\ese 
The states $s_{t}$'s model local departures from the state specific means $\mu_{c}$'s but are globally shared across all states. 
The moment restriction ensures that the marginal mean of each latent state $c$ is still $\mu_{c}$.  
The model is similar to that in \cite{Yau_etal:2011}, 
where they did not have any moment restriction on the globally shared components $\mu_{s}$ 
but assumed one local mean $\mu_{c}$ to be exactly known to identify the state specific means. 
We assign the priors
$\mu_{c} \sim \Normal(\mu_{0},\sigma_{0}^2)$, 
$\bpi_{\eta} = (\pi_{1},\dots,\pi_{S})\trans \sim \Dir(\alpha_{\eta}/S,\dots,\alpha_{\eta}/S), 
\eta_{s} \sim \Normal(\mu_{\eta,0},\sigma_{\eta,0}^{2})$, and 
$\sigma_{s}^{2} \sim \IG(a_{0},b_{0})$.

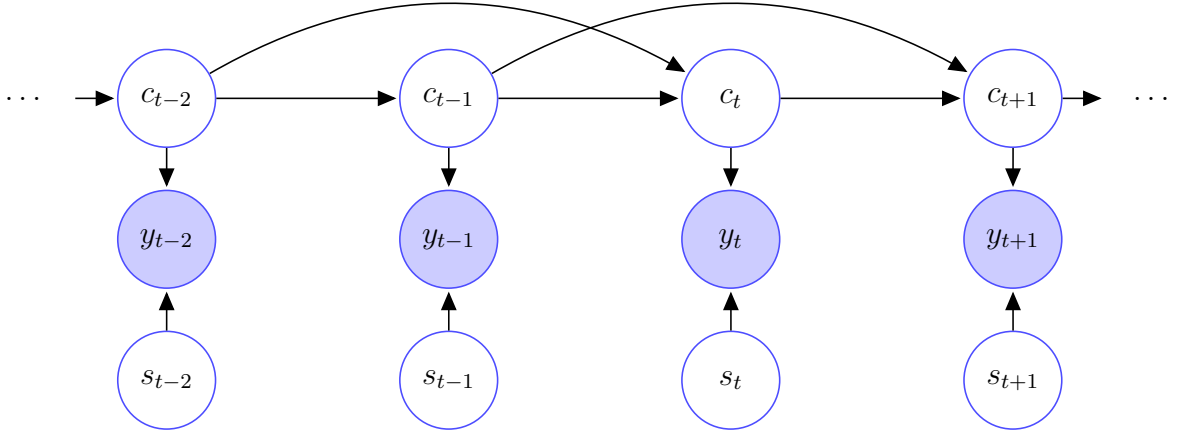
\begin{figure}[ht!]
{
\begin{tikzpicture}[scale=1.5,->,>=triangle 45,shorten >=1pt,auto,node distance=2.8cm,semithick]

  \node[style={draw=blue!70,circle}, minimum size=1.3cm] (c_t-2) at (0.5,0) {$c_{t-2}$};
  \node[style={draw=blue!70,circle}, minimum size=1.3cm] (c_t-1) at (3,0) {$c_{t-1}$};
  \node[style={draw=blue!70,circle}, minimum size=1.3cm] (c_t) at (5.5,0) {$c_{t}$};
  \node[style={draw=blue!70,circle}, minimum size=1.3cm] (c_t+1) at (8,0) {$c_{t+1}$};
  
  \node[style={draw=blue!70,circle}, minimum size=1.3cm] (s_t-2) at (0.5,-2.5) {$s_{t-2}$};
  \node[style={draw=blue!70,circle}, minimum size=1.3cm] (s_t-1) at (3,-2.5) {$s_{t-1}$};
  \node[style={draw=blue!70,circle}, minimum size=1.3cm] (s_t) at (5.5,-2.5) {$s_{t}$};
  \node[style={draw=blue!70,circle}, minimum size=1.3cm] (s_t+1) at (8,-2.5) {$s_{t+1}$};

  \node[style={draw=blue!70,circle,fill=blue!20}, minimum size=1.3cm] (y_t-2) at (0.5,-1.25) {$y_{t-2}$};
  \node[style={draw=blue!70,circle,fill=blue!20}, minimum size=1.3cm] (y_t-1) at (3,-1.25) {$y_{t-1}$};
  \node[style={draw=blue!70,circle,fill=blue!20}, minimum size=1.3cm] (y_t) at (5.5,-1.25) {$y_{t}$};
  \node[style={draw=blue!70,circle,fill=blue!20}, minimum size=1.3cm] (y_t+1) at (8,-1.25) {$y_{t+1}$};

  \node[minimum size=1.3cm] (z_1_t-2) at (-0.75,0) {$\dots$};
  \node[minimum size=1.3cm] (z_1_t+2) at (9.25,0) {$\dots$};

  \path (c_t-3) edge (c_t-2);
  \path (c_t-2) edge[bend left] (c_t);
  \path (c_t-2) edge (c_t-1);
  \path (c_t-1) edge (c_t);
  \path (c_t-1) edge[bend left] (c_t+1);
  \path (c_t) edge (c_t+1);
  \path (c_t+1) edge (c_t+2);
  
  \path (s_t-2) edge (y_t-2);
  \path (s_t-1) edge (y_t-1);
  \path (s_t) edge (y_t);
  \path (s_t+1) edge (y_t+1);
  
  \path (c_t-2) edge (y_t-2);
  \path (c_t-1) edge (y_t-1);
  \path (c_t) edge (y_t);
  \path (c_t+1) edge (y_t+1);
  
\end{tikzpicture}
}
\caption{Graphical model depicting the dependence structure of a second order hidden Markov model with translated mixtures as emission distributions.}
\label{fig: graph mixture emissions 2}
\end{figure}

\subsection{Likelihood Factorization} \label{sec: likelihood factorization}
Collecting all potential predictors of $c_{t}$ in $\bw_{t}=(w_{1,t},\ldots,w_{q,t})\trans$ with $w_{j,t} = c_{t-j}$ for $j=1,\ldots,q$ and $t=t^{\star},\dots,T$, where $t^{\star}=(q+1)$, 
the joint distribution of $\by=\{\by_{t}:t=1,\dots,T\}$, $\bc=\{c_{t}: t=t^{\star},\dots,T\}$ and $\bz=\{z_{j,t}: t=t^{\star},\dots,T,j=1,\dots,q\}$ admits the following factorization
\be
&&\hspace{-1cm} p(\by,\bc,\bz \mid \blambda_{\bk},\bpi_{\bk},\bk)   
	= \prod_{t=t^{\star}}^{T} \left\{p(c_{t} \mid \blambda_{\bz_{t}}) \prod_{j=1}^{q} p(z_{j,t} \mid w_{j,t},\bpi_{k_{j}}^{(j)},k_{j})\right\}   
				\prod_{t=1}^{T} f(\by_{t} \mid \bbeta_{c_{t}},\bpsi)       \nonumber\\
&& = \prod_{t=1}^{T} f(\by_{t} \mid \bbeta_{c_{t}},\bpsi)  \prod_{t=t^{\star}}^{T} \{p(c_{t} \mid \blambda_{\bz_{t}})  p(\bz_{t} \mid \bw_{t},\bpi_{\bk},\bk)\}    \nonumber\\
&& = p(\by\mid\bc,\bbeta,\bpsi)  p(\bc \mid \bz,\blambda_{\bk},\bk)  \prod_{j=1}^{q}p(\bz_{j} \mid \bw_{j}, \bpi_{k_{j}}^{(j)}, k_{j})    \nonumber\\
&& = p(\by\mid\bc,\bbeta,\bpsi)  p(\bc \mid \bz,\blambda_{\bk},\bk)  p(\bz \mid \bw, \bpi_{\bk}, \bk),  \label{eq: p(c,z,zstar) factorization}
\ee
Here $\bk=\{k_{j}: j=1,\dots,q\}$, $\blambda_{\bk}=\{\lambda_{h_{1},\dots,h_{q}}(c): c=1,\dots,C, h_{j}=1,\dots,k_{j}, j=1,\dots,q\}$, 
$\bpi_{k_{j}}^{(j)}(w_{j})=\{\pi_{h_{j}}^{(j)}(w_{j}): h_{j}=1,\dots,k_{j}\}$, 
$\bpi_{k_{j}}^{(j)}=\{\bpi_{k_{j}}^{(j)}(w_{j}): w_{j}=1,\dots,C\}$, 
$\bpi_{\bk}=\{\bpi_{k_{j}}^{(j)}: j=1,\dots,q\}$. 
Also, $\bz_{t}=\{z_{j,t}: j=1,\dots,q\}$ for all $t=t^{\star},\dots,T$, $\bz_{j}=\{z_{j,t}: t=t^{\star},\dots,T\}$ for $j=1,\dots,q$ and $\bw_{j}=\{w_{j,t}: t=t^{\star},\dots,T\}$. 
The subscripts $\bk$ and $k_{j}$ signify that the dimensions of the associated parameters depend on them.  
In what follows, the subscript $\bk$ may sometimes be dropped from $\blambda_{\bk}$ to highlight that, 
unlike $\bpi_{\bk}$, the support of the core probability vectors comprising $\blambda_{\bk}$ does not depend on $\bk$. 
The conditional independence relationships encoded in the factorization are used in deriving MCMC algorithms to draw samples from the posterior in Section \ref{sec: posterior computation}.

\subsection{{Posterior Consistency}} \label{sec: posterior consistency}
Consistency of the posterior of Bayesian first order HMMs under the frequentist assumption 
that there is a true fixed parameter has been studied in \cite{vernet2015rates, vernet2015posterior, gassiat2014posterior}. 
Asymptotic treatment of the posterior is facilitated under the assumptions of stationarity and uniform ergodicity of the underlying Markov chain, 
and some additional mild assumptions on the emission distributions. 
Specifically, it is assumed that the true transition probabilities as well as their priors are all bounded below by some positive number $\underline{p}$ 
\citep{vernet2015rates, vernet2015posterior}. 
Under similar assumptions on $p(c_{t} \mid \bc_{(t-q):(t-1)})$, such results can be extended to HOHMMs. 


In the following, we write $\pi_{c_{1},\dots,c_{q}}=\pi(c_{1},\dots,c_{q})$, $p_{c_{q+1} \mid c_{1},\dots,c_{q}} = p(c_{q+1} \mid c_{1},\dots,c_{q})$ and $f(\cdot \mid c)=f_{c}(\cdot)$. 
With some abuse of notation, we let $\btheta=(\bp,\b1f) \in \calP \times \F = \Theta$ collect the parameters of the model, 
where $\calP$ 
denotes the space of transition probabilities $\bp$
and $\F$ 
denotes the space of emission distributions $\b1f=(f_{1},\dots,f_{C})\trans$ with respect to some reference measure $\lambda$ on the observation space $\Y \subset \rR^{d}$ for some positive integer $d$.  
Let $\Pi = \Pi_{P} \times \Pi_{F}$, where $\Pi_{P}$ and $\Pi_{F}$ denote the priors on the transition probabilities and the emission distribution, respectively. 
Let $\Pi(\cdot \mid \by_{1:T})$ denote the corresponding posterior based on an observed sequence $\by_{1:T}$ of length $T$. 
Let 
\bse
f_{1:\ell}^{\star}(\by_{1},\dots,\by_{\ell} \mid \btheta) = \sum_{c_{1}=1}^{C}\dots\sum_{c_{\ell}=1}^{C} \pi_{c_{1},\dots,c_{q}} \prod_{t=q+1}^{\ell} p_{c_{t} \mid c_{(t-q)},\dots, c_{(t-1)}} \prod_{t=1}^{\ell} f_{c_{t}}(\by_{t}),
\ese 
denote the $\ell\th$ order marginal distribution of a stationary HOHMM. 
Let  
\bse
& D_{\ell}(\btheta_{0}, \btheta) = \int \abs{f_{1:\ell}^{\star}(\by_{1},\dots,\by_{\ell} \mid \btheta_{0}) - f_{1:\ell}^{\star}(\by_{1},\dots,\by_{\ell} \mid \btheta)} \lambda(d\by_{1})\cdots\lambda(d\by_{\ell}).
\ese 
For $\b1f,\b1f_{0} \in \F$, let $d(\b1f,\b1f_{0}) = \max_{c} \int {\abs{f_{c}(\by)-f_{c,0}(\by)}} \lambda(d\by)$ 
and $\N(\delta,\F,d)$ is the $\delta$-covering number of the set $\F$ with respect to the distance metric $d$. 
Let $\calP(\underline{p})$ denote the space of transition probability tensors supported on the compact set $\underline{p} \leq p(c_{t}\mid c_{(t-q),\dots,(t-1)}) \leq 1-(C-1)\underline{p}$ for some $0<\underline{p}<1/C$.
In the proposed tensor factorized formulation (\ref{eq: TFM1}), such a restriction can be imposed by assigning truncated Dirichlet priors on $\blambda_{h_{1},\dots,h_{q}}$.

\begin{Assmp} \label{assmp: 1}
\begin{enumerate}[label=\alph*.]
\item The true data generating process is a stationary HOHMM of maximal order $q$ with parameter $\btheta_{0}=(\bp_{0},\b1f_{0}) \in \Theta(\underline{p}) = \calP(\underline{p}) \times \F$. 
\item For all $\epsilon>0$, there exists $\Theta_{\epsilon}=\calP_{\epsilon} \times \F_{\epsilon} \subset \Theta(\underline{p})$ 
such that for all $(\bp,\b1f) \in \Theta_{\epsilon}$ 
	\begin{enumerate}
	\item $\Pi(\Theta_{\epsilon})=\Pi_{P}(\calP_{\epsilon}) \Pi_{F}(\F_{\epsilon})>0$,
	\item $\max_{\bc_{(t-q):t}}\abs{p_{c_{t} \mid c_{(t-q)}, \dots, c_{(t-1)}}-p_{c_{t} \mid c_{(t-q)},\dots, c_{(t-1)},0}} < \epsilon$,
	\item $\max_{c} \int f_{c,0}(\by)\max _{k} \log \frac{f_{k,0}(\by)}{f_{k}(\by)} \lambda(d\by) < \epsilon$,
	\item $\sum_{c}f_{c}(\by)>0$ whenever $\sum_{c}f_{c,0}(\by)>0$, 
	\item $\sup_{\{\by: \sum_{c}f_{c,0}(\by)>0\}} \max_{k} f_{k}(\by) < \infty$,
	\item $\sum_{c} \int f_{c,0}(\by) \abs{\log \{\sum_{k} f_{k}(\by)\}} \lambda(d\by) < \infty$,
	\end{enumerate}
\item For all $T$ and all $\epsilon>0$, there exists $\F_{T}\subset \F$ such that  
	\begin{enumerate}
	\item $\Pi_{F}(\F_{T}^{c}) \leq \exp(-T\beta_{1})$ for some $\beta_{1}>0$, 
	\item $\sum_{T=1}^{\infty} \N\left\{\frac{\epsilon}{36\ell},\F_{T},d(\cdot,\cdot)\right\} \exp\left(-\frac{T\epsilon^{2}C^{2}\underline{p}^{2}}{32\ell}\right) < \infty.$ 
	\end{enumerate}
\end{enumerate}
\end{Assmp}

\begin{Thm} \label{thm: consistency}
Under Assumptions \ref{assmp: 1}, for $\ell \geq q$, for any $\epsilon>0$, 
\bse
P_{\btheta_{0}} \left[\lim_{T \to \infty}  \Pi\{\btheta: D_{\ell}(\btheta_{0},\btheta) < \epsilon \mid \by_{1:T}\} =1 \right] =1.
\ese 
\end{Thm}

The proof of the theorem and some additional convergence results are discussed in Section \ref{appendix: proof of consistency} in the Supplementary Materials.  

%

\subsection{Prediction} \label{sec: predictions}
For a $q\th$ order HMM with state space $\C$, 
transition probabilities $p(c_{t} \mid c_{t-q},\ldots,c_{t-1})$
and emission distributions $\{f(y \mid c): c\in \C\}$, 
the $r$-step ahead density is 
\bse
&&\hspace{-1cm} f_{pred,T+r}(y_{})=\sum_{c_{T+r}} \sum_{c_{T+r-1}} \dots \sum_{c_{T+1}} f(y_{} \mid c_{T+r}) p(c_{T+r} \mid \bc_{(T+r-q):(T+r-1)}) \dots p(c_{T+1} \mid \bc_{(T+1-q):T}).
\ese
With stationary distribution $\bpi=\{\pi( c_{t-q+1},\ldots,c_{t}): c_{j} \in \C, j=t-q+1,\ldots,t \}$, for $r \to \infty$ we then have 
\bse
\sum_{c_{T+r-q}} \dots \sum_{c_{T+1}}  p(c_{T+r} \mid \bc_{(T+r-q):(T+r-1)}) \dots p(c_{T+1} \mid \bc_{(T+1-q):T}) \to \pi(\bc_{(T+r-q+1):(T+r)}).
\ese
The marginal probabilities of occurrences of individual states $i\in \C$, denoted with slight abuse of notation also by $\pi(i)$, may be obtained from $\bpi$ 
by fixing the last (or any other) element in $\bpi$ at $i$ and then summing across the values of the remaining entries.  
That is, $\pi(i) = \sum_{c_{t-q+1},\dots,c_{t-1}}\pi(c_{t-q+1},\dots,c_{t-1},i)$. 
Likewise, for any $(i,j)\in \C^{2}$, we have $\pi(i,j) = \sum_{c_{t-q+1},\dots,c_{t-2}}\pi(c_{t-q+1},\dots,c_{t-2},i,j)$. 
This implies, as $r \to \infty$ 
\bse
\sum_{c_{T+r-1}} \dots \sum_{c_{T+1}}  p(c_{T+r} \mid \bc_{(T+r-q):(T+r-1)}) \dots p(c_{T+1} \mid \bc_{(T+1-q):T}) \\
\to \sum_{c_{T+r-1}} \dots \sum_{c_{T+r-q+1}}  \pi(\bc_{(T+r-q+1):(T+r)})=\pi(c_{T+r}).
\ese
Hence, we have 
\vspace{-2ex}
\bse
f_{pred,T+r}(y_{})   \to   \sum_{c_{}}\pi(c_{})f(y_{} \mid c_{}).
\ese

Next, consider a first order HMM, characterized by the transition probabilities $\{P(j\mid i)=\pi(i,j) / \pi(j): i,j\in \C\}$, stationary distribution $\{\pi(i): i\in \C\}$ and emission distributions $\{f(y \mid c): c\in \C\}$. 
The $r$-step ahead density then approaches the same limit as $r\to \infty$. 
That is, we have 
\bse
&&\hspace{-1cm} f_{pred,T+r}(y_{})=\sum_{c_{T+r}} \sum_{c_{T+r-1}} \dots \sum_{c_{T+1}} f(y_{} \mid c_{T+r}) P(c_{T+r} \mid c_{T+r-1}) \dots P(c_{T+1} \mid c_{T})    \to   \sum_{c_{}}\pi(c_{})f(y_{} \mid c_{}).
\ese

As will be seen in Section \ref{sec: simulation experiments}, significant gains in efficiency in estimating several steps ahead predictive densities can be achieved through modeling higher order dynamics when such lags are truly present. 
As the number of steps ahead is increased, the performances of higher and comparable first order HMMs in estimating the predictive densities, will, however, tend to be similar. 
In both cases, as $r$ increases, the error in estimating $f_{pred,T+r}(y_{})$ will also tend to stabilize.

\subsection{Unknown Maximal Order and Unknown State Space} \label{sec: unknown order and state space}
In Sections \ref{sec: transition}-\ref{sec: predictions}, 
we assumed the maximal order and the size of the state space to be known. 
In practical applications, one or both of these quantities are often unknown. 
In this section, we devise practical strategies to relax these assumptions. 

We first relax the assumption of known maximal order, 
letting $q$ to be a known upper bound on the maximal order, 
and using the following prior. 
\be
p(k_{1},\dots,k_{q}) &\propto& 1\left\{\sum_{j=1}^{q}k_{j}>q\right\} \prod_{j=1}^{q} p_{0,j}(k),    \label{eq: prior on k_{j}'s 3}\\
p_{0,j}(k) &\propto& \exp(-\varphi jk),~~~j=1,\dots,q,~~~k=1,\dots,C.     \label{eq: prior on k_{j}'s 4}
\ee
The restriction $k_{q}>1$ is now replaced by $\sum_{j=1}^{q}k_{j}>q$ which ensures that at least one lag is important 
and the transition matrix corresponding to the true order has full rank. 
The proposed methodology automatically selects the important lags. 

To relax the assumption of known state space, we look toward Bayesian nonparametric models 
that can accommodate countably infinitely many states in the prior 
and allow the number of states required to model the data 
to be sampled and inferred from the posterior. 
Such models also accommodate the possibility that additional latent states may be required to allow the model to grow in complexity as more data points become available. 
To this end, the finite state space model for the latent sequence $\{c_{t}\}$ proposed in Section \ref{sec: transition} is extended to
\be
&& (c_{t}\mid z_{j,t}=h_{j},j=1,\dots,q)  \sim \blambda_{h_{1},\dots,h_{q}}, \label{eq: iHOHMM1}\\
&& \blambda_{h_{1},\dots,h_{q}}  \sim \DP(\alpha,\blambda_{0}),  ~~~~~~ \blambda_{0} \sim \SB(\alpha_{0}),  \label{eq: iHOHMM2}  \\
&& (z_{j,t}\mid c_{t-j})  \sim \bpi_{k_{j}}^{(j)}(c_{t-j}),  ~~~~~~ \bpi_{k_{j}}^{(j)}(c_{t-j}) \sim \Dir(\gamma_{j},\dots,\gamma_{j}),  \label{eq: iHOHMM3} \\
&& p(k_{1},\dots,k_{q}) \propto 1\left\{\sum_{j=1}^{q}k_{j}>q\right\} \prod_{j=1}^{q} p_{0,j}(k),~~~~~~ p_{0,j}(k_{j}) ~ \propto ~ \exp(-\varphi jk_{j}).    \label{eq: iHOHMM4}
\ee
Here $h_{j}=1,\dots,k_{j}; k_{j}=1,\dots,\infty; j=1,\dots,q$; $\DP(\alpha,\blambda_{0})$ denotes a Dirichlet process prior \citep{ferguson:1973} with concentration parameter $\alpha$ and base probability measure $\blambda_{0}$; 
and $\blambda_{0} \sim\SB(\alpha_{0})$ denotes the stick-breaking construction \citep{sethuraman:1994} of $\blambda_{0}=\{\lambda_{0}(1),\lambda_{0}(2),\dots\}$ as 
\bse
\lambda_{0}(\ell)=v_{\ell}\prod_{m=1}^{\ell-1}(1-v_{m}), ~v_{\ell}\sim \Beta(1,\alpha_{0}),~ \ell=1,2,\dots.
\ese
Equation (\ref{eq: iHOHMM2}) defines an HDP prior on the probability distributions $\blambda$. 
%
%
In the special case of a first order HMM, with $q=1$, the model reduces to 
\be
&& (c_{t}\mid z_{t}=h) \sim \blambda_{h}, \\
&& \blambda_{h} \sim \DP(\alpha,\blambda_{0}) ~\hbox{for}~h=1,\dots,k, ~~~~~~\blambda_{0} \sim \SB(\alpha_{0}) \\
&& (z_{t} \mid c_{t-1}) \sim \bpi_{k}(c_{t-1}), ~~~~~~\bpi_{k}(c_{t-1}) \sim \Dir(\gamma_{j},\dots,\gamma_{j}),  \\
&& p_{0}(k) ~ \propto ~ \exp(-\varphi k),~~~k=2,\dots, \infty.
\ee 
The HDP-HMM of \cite{teh_etal:2006} is obtained as a further special case if we let $k=\infty$, 
and $\pi_{h}(c)=1$ if $h=c$ and $0$ otherwise for all $h=1,\dots,\infty$. 
The proposed model thus generalizes the HDP-HMM in at least two directions. 
First, it models higher order transition dynamics that can also accommodate lag gaps. 
Second, even in the special first order setting, the soft allocation feature of the model, as opposed to the hard clustering in HDP-HMMs, 
enables better representation of the dynamics, resulting in improved estimation and prediction performance. 
See Section \ref{sec: simulation experiments}.

For moderate to large values of $C$, the finite dimensional prior $\Dir(\alpha/C,\dots,\alpha/C)$ provides a weak limit approximation to the infinite dimensional Dirichlet process prior \citep{ishwaran_zarepour:2002} in the sense that if 
$G_{C} = \sum_{\ell=1}^{C} \lambda_{\ell} \delta_{\theta_{k}}$ with $\blambda \sim \Dir(\alpha/C,\dots,\alpha/C)$ and $\theta_{k}\sim G_{0}$,
then, for any measurable function $g$ integrable with respect to $G_{0}$,  
$\int g(\theta) dG_{C}(\theta) \overset{d}{\to} \int g(\theta) dG(\theta)$ as $C\to \infty$,
where $G \sim \DP(\alpha,G_{0})$.
The finite state-space HOHMM model proposed in Section \ref{sec: transition} thus provides an excellent practical basis 
for approximate inference on integrable functionals of the infinite dimensional model (\ref{eq: iHOHMM1})-(\ref{eq: iHOHMM4}). 
In effect, as in the case of maximal order, having a known finite upper bound on the state space size suffices. 

\section{Posterior Computation} \label{sec: posterior computation}

In this article, inference about the proposed HOHMM is based on samples drawn from the posterior using MCMC algorithms. 
In our proposed HOHMM, the values of $k_{j}$'s, being crucial in controlling the model size and acting as lag selection indicators, 
are of inferential importance.   
Varying values of $k_{j}$'s, however, result in varying dimensional models, 
posing significant computational challenges. 
Dynamic message passing algorithms, such as the forward-backward sampler, are popular strategies for inference in first order HMMs. 
See \cite{Rabiner:1989} for a review and \cite{Scott:2002} for Bayesian MCMC based adaptations. 
More conventional strategies can be found in \cite{fruhwirth2001markov}. 
Such algorithms cannot, however, be straightforwardly adapted to the HOHMM setting.  
It is not clear how message passing strategies can be adapted to include inferences about the $k_{j}$'s. 
Even if the $k_{j}$'s are known, when higher order lags are present, 
computing forward or backward messages would require summing across all possible past histories 
comprising all possible combinations of values of important lags for each state at each time stamp at each iteration. 
This involves a prohibitively large number of operations.

We address these challenges by designing a two-stage MCMC algorithm.  
In the first stage, we sample the $k_{j}$'s from the posterior of a coarser version of the proposed model. 
This coarser version is itself fully capable of modeling any transition probability tensor with the $k_{j}$'s still being interpretable as lag selection indicators. 
In the second stage, we sample from the posterior keeping the $k_{j}$'s fixed. 
In what follows, the notation $\bzeta$ is used to collect all parameters and data points that are not explicitly mentioned.


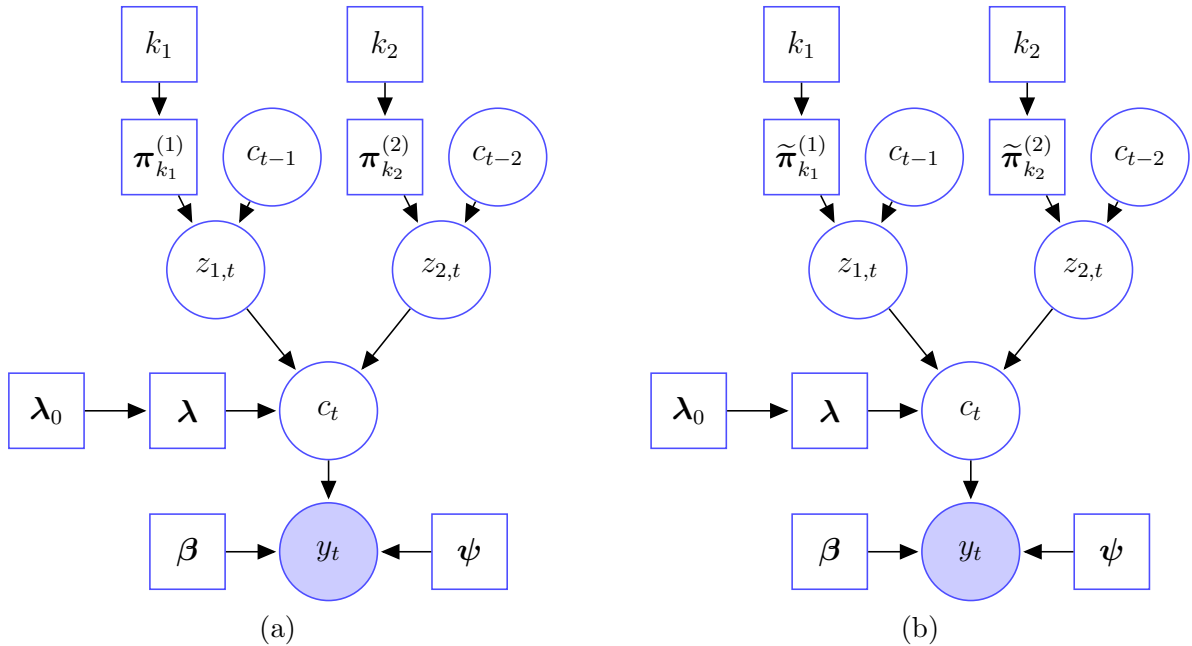
\begin{figure}[ht!]
\subfloat[][]
{
\centering
\begin{tikzpicture}[scale=1.5,->,>=triangle 45,shorten >=1pt,auto,node distance=2.8cm,semithick]

  \node[style={draw=blue!70,rectangle}, minimum size=1cm] (k_1) at (0,1) {$k_{1}$};
  \node[style={draw=blue!70,rectangle}, minimum size=1cm] (k_2) at (2,1) {$k_{2}$};

  \node[style={draw=blue!70,rectangle}, minimum size=1cm] (pi_1) at (0,0) {$\bpi_{k_{1}}^{(1)}$};
  \node[style={draw=blue!70,rectangle}, minimum size=1cm] (pi_2) at (2,0) {$\bpi_{k_{2}}^{(2)}$};

  \node[style={draw=blue!70,circle}, minimum size=1.3cm] (w_1_t) at (1,0) {$c_{t-1}$};
  \node[style={draw=blue!70,circle}, minimum size=1.3cm] (w_2_t) at (3,0) {$c_{t-2}$};

  \node[style={draw=blue!70,circle}, minimum size=1.3cm] (z_1_t) at (0.5,-1) {$z_{1,t}$};
  \node[style={draw=blue!70,circle}, minimum size=1.3cm] (z_2_t) at (2.5,-1) {$z_{2,t}$};

  \node[style={draw=blue!70,rectangle}, minimum size=1cm] (lambda) at (0.25,-2.25) {$\blambda$};
  \node[style={draw=blue!70,rectangle}, minimum size=1cm] (lambda_0) at (-1,-2.25) {$\blambda_{0}$};

  \node[style={draw=blue!70,circle}, minimum size=1.3cm] (c_t) at (1.5,-2.25) {$c_{t}$};

  \node[style={draw=blue!70,rectangle}, minimum size=1cm] (beta) at (0.25,-3.5) {$\bbeta$};
  
  \node[style={draw=blue!70,rectangle}, minimum size=1cm] (psi) at (2.75,-3.5) {$\bpsi$};

  \node[style={draw=blue!70,circle,fill=blue!20}, minimum size=1.3cm] (y_t) at (1.5,-3.5) {$y_{t}$};

  \path (k_1) edge (pi_1);
  \path (k_2) edge (pi_2);

  \path (pi_1) edge (z_1_t);
  \path (pi_2) edge (z_2_t);

  \path (z_1_t) edge (c_t);
  \path (z_2_t) edge (c_t);

  \path (w_1_t) edge (z_1_t);
  \path (w_2_t) edge (z_2_t);

  \path (lambda_0) edge (lambda);
  \path (lambda) edge (c_t);
  \path (c_t) edge (y_t);
  \path (beta) edge (y_t);
  \path (psi) edge (y_t);

\end{tikzpicture}
\label{fig: graph detailed (b)}
}
\hspace{1cm}
\subfloat[][]
{
\centering
\begin{tikzpicture}[scale=1.5,->,>=triangle 45,shorten >=1pt,auto,node distance=2.8cm,semithick]

  \node[style={draw=blue!70,rectangle}, minimum size=1cm] (k_1) at (0,1) {$k_{1}$};
  \node[style={draw=blue!70,rectangle}, minimum size=1cm] (k_2) at (2,1) {$k_{2}$};

  \node[style={draw=blue!70,rectangle}, minimum size=1cm] (pi_1) at (0,0) {$\wt\bpi_{k_{1}}^{(1)}$};
  \node[style={draw=blue!70,rectangle}, minimum size=1cm] (pi_2) at (2,0) {$\wt\bpi_{k_{2}}^{(2)}$};

  \node[style={draw=blue!70,circle}, minimum size=1.3cm] (w_1_t) at (1,0) {$c_{t-1}$};
  \node[style={draw=blue!70,circle}, minimum size=1.3cm] (w_2_t) at (3,0) {$c_{t-2}$};

  \node[style={draw=blue!70,circle}, minimum size=1.3cm] (z_1_t) at (0.5,-1) {$z_{1,t}$};
  \node[style={draw=blue!70,circle}, minimum size=1.3cm] (z_2_t) at (2.5,-1) {$z_{2,t}$};

  \node[style={draw=blue!70,rectangle}, minimum size=1cm] (lambda) at (0.25,-2.25) {$\blambda$};
  \node[style={draw=blue!70,rectangle}, minimum size=1cm] (lambda_0) at (-1,-2.25) {$\blambda_{0}$};

  \node[style={draw=blue!70,circle}, minimum size=1.3cm] (c_t) at (1.5,-2.25) {$c_{t}$};

  \node[style={draw=blue!70,rectangle}, minimum size=1cm] (beta) at (0.25,-3.5) {$\bbeta$};
  
  \node[style={draw=blue!70,rectangle}, minimum size=1cm] (psi) at (2.75,-3.5) {$\bpsi$};

  \node[style={draw=blue!70,circle,fill=blue!20}, minimum size=1.3cm] (y_t) at (1.5,-3.5) {$y_{t}$};

  \path (k_1) edge (pi_1);
  \path (k_2) edge (pi_2);

  \path (pi_1) edge (z_1_t);
  \path (pi_2) edge (z_2_t);

  \path (z_1_t) edge (c_t);
  \path (z_2_t) edge (c_t);

  \path (w_1_t) edge (z_1_t);
  \path (w_2_t) edge (z_2_t);

  \path (lambda_0) edge (lambda);
  \path (lambda) edge (c_t);
  \path (c_t) edge (y_t);
  \path (beta) edge (y_t);
  \path (psi) edge (y_t);

\end{tikzpicture}
\label{fig: graph detailed (a)}
}
\caption{Graphical model depicting the local dependency structure in a second order HMM at time point $t$.
The proposed model (a) and its approximated version (b) 
that forms the basis of Algorithm \ref{algo: approx MCMC} described in Section \ref{sec: posterior computation}. 
The difference between the two models lies in how the latent variables $z_{j,t}$'s are allocated to different latent clusters. 
The original model implements a soft clustering of $z_{j,t} \sim \bpi_{k_{j}}^{(j)}(c_{t-j})$ with $\bpi_{k_{j}}^{(j)}(c_{t-j}) \sim \Dir(\gamma_{j},\dots,\gamma_{j})$ for all $j, c_{t-j}$.  
The approximate version implements a hard clustering of $z_{j,t} \sim \wt\bpi_{k_{j}}^{(j)}(c_{t-j})$ with $\wt\pi_{h_{j}}^{(j)}(c_{t-j}) \in \{0,1\}$ for all $j,h_{j}$ and $c_{t-j}$. 
}
\end{figure}


\begin{algorithm} 
\caption{}
\label{algo: approx MCMC}
\begin{algorithmic}[1]

\Algphase{Updating the Latent State Sequence $\bc$ and the Latent Variables $\bz$}
\State
Given the current values $\bc$ and $\bz$, propose new values $\bc^{new}$ and $\bz^{new}$ according to 
$Q(\bc^{new},\bz^{new} \mid \bc,\bz,\bzeta) = Q(\bc^{new} \mid \bz,\bzeta)    Q(\bz^{new} \mid \bc^{new},\bzeta)$, 
where $Q(\bc \mid \bz,\bzeta) \propto \prod_{t} \blambda_{\bz_{t}}(c_{t}) p(\by_{t}\mid c_{t})$ 
and $Q(\bz \mid \bc, \bzeta) = \prod_{j=1}^{q} \wt\pi_{z_{j,t+j}}^{(j)}(c_{t})$. 
Accept the proposed values $\bc^{new}$ and $\bz^{new}$ with probability
\bse
\min \left\{  \left[\frac{\prod_{t} \blambda_{\bz_{t}^{new}}(c_{t}^{new}) }{\prod_{t} \blambda_{\bz_{t}}(c_{t})}     \frac{\prod_{t} \blambda_{\bz_{t}^{new}}(c_{t})}{\prod_{t} \blambda_{\bz_{t}}(c_{t}^{new})}\right]^{1/\T(m)}, 1\right\},
\ese 
with $\T_{0}$ and $\T(m)=\max\{\T_{0}^{1-m/m_{0}},1\}$ denoting the initial and the current annealing temperature, 
$m$ the current iteration number and $m_{0}$ the iteration number at which the temperature reduces to one. 

\Algphase{Updating $\bk$, the Cluster Mappings $\bpi_{\bk}$ and the Latent Variables $\bz$}
Given the current values of $k_{j}$ and the current clusters $\C=\{\C_{j,r}: j=1,\dots,q, r=1,\dots,k_{j}\}$, do the following for $j=1,\dots,q$.
\State
If $k_{j}<C$, propose to increase $k_{j}$ to $(k_{j}+1)$. 
If $k_{j}>1$, propose to decrease $k_{j}$ to $(k_{j}-1)$.
For $1<k_{j}<C$, the moves are proposed with equal probabilities. 
For $k_{j}=1$, the increase move is selected with probability $1$.
For $k_{j}=C$, the decrease move is selected with probability $1$. 
\State
If an increase move is proposed, randomly split a cluster into two. 
If a decrease move is proposed, randomly merge two clusters into a single one. 
\State
Accept the proposed moves with acceptance rates based on the approximated marginal likelihood (\ref{eq: marginal likelihood}) 
if $\sum_{\ell}k_{\ell}>q$.
Set the latent variables $\bz$ at the cluster allocation variables determined by the cluster mappings $z_{j,t+j} \sim \wt\bpi_{k_{j}}^{(j)}(c_{t})$.

\Algphase{Updating the Mixture Weights $\wt\bpi_{\bk}$}
\State
The parameters $\wt\bpi_{k_{j}}^{(j)}(w_{j})$ are determined by the cluster mappings. 

\Algphase{Updating the Transition Distribution Parameters $\blambda_{\bk}$ and $\blambda_{0}$}
\State
Sample the parameters $\blambda_{h_{1},\dots,h_{q}}$ and $\blambda_{0}$ as in Algorithm \ref{algo: MCMC}.

\Algphase{Updating the Parameters of the Emission Distribution}
\State
Sample the parameters $\bpsi$ and $\bbeta$ as in Algorithm \ref{algo: MCMC}.  
\end{algorithmic}
\end{algorithm}


\begin{algorithm} 
\caption{}
\label{algo: MCMC}
\begin{algorithmic}[1]
\vspace{0.2cm}

\Algphase{Updating the Latent State Sequence $\bc$}
\State
Sample the $c_{t}$'s from their multinomial full conditionals 
\bse
&&\hspace{-1cm} \textstyle p(c_{t}\mid \bzeta) \propto \blambda_{z_{1,t},\dots,z_{q,t}}(c_{t}) ~ \prod_{j=1}^{q}\pi^{(j)}_{z_{j,t+j}}(c_{t}) ~ f(\by_{t} \mid \bbeta_{c_{t}},\bpsi)).  
\ese

\Algphase{Updating the Second Level Latent Variables $\bz$}
\State
Sample the $z_{j,t}$'s from their multinomial full conditionals 
\bse
p(z_{j,t}=h \mid \bzeta,z_{\ell,t}=h_{\ell}, \ell \neq j) \propto \pi_{h}^{(j)}(w_{j,t}) \lambda_{h_{1},\dots,h_{j-1},h,h_{j+1},\dots,h_{q}}(c_{t}). 
\ese

\Algphase{Updating the Mixture Weights $\bpi_{\bk}$}
\State
Let $n_{j,w_{j}}(h_{j}) = \sum_{t}1\{w_{j,t}=w_{j},z_{j,t}=h_{j}\}$. 
Sample $\bpi_{k_{j}}^{(j)}(w_{j})$ as
\bse
\{\pi_{1}^{(j)}(w_{j}),\dots,\pi_{k_{j}}^{(j)}(w_{j})\} \mid \bzeta \sim \Dir\{\gamma_{j}+{n}_{j,w_{j}}(1),\dots,\gamma_{j}+{n}_{j,w_{j}}(k_{j})\}.
\ese

\Algphase{Updating the Transition Distribution Parameters $\blambda_{\bk}$ and $\blambda_{0}$} 
\State
Let  ${n}_{h_{1},\dots,h_{q}}(c) = \sum_{t}1\{z_{1,t}=h_{1}, \dots, z_{q,t}=h_{q}, c_{t}=c\}$. Sample the $\blambda_{h_{1},\dots,h_{q}}$'s as
\bse
\{\lambda_{h_{1},\dots,h_{q}}(1),\dots,\lambda_{h_{1},\dots,h_{q}}(C)\} \mid \bzeta \sim \Dir\{\alpha \lambda_{0}(1)+n_{h_{1},\dots,h_{q}}(1),\dots,\alpha \lambda_{0}(C)+n_{h_{1},\dots,h_{q}}(C)\}.
\ese

\State
For $\ell=1,\dots,n_{h_{1},\dots,h_{q}}(c)$, sample an auxiliary variable $x_{\ell}$ as
\bse
x_{\ell} \mid \bzeta \sim \Bern\left\{\frac{\alpha\lambda_{0}(c)}{\ell-1+\alpha\lambda_{0}(c)}\right\}.
\ese
Set $m_{h_{1},\dots,h_{q}}(c)=\sum_{\ell}x_{\ell}$. 

\State
Set $m_{0}(c)=\sum_{(h_{1},\dots,h_{q})}m_{h_{1},\dots,h_{q}}(c)$.
Sample $\blambda_{0}$ as
\bse
\{\lambda_{0}(1),\dots,\lambda_{0}(C)\} \mid \bzeta \sim \Dir\{\alpha_{0}/C+m_{0}(1),\dots,\alpha_{0}/C+m_{0}(C)\}. 
\ese

\Algphase{Updating the Parameters of the Emission Distribution}
\State
Sample the global parameters $\bpsi$ from their full conditionals 
\bse
\textstyle p(\bpsi \mid \bzeta) \propto p_{0}(\bpsi) \prod_{t=1}^{T}f(\by_{t} \mid \bbeta_{c_{t}},\bpsi).
\ese

\State
Sample the cluster specific parameters $\bbeta$ from their full conditionals 
\bse
\textstyle p(\bbeta_{c} \mid \bzeta) \propto p_{0}(\bbeta_{c}) \prod_{\{t:c_{t}=c\}}f(\by_{t} \mid \bbeta_{c},\bpsi).
\ese
\end{algorithmic}
\end{algorithm}

\begin{figure}[ht!]
\subfloat[Propose $\bc^{new}$ according to $(c_{t}^{new} \mid \bz_{t}) \propto \blambda_{\bz_{t}}(c_{t}^{new})p(\by_{t} \mid c_{t}^{new})$.]
{
\begin{tikzpicture}[scale=1.5,->,>=triangle 45,shorten >=1pt,auto,node distance=2.8cm,semithick]

  \node[style={draw=blue!70,circle}, minimum size=1.3cm] (c_t-2) at (0.5,0) {$c_{t-2}^{new}$};
  \node[style={draw=blue!70,circle}, minimum size=1.3cm] (c_t-1) at (3,0) {$c_{t-1}^{new}$};
  \node[style={draw=blue!70,circle}, minimum size=1.3cm] (c_t) at (5.5,0) {$c_{t}^{new}$};
  \node[style={draw=blue!70,circle}, minimum size=1.3cm] (c_t+1) at (8,0) {$c_{t+1}^{new}$};
  
  \node[style={draw=blue!70,circle,fill=blue!20}, minimum size=1.3cm] (y_t-2) at (0.5,-1.25) {$y_{t-2}$};
  \node[style={draw=blue!70,circle,fill=blue!20}, minimum size=1.3cm] (y_t-1) at (3,-1.25) {$y_{t-1}$};
  \node[style={draw=blue!70,circle,fill=blue!20}, minimum size=1.3cm] (y_t) at (5.5,-1.25) {$y_{t}$};
  \node[style={draw=blue!70,circle,fill=blue!20}, minimum size=1.3cm] (y_t+1) at (8,-1.25) {$y_{t+1}$};

  \node[minimum size=1.3cm] (z_1_t-2) at (-0.75,0) {$\dots$};
  \node[style={draw=blue!70,circle,fill=blue!20}, minimum size=1.3cm] (z_2_t-1) at (0.5,1.25) {$z_{2,t-1}$};
  \node[style={draw=blue!70,circle,fill=blue!20}, minimum size=1.3cm] (z_1_t-1) at (1.75,0) {$z_{1,t-1}$};
  \node[style={draw=blue!70,circle,fill=blue!20}, minimum size=1.3cm] (z_2_t) at (3,1.25) {$z_{2,t}$};
  \node[style={draw=blue!70,circle,fill=blue!20}, minimum size=1.3cm] (z_1_t) at (4.25,0) {$z_{1,t}$};
  \node[style={draw=blue!70,circle,fill=blue!20}, minimum size=1.3cm] (z_2_t+1) at (5.5,1.25) {$z_{2,t+1}$};
  \node[style={draw=blue!70,circle,fill=blue!20}, minimum size=1.3cm] (z_1_t+1) at (6.75,0) {$z_{1,t+1}$};
  \node[style={draw=blue!70,circle,fill=blue!20}, minimum size=1.3cm] (z_2_t+2) at (8,1.25) {$z_{2,t+2}$};
  \node[minimum size=1.3cm] (z_1_t+2) at (9.25,0) {$\dots$};

  \path (z_1_t-2) edge (c_t-2);
  \path (z_2_t-1) edge (c_t-1);
  \path (z_1_t-1) edge (c_t-1);
  \path (z_2_t) edge (c_t);
  \path (z_1_t) edge (c_t);
  \path (z_2_t+1) edge (c_t+1);
  \path (z_1_t+1) edge (c_t+1);

  \path[gray!40] (c_t-2) edge (z_1_t-1);
  \path[gray!40] (c_t-1) edge (z_1_t);
  \path[gray!40] (c_t) edge (z_1_t+1);
  \path[gray!40] (c_t+1) edge (z_1_t+2);
  
  \path[gray!40] (c_t-2) edge (z_2_t);
  \path[gray!40] (c_t-1) edge (z_2_t+1);
  \path[gray!40] (c_t) edge (z_2_t+2);

  \path (c_t-2) edge (y_t-2);
  \path (c_t-1) edge (y_t-1);
  \path (c_t) edge (y_t);
  \path (c_t+1) edge (y_t+1);
  
\end{tikzpicture}
\label{fig: MCMC proposals a}
}

\subfloat[Propose $\bz^{new}$ according to $z_{j,t+j}^{new} \sim \bpi_{k_{j}}^{(j)}(c_{t}^{new})$.]
{
\begin{tikzpicture}[scale=1.5,->,>=triangle 45,shorten >=1pt,auto,node distance=2.8cm,semithick]

  \node[style={draw=blue!70,circle,fill=blue!20}, minimum size=1.3cm] (c_t-2) at (0.5,0) {$c_{t-2}^{new}$};
  \node[style={draw=blue!70,circle,fill=blue!20}, minimum size=1.3cm] (c_t-1) at (3,0) {$c_{t-1}^{new}$};
  \node[style={draw=blue!70,circle,fill=blue!20}, minimum size=1.3cm] (c_t) at (5.5,0) {$c_{t}^{new}$};
  \node[style={draw=blue!70,circle,fill=blue!20}, minimum size=1.3cm] (c_t+1) at (8,0) {$c_{t+1}^{new}$};
  
  \node[style={draw=blue!70,circle,fill=blue!20}, minimum size=1.3cm] (y_t-2) at (0.5,-1.25) {$y_{t-2}$};
  \node[style={draw=blue!70,circle,fill=blue!20}, minimum size=1.3cm] (y_t-1) at (3,-1.25) {$y_{t-1}$};
  \node[style={draw=blue!70,circle,fill=blue!20}, minimum size=1.3cm] (y_t) at (5.5,-1.25) {$y_{t}$};
  \node[style={draw=blue!70,circle,fill=blue!20}, minimum size=1.3cm] (y_t+1) at (8,-1.25) {$y_{t+1}$};

  \node[minimum size=1.3cm] (z_1_t-2) at (-0.75,0) {$\dots$};
  \node[style={draw=blue!70,circle}, minimum size=1.3cm] (z_2_t-1) at (0.5,1.25) {$z_{2,t-1}^{new}$};
  \node[style={draw=blue!70,circle}, minimum size=1.3cm] (z_1_t-1) at (1.75,0) {$z_{1,t-1}^{new}$};
  \node[style={draw=blue!70,circle}, minimum size=1.3cm] (z_2_t) at (3,1.25) {$z_{2,t}^{new}$};
  \node[style={draw=blue!70,circle}, minimum size=1.3cm] (z_1_t) at (4.25,0) {$z_{1,t}^{new}$};
  \node[style={draw=blue!70,circle}, minimum size=1.3cm] (z_2_t+1) at (5.5,1.25) {$z_{2,t+1}^{new}$};
  \node[style={draw=blue!70,circle}, minimum size=1.3cm] (z_1_t+1) at (6.75,0) {$z_{1,t+1}^{new}$};
  \node[style={draw=blue!70,circle}, minimum size=1.3cm] (z_2_t+2) at (8,1.25) {$z_{2,t+2}^{new}$};
  \node[minimum size=1.3cm] (z_1_t+2) at (9.25,0) {$\dots$};

  \path[gray!40] (z_1_t-2) edge (c_t-2);
  \path[gray!40] (z_2_t-1) edge (c_t-1);
  \path[gray!40] (z_1_t-1) edge (c_t-1);
  \path[gray!40] (z_2_t) edge (c_t);
  \path[gray!40] (z_1_t) edge (c_t);
  \path[gray!40] (z_2_t+1) edge (c_t+1);
  \path[gray!40] (z_1_t+1) edge (c_t+1);

  \path (c_t-2) edge (z_1_t-1);
  \path (c_t-1) edge (z_1_t);
  \path (c_t) edge (z_1_t+1);
  \path (c_t+1) edge (z_1_t+2);
  
  \path (c_t-2) edge (z_2_t);
  \path (c_t-1) edge (z_2_t+1);
  \path (c_t) edge (z_2_t+2);

  \path[gray!40] (c_t-2) edge (y_t-2);
  \path[gray!40] (c_t-1) edge (y_t-1);
  \path[gray!40] (c_t) edge (y_t);
  \path[gray!40] (c_t+1) edge (y_t+1);

\end{tikzpicture}
\label{fig: MCMC proposals b}
}
\caption{Graphical model showing the mechanisms to propose new values of $\bc$ and $\bz$ in the Metropolis-Hastings step 
of the approximate sampler described in Section \ref{sec: posterior computation}.
Starting with Figure \ref{fig: graph 2}(b), the lighter edges above are ignored in the construction of the proposals. 
}
\label{fig: MCMC proposals}
\end{figure}
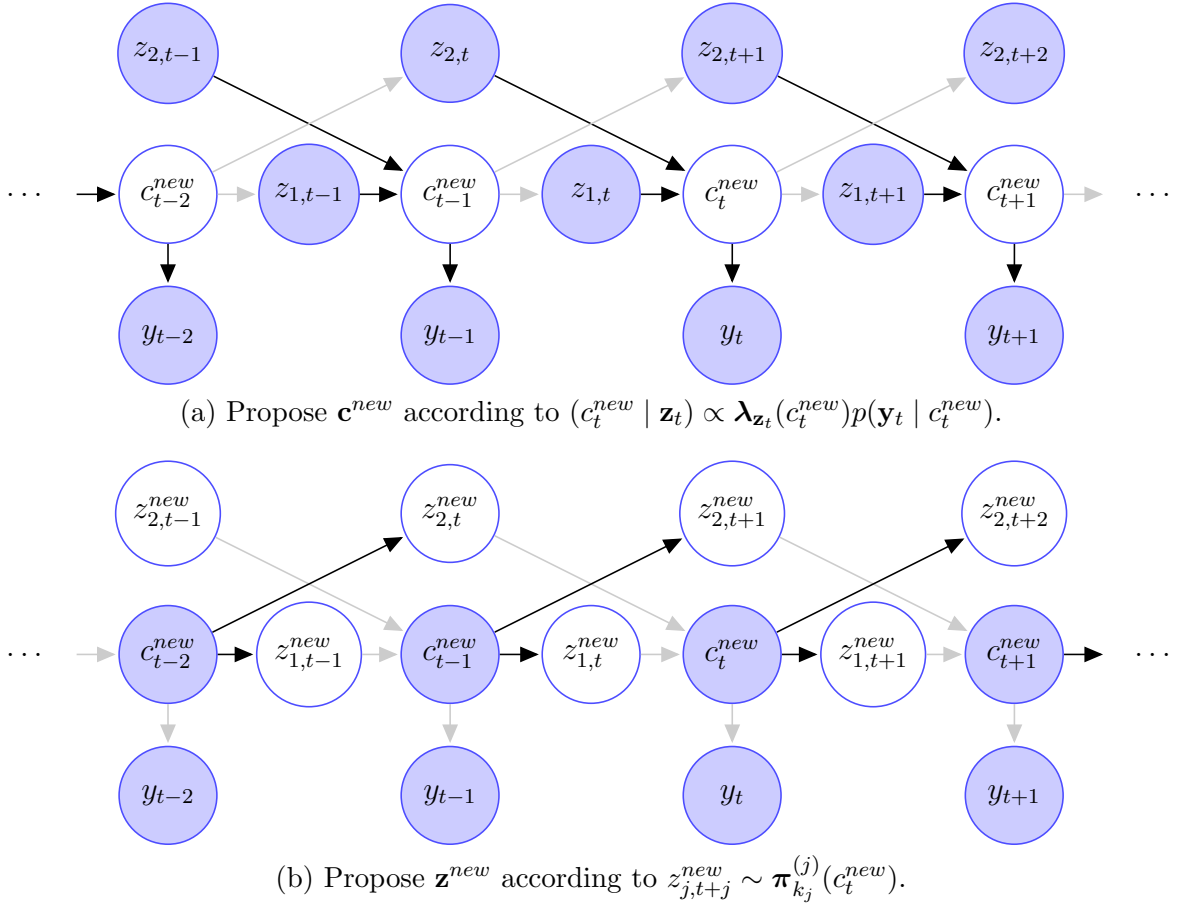

The first stage proceeds as follows. 
The mixture probabilities are now denoted by $\wt\bpi_{k_{j}}$, for reasons to become obvious shortly. 
Given the current values of $\bk$ and $\bc$, 
we partition the levels of $w_{j,t} = c_{t-j}$ into $k_{j}$ clusters $\{\C_{j,r}: r=1,\dots,k_{j}\}$ with each cluster $\C_{j,r}$ assumed to correspond to its own latent class $h_{j}=r$. 
The cluster mappings are then defined as $\wt\pi_{h_{j}}^{(j)}(c_{t-j})=1$ for $h_{j}=r$ and $\wt\pi_{h_{j}}^{(j)}(c_{t-j})=0$ otherwise. 
This imposes restrictions on soft allocation of the $z_{t,j}$'s, forcing the coarser hard allocation instead. 
With $\blambda_{h_{1},\dots,h_{q}} \sim \Dir\{\alpha\lambda_{0}(1),\dots,\alpha\lambda_{0}(C)\}$ marginalized out, 
conditional on the cluster configurations $\C=\{\C_{j,r}: j=1,\dots,q, r=1,\dots,k_{j}\}$, we then have 
\be
&& p(\bc \mid \C, \bzeta) =  \prod_{(h_{1},\dots,h_{q})}    \frac{\beta\{\alpha\lambda_{0}(1)+n_{h_{1},\dots,h_{q}}(1), \dots, \alpha\lambda_{0}(C)+n_{h_{1},\dots,h_{q}}(C)\}}{\beta\{\alpha\lambda_{0}(1),\dots,\alpha\lambda_{0}(C)\}}, \label{eq: marginal likelihood}
\ee
where $n_{h_{1},\dots,h_{q}}(c) = \sum_{t=t^{\star}}^{T}1\{c_{t}=c,w_{1,t}\in\C_{1,h_{1}},\dots,w_{q,t}\in\C_{m,h_{q}}\}$.
We then use an SSVS approach \citep{george_mcculloch:1997} based on the approximated marginal likelihood (\ref{eq: marginal likelihood}) to sample the $k_{j}$'s from their posterior. 
Conditional on $\bk$ and the current cluster mappings, 
we then update $\bc$ and $\bz$ using a Metropolis-Hastings step. 
We chose the proposal distributions that mimic their full conditionals 
and used simulated annealing to facilitate convergence. 
See Figure \ref{fig: MCMC proposals} and Algorithm \ref{algo: MCMC} for the second stage.
The parameters $\blambda$, $\blambda_{0}$, $\bbeta$ and $\bpsi$ are updated as in the second stage described in Algorithm \ref{algo: MCMC}.

Conditional on $\bk$, the elements of $\bc$, $\bz$ and $\bpi$ all have either multinomial or Dirichlet full conditionals and hence can be straightforwardly updated. 
Conditional on $\bc$ and $\bz$, the transition distributions $\blambda$ and $\blambda_{0}$ can be updated adapting existing computational machineries for sampling from the posterior in HDP models. 
The full conditionals of the parameters characterizing the emission distribution depend on the choice of the parametric family used to model the emission distribution 
but are usually straightforward to compute, often simplified with conjugate priors on the emission parameters. 
For fixed $k_{j}$'s, the parameters of the model can thus be easily sampled from the posterior. 
One such algorithm for finite state space HOHMM, with the auxiliary variable sampler for HDP \citep{teh_etal:2006} adapted to our setting, is outlined in Algorithm \ref{algo: MCMC}. 
A Chinese restaurant franchise process analog used to derive this algorithm is presented in the Supplementary Materials. 

Additionally, to make the approach even more data adaptive, hyper-priors can be assigned to the prior hyper-parameters $\alpha$, $\varphi$ etc. 
and they can also be sampled from the posterior. 
Priors and full conditionals used to update these hyper-parameters are outlined in the Supplementary Materials.


Specifics of the full conditionals of the emission distribution parameters in steps 7 and 8 of Algorithm \ref{algo: MCMC} depend on the choice of the family and the associated priors. 
The full conditionals are straightforward for the parametric Normal and Poisson families, and hence are omitted.  
%
%
For the translated normal family, the full conditionals of $\mu_{c}$, $\bpi_{\eta}$, $\sigma_{s}^{2}$ and $s_{t}$ are given by  
\bse
&& \textstyle (\mu_{c} \vert \bzeta) \sim \Normal(\mu_{c,T},\sigma_{c,T}^{2}),\\
&& \textstyle (\sigma_{s}^{2} \vert \bzeta) \sim \IG\big\{a_{0}+n_{s}/2, b_{0}+\sum_{t: s_{t}=s}(y_{t}-\mu_{c_{t}}-\eta_{s})^{2}/2\big\}, \\
&& (\bpi_{\eta} \vert \bzeta) \sim \Dir(\alpha_{\eta}/S+n_{1},\dots,\alpha_{\eta}/S+n_{S}),\\
&& p(s_{t}=s \vert  \bzeta) \propto \pi_{s} \times \Normal(y_{t}\vert\mu_{c_{t}}+\eta_{s},\sigma_{s}^{2}),
\ese
where $n_{s} = \sum_{t}1(s_{t}=s)$, 
$\sigma_{c,T}^{-2}=(\sigma_{0}^{-2}+\sum_{t:c_{t}=c}\sigma_{s_{t}}^{-2})$ and 
$\mu_{c,T}=\sigma_{c,T}^{-2}(\mu_{0}\sigma_{0}^{-2}+\sum_{t:c_{t}=c}y_{t}\sigma_{s_{t}}^{-2})$. 
Without the mean restriction, 
the posterior full conditional of $\etam^{S \times 1} = (\eta_{1},\dots,\eta_{S})\trans$ is given by
\vspace{-4ex}\\
\be
\MVN_{S} \left\{\left(\begin{array}{c}
\eta_{1}^{0} \\
\eta_{2}^{0}\\
\vdots\\
\eta_{S}^{0}
\end{array} \right),
\left(\begin{array}{c c c c}
\sigma_{11}^{0} & 0 & \dots & 0\\
0 & \sigma_{22}^{0} & \dots & 0\\
\vdots\\
0 & 0 & \dots & \sigma_{SS}^{0}
\end{array} \right)\right\} \equiv \MVN_{S} (\etam^{0},\bSigma^{0}),   \label{eq: joint mvt Normal posterior of the mean vector}
\ee
\vspace{-3ex}\\
where 
$\sigma_{ss}^{0} = (\sigma_{\eta,0}^{-2}+n_{s}\sigma_{s}^{-2})^{-1}$, 
$\eta_{s}^{0} =  \sigma_{ss}^{0} \left\{\sigma_{s}^{-2}\textstyle\sum_{t:s_{t}=s}(y_{t}-\mu_{c_{t}})+\sigma_{\eta,0}^{-2}\mu_{\eta,0}\right\}$. 
The posterior full conditional of $\etam$ under the mean restriction can then be obtained easily by further conditioning (\ref{eq: joint mvt Normal posterior of the mean vector}) by $\eta_{R} = \sum_{s=1}^{S}\pi_{s} \eta_{s} = 0$ 
and is given by 
\vspace{-5ex}\\
\be
(\etam \mid \eta_R = 0, \bzeta)  \sim \MVN_{S}\{\etam^{0} - \bsigma_{1,R}^{0}(\sigma_{R,R}^{0})^{-1}\eta_R^{0},\bSigma^{0} - \bsigma_{1,R}^{0}(\sigma_{R,R}^{0})^{-1}\bsigma_{R,1}^{0}\}, \label{eq: conditional mvt Normal posterior of the mean vector}
\ee
\vspace{-5ex}\\
where $\eta_{R}^{0} = \sum_{s=1}^{S}\pi_{s} \eta_{0} = E(\eta_{R})$, 
$\sigma_{s,S+1} = \pi_{s}\sigma_{ss}^{0} = \cov(\eta_{s},\eta_{R})$, 
$\sigma_{R,R}^{0} = \sum_{s=1}^{S}\pi_{s}^{2}\sigma_{ss}^{0} = \cov(\eta_{R})$, 
and $\bsigma_{R,1}^{0} = ( \sigma_{1,S+1}, \sigma_{2,S+1}, \dots, \sigma_{S,S+1} )=\bsigma_{1,R}^{0\text{T}}$.
To sample from this singular density,
we can first sample from the non-singular distribution of $\{(\eta_{1},\eta_{2},\dots,\eta_{S-1})\trans \mid \eta_{R} = 0\}$, 
which can also be trivially obtained from (\ref{eq: conditional mvt Normal posterior of the mean vector}), and then set $\eta_{S} = - \sum_{s=1}^{S-1}\pi_{s}\eta_{s}/\pi_{S}$.

\section{Simulation Experiments} \label{sec: simulation experiments}

We designed simulation experiments to evaluate the performance of our method in a wide range of scenarios. 

For the latent state dynamics, we considered the cases 
(A) $[3, \{1\}]$, 
(B) $[3, \{1,2,3\}]$, 
(C) $[3, \{1,2,4\}]$,  
(D) $[3, \{1,3,5\}]$,    
(E) $[3, \{1,4,8\}]$,
where $[C_{0}, \{i_{1},\dots,i_{r}\}]$ means that the latent sequence has $C_{0}$ categories and $\{c_{t-i_{1}},\dots,c_{t-i_{r}}\}$ are the important lags.  
In each case, we considered two sample sizes $T=500, 1000$. 
To generate the true transition probability tensors, for each combination of the true lags, 
we first generated the probability of the first response category as $f(u_{1})=u_{1}^{2}/\{u_{1}^{2}+(1-u_{1})^{2}\}$ with $u_{1} \sim \Unif(0,1)$. 
The probabilities of the remaining categories are then generated via a stick-breaking type construction as $f(u_{2}) \{1-f(u_{1})\}$ with $u_{2}\sim \Unif(0,1)$ and so on, 
until the next to last category $(C-1)$ is reached. 
The hyper-parameters were set at 
$\alpha_{0}=1$, 
and $\gamma_{j}=1/C$ for all $j$.  
In each case, we set the maximal number of states at $C=10$ and the maximal lag at $q=10$.

We considered (1) Normal, (2) Poisson and (3) translated mixtures of Normals emission distributions. 
For the Gaussian case $f(y\mid c_{t}=c) = \Normal(y \mid \mu_{c},\sigma_{c}^{2})$, 
we set $\mu_{c}=-2,0,2$ for $c=1,2,3$, respectively, and $\sigma_{c}^{2}=0.5^{2}$ for all $c$.
While the $\sigma_{c}^{2}$'s were all equal and could be treated as a global parameter,
we allowed the component specific variances to be different in the fitted model.
The hyper-parameters of the priors were set at $\mu_{0}=\overline{\by}, \sigma_{0}^{2}=3\var(\by), a_{0}=b_{0}=1$.  
For Poisson emission distributions $f(y\mid c_{t}=c) = \Poi(y \mid \mu_{c})$, we let $\mu_{c}=1,8,15$ for $c=1,2,3$, respectively.
The hyper-parameters of the Gamma prior on $\mu_{c}$ were chosen such that $E(\mu_{c})=a_{0}b_{0}=\overline{\by}, \var(\mu_{c})=a_{0}b_{0}^{2}=2\var(\by)$.  
For translated Gaussian mixture emission distributions $f(y\mid c_{t}=c) =  \sum_{s=1}^{S}\pi_{s}  \Normal(y \mid \mu_{c}+\eta_{s},\sigma_{s}^{2})$, 
we set $\mu_{c}=-4,0,4$ for $c=1,2,3$; $\pi_{s}=0.2,0.5,0.3,\eta_{s}=-2,0,1.33$ for $s=1,2,3$; and $\sigma_{s}^{2}=0.5^{2}$ for all $s$.
As in the case of simpler Gaussian emissions, 
even though $\sigma_{s}^{2}$'s were all equal in the true data generating mechanism, they were allowed to be different in the fitted model.

In each case, we initialized the latent states $\bc$ applying a k-means clustering algorithm to $\by$ with $k=C=5$ states.  
With $k_{1}=2$ and $k_{j}=1$ for $j=2,\dots,10$, initially only the first lag was chosen to be important. 
The parameters $\blambda,\blambda_{0},\bz,\bpi$ etc. were then initialized by randomly sampling from the prior generative model. 
For Normal emission distributions, $(\mu_{c},\sigma_{c}^{2})$'s were set at the corresponding empirical values; 
and for Poisson and translated Normal emissions, $\mu_{c}$'s were set at the corresponding empirical means. 
For translated Normal emissions, the indices $\bs$ and $(\mu_{s},\sigma_{s}^{2})$'s were likewise set using a k-means algorithm applied to $(y_{t}-\mu_{c_{t}})$ with $k=S=5$. 
In each case, the mean parameters associated with the remaining $5$ states were spread over the range of $\by$.  
For Normal emissions, the remaining $\sigma_{c}^{2}$'s were set at $\var(\by)$. 
In numerical experiments, we found the results to be very robust to the choice of the prior hyper-parameters and parameter and latent variable initializations. 

We coded in MATLAB\footnote{Codes implementing our method will be included as part of the Supplementary Materials once the paper is accepted for publication.}.
For the case (D1) described above, with $T=500$ data points, 
$5,000$ MCMC iterations required approximately $30$ minutes on an ordinary laptop.
For the $m\th$ iteration of the first algorithm, the annealing temperature was set at $\T(m)=\max\{\T_{0}^{1-m/m_{0}},1\}$ with $\T_{0}=1000$ and $m_{0}=1000$.
In each case, we discarded the first $2000$ iterations as burn-in. 
The remaining samples were thinned by retaining every $5\th$ sample after the burn-in to reduce auto-correlation.
The resulting samples showed good mixing and convergence behavior in diagnostic checks and produced stable estimates of the parameters of interest.

\begin{table}[ht!]
\begin{center}
\footnotesize
\begin{tabular}{|c|c|c c c|c c c|}
\hline
 \multirow{3}{40pt}{True Dynamics} 	& \multirow{3}{35pt}{Sample Size}	& \multicolumn{6}{|c|}{Median ISE $\times 100$} \\ \cline{3-8}
							& 		& \multicolumn{3}{|c|}{HDP-HMM} 	& \multicolumn{3}{|c|}{CTF-HOHMM} 	\\ \cline{3-8}
							&		& One 	& Two 	& Three 	& One 		& Two		&Three 	  	\\  \hline\hline
\multicolumn{8}{|c|}{Normal Emission Distribution} \\ \hline \hline
 \multirow{2}{*}{(A) $3, \{1\}$}		& 500	& 0.37	& 0.20 	& 0.23 	& 0.37~(0.33)	& 0.19 (0.17) 	& 0.20 (0.19)   	\\
							& 1000	& 0.16	& 0.13	& 0.11	& 0.11~(0.11)	& 0.11 (0.11) 	& 0.09 (0.10)    \\\cline{1-8}

 \multirow{2}{*}{(B) $3, \{1,2,3\}$}	& 500	& 10.96	& 8.46	& 4.21 	& 0.82		 & 0.67 & 0.65\\
							& 1000	& 4.47	& 5.04 	& 3.26 	& 0.32		& 0.31 & 0.20 \\\cline{1-8}

 \multirow{2}{*}{(C) $3, \{1,2,4\}$}	& 500	& 11.39	& 8.46 	& 6.66	& 0.88		& 0.80 & 0.65 \\
							& 1000	& 9.52	& 6.51 	& 5.98 	& 0.31		& 0.30 & 0.29 \\\cline{1-8}

 \multirow{2}{*}{(D) $3, \{1,3,5\}$}	& 500	& 10.91	& 11.14	& 9.38 	& 0.71		& 0.69 & 0.57 \\
							& 1000	& 15.60	& 11.24	& 9.21 	& 0.30		& 0.35 & 0.32 \\ \hline
\hline
\multicolumn{8}{|c|}{Poisson Emission Distribution} \\ \hline \hline
 \multirow{2}{*}{(A) $3, \{1\}$}		& 500	& 0.15	& 0.22 	& 0.10 	& 0.09~(0.07)	& 0.07~(0.05) 	& 0.05~(0.04)   	\\
							& 1000	& 0.30	& 0.26	& 0.17	& 0.09~(0.04)		&  0.05~(0.04) 	& 0.03 (0.03)    \\\cline{1-8}

 \multirow{2}{*}{(B) $3, \{1,2,3\}$}	& 500	& 3.02	& 1.92 	& 1.77 	& 1.17	& 0.61 	& 0.65 	\\
							& 1000	& 1.99	& 1.65  	& 1.32	& 0.32	& 0.14  	& 0.09 \\\cline{1-8}

 \multirow{2}{*}{(C) $3, \{1,2,4\}$}	& 500	& 2.58	& 1.75 	& 1.62	& 0.57	& 0.47 	& 0.45 \\
							& 1000	& 3.61	& 2.41 	& 1.50  	& 0.72	& 0.51	& 0.32 \\\cline{1-8}

 \multirow{2}{*}{(D) $3, \{1,3,5\}$}	& 500	& 4.72	& 1.71	& 2.71 	& 2.41	& 1.26	& 0.78 \\
							& 1000	& 5.08	& 2.37	& 3.24 	& 1.03	& 0.76 	& 0.53  \\ \hline
\hline
\multicolumn{8}{|c|}{Translated Mixture Normal Emission Distribution} \\ \hline \hline
 \multirow{2}{*}{(A) $3, \{1\}$}		& 500	& 0.35	& 0.27 	& 0.27 	& 0.39~(0.35)	&  0.29~(0.29) 	&  0.27~(0.27)   \\
							& 1000	& 0.21	& 0.15	& 0.11	& 0.23~(0.22)	&  0.20~(0.15) 	&  0.12~(0.12)   \\\cline{1-8}

 \multirow{2}{*}{(B) $3, \{1,2,3\}$}	& 500	& 5.39	& 3.40	& 2.99 	& 1.25	& 1.09 	& 0.85 \\
							& 1000	& 4.09	& 3.84 	& 2.60 	& 0.53	& 0.47  	& 0.31 \\\cline{1-8}

 \multirow{2}{*}{(C) $3, \{1,2,4\}$}	& 500	& 6.55	& 3.88 	& 3.37	& 1.44	& 1.401	& 1.06 \\
							& 1000	& 3.51	& 3.00 	& 2.84 	& 0.58	& 0.46 	& 0.43 \\\cline{1-8}

 \multirow{2}{*}{(D) $3, \{1,3,5\}$}	& 500	& 7.80	& 5.15 	& 2.81 	& 3.06	& 1.76 	& 1.21 \\
							& 1000	& 7.20	& 3.77 	& 3.19 	& 0.53	& 0.59  	& 0.47 \\\cline{1-8}
\hline
\end{tabular}
\caption{\baselineskip=10pt Median ISEs in estimating one, two and three step ahead predictive densities 
for the conditional tensor factorization (CTF) based HOHMM compared with the HDP-HMM. 
In the first column, $C_{0}, \{i_{1},\dots,i_{r}\}$ means that the latent sequence truly has $C_{0}$ categories and $\{c_{t-i_{1}},\dots,c_{t-i_{r}}\}$ are the true important lags. 
In the rows corresponding to the first order case (A) $3,\{1\}$, 
the numbers within parenthesis in the CTF-HOHMM columns show the estimated median MISEs with the maximal order set at $q=1$. 
In all other cases, $q=5$.
See Section \ref{sec: simulation experiments} for additional details.
}
\label{tab: MISEs}
\end{center}
\end{table}

\begin{table}[ht!]
\begin{center}
\footnotesize
\begin{tabular}{|c|c|c|c|}
\hline
 \multirow{2}{40pt}{True Dynamics} 	& \multirow{2}{35pt}{Sample Size}	& \multicolumn{2}{|c|}{Median Hamming Distance $\times 100$} \\ \cline{2-4}
							& 		&{HDP-HMM} 	& {CTF-HOHMM} 	\\ \hline \hline
\multicolumn{4}{|c|}{Normal Emission Distribution} \\ \hline \hline
 \multirow{2}{*}{(A) $3, \{1\}$}		& 500	& 3.77	&  2.56~(2.23)    	\\
							& 1000	& 2.57	&  2.10~(2.00)	\\\cline{1-4}

 \multirow{2}{*}{(B) $3, \{1,2,3\}$}	& 500	& 19.02	& 2.43	\\
							& 1000	& 16.37	& 2.24 	\\\cline{1-4}

 \multirow{2}{*}{(C) $3, \{1,2,4\}$}	& 500	& 20.46	& 2.50	\\
							& 1000	& 19.82	& 2.19 	\\\cline{1-4}

 \multirow{2}{*}{(D) $3, \{1,3,5\}$}	& 500	& 16.45	& 2.33	\\
							& 1000	& 12.01	& 2.30	\\ \hline
\hline
\multicolumn{4}{|c|}{Poisson Emission Distribution} \\ \hline \hline
 \multirow{2}{*}{(A) $3, \{1\}$}		& 500	& 11.01	& 8.72 (8.10)	 \\
							& 1000	& 12.78	& 8.12 (8.01)	\\\cline{1-4}

 \multirow{2}{*}{(B) $3, \{1,2,3\}$}	& 500	& 23.45	& 12.66 	\\
							& 1000	& 21.50	& 8.92  	\\\cline{1-4}

 \multirow{2}{*}{(C) $3, \{1,2,4\}$}	& 500	& 23.29	& 10.95	 \\
							& 1000	& 22.11	& 8.91 	\\\cline{1-4}

 \multirow{2}{*}{(D) $3, \{1,3,5\}$}	& 500	& 21.35	& 14.26	\\
							& 1000	& 21.03	& 10.96	\\ \hline
\hline
\multicolumn{4}{|c|}{Translated Mixture Normal Emission Distribution} \\ \hline \hline
 \multirow{2}{*}{(A) $3, \{1\}$}		& 500	& 4.83	&  7.31~(5.80)	\\
							& 1000	& 4.38	&  6.81~(5.58)	\\\cline{1-4}

 \multirow{2}{*}{(B) $3, \{1,2,3\}$}	& 500	& 19.21	& 7.02	\\
							& 1000	& 24.01	& 6.92 	\\\cline{1-4}

 \multirow{2}{*}{(C) $3, \{1,2,4\}$}	& 500	& 17.36	& 10.38 	\\
							& 1000	& 20.59	& 5.93 	\\\cline{1-4}

 \multirow{2}{*}{(D) $3, \{1,3,5\}$}	& 500	& 14.09	& 15.88	\\
							& 1000	& 13.81	& 5.32 	\\\cline{1-4}
\hline
\end{tabular}
\caption{\baselineskip=10pt Median Normalized Hamming distances between the true and the estimated state sequences 
for the conditional tensor factorization (CTF) based HOHMM and the HDP-HMM. 
In the first column, $C_{0}, \{i_{1},\dots,i_{r}\}$ means that the latent sequence truly has $C_{0}$ categories and $\{c_{t-i_{1}},\dots,c_{t-i_{r}}\}$ are the true important lags. 
In the rows corresponding to the first order case (A) $3,\{1\}$, 
the numbers within parenthesis in the CTF-HOHMM columns show the estimated median MISEs with the maximal order set at $q=1$. 
In all other cases, $q=5$.
See Section \ref{sec: simulation experiments} for additional details.
}
\label{tab: Hamming Ds}
\end{center}
\end{table}

\begin{figure}[!ht]
\begin{center}
\includegraphics[height=10cm, width=15cm, trim=0cm 0cm 0cm 0cm, clip=true]{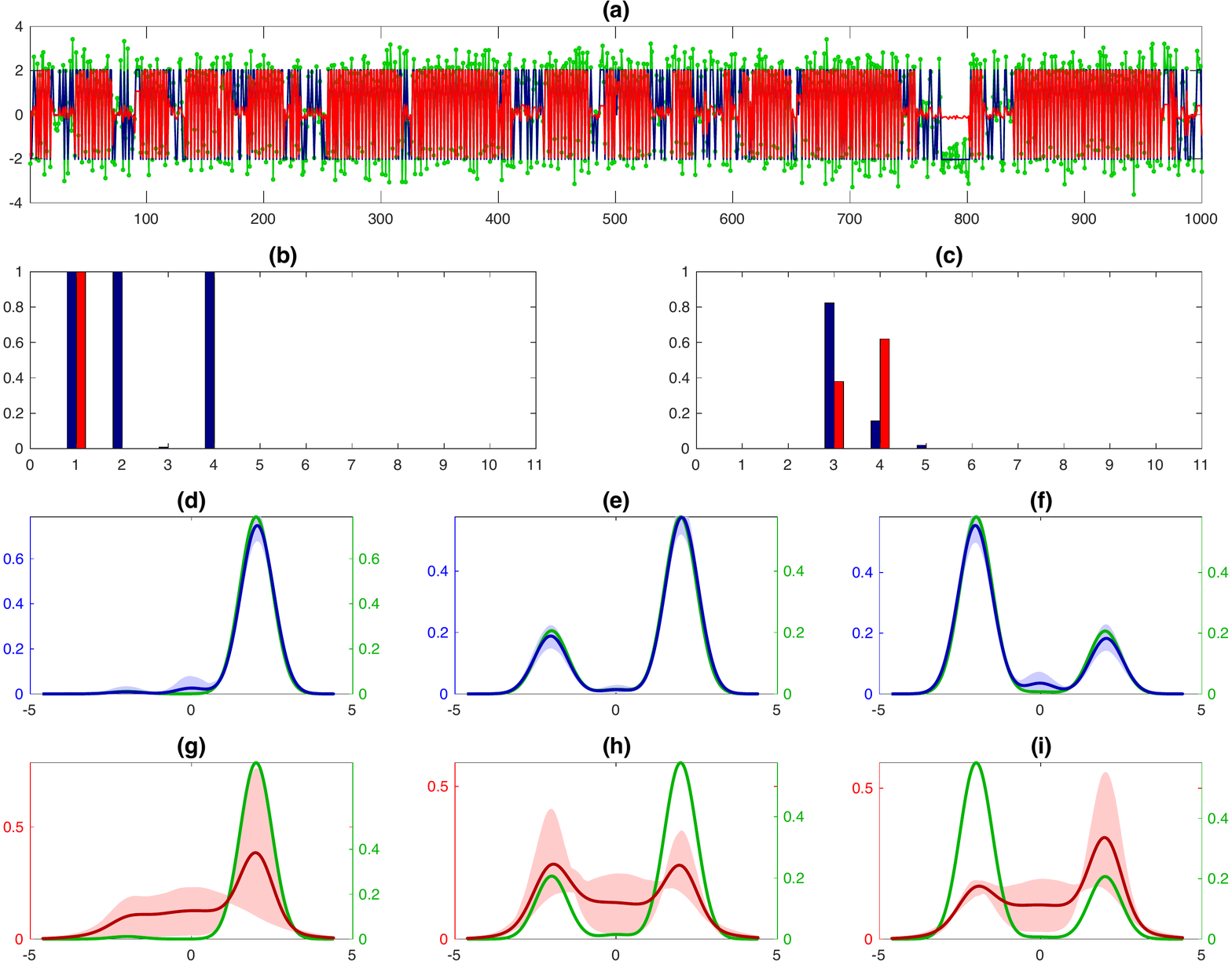}
\end{center}
\caption{\baselineskip=10pt 
Results for a synthetic dataset in the (D1) case described in Section \ref{sec: simulation experiments} (transition dynamics $3 \{1,2,4\}$ and Normal emission densities):  
CTF-HOHMM in blue and HDP-HMM in red 
(a) posterior means super-imposed over the observed time series in green; 
(b) the inclusion probabilities of different lags; 
(c) the distribution of different number of states; 
(d), (e), (f) estimated one, two and three steps ahead predictive densities, respectively, and their 90\% credible intervals by the CTF-HOHMM; 
(g), (h), (i) estimated one, two and three steps ahead predictive densities, respectively, and their 90\% credible intervals by the HDP-HMM. 
The true predictive densities are shown in green. 
}
\label{fig: simulation D1}
\end{figure}

We evaluated the performance of our proposed model and the HDP-HMM in estimating one, two and three-step ahead predictive densities. 
For an HOHMM of maximal order $q$, the r-step ahead predictive density is given by
\bse
&&\hspace{-1cm} {f}_{pred,T+r}(y_{}|\by_{1:T}) = E_{P(\tiny{\bzeta},\bc \mid \by_{1:T})} p(y_{} \mid \bc,\bzeta) = \int p(y_{} \mid \bc,\bzeta)  dP(\bzeta,\bc \mid \by_{1:T})   \\
&&\hspace{-1cm} = E_{P(\tiny{\bzeta},\bc \mid \by_{1:T})} \left[ \sum_{c_{T+r}} \sum_{c_{T+r-1}} \dots \sum_{c_{T+1}} f(y_{} \mid c_{T+r},\bzeta) p(c_{T+r} \mid \bc_{(T+r-q):(T+r-1)},\bzeta) \dots p(c_{T+1} \mid \bc_{(T+1-q):T},\bzeta) \right].
\ese
Based on $M$ samples $\{(\bc_{1:T}^{(m)},\bzeta^{(m)})\}_{m=1}^{M}$ drawn from the posterior, $f_{pred,T+r}(y_{}|\by_{1:T})$ can be estimated as   
\bse
&&\hspace{-1cm} \wh{f}_{pred,T+r}(y_{}|\by_{1:T}) = M^{-1} \sum_{m=1}^{M} \sum_{c_{T+r}} \sum_{c_{T+r-1}} \dots \sum_{c_{T+1}} f(y_{} \mid c_{T+r},\bzeta^{(m)}) \\
&& ~~~~~~~~~~~~~~~p(c_{T+r} \mid \bc_{(T+r-q):(T+r-1)}^{(m)},\bzeta^{(m)}) \dots p(c_{T+1} \mid \bc_{(T+1-q):T}^{(m)},\bzeta^{(m)}),
\ese
where $\bc_{(T+r-q):(T+r-1)}^{(m)}=(c_{T+r-q}^{(m)},c_{T+r-q+1}^{(m)},\dots,c_{T}^{(m)},c_{T+1},\dots,c_{T+r-1})$ for all $(r,m)$. 
The corresponding true density is given by 
\bse
&&\hspace{-1cm} f_{pred,T+r,0}(y_{})=\sum_{c_{T+r,0}} \sum_{c_{T+r-1,0}} \dots \sum_{c_{T+1,0}} f_{0}(y_{} \mid c_{T+r,0}) \\
&& ~~~~~~~~~~~~~~~p_{0}(c_{T+r,0} \mid \bc_{(T+r-q):(T+r-1),0}) \dots p_{0}(c_{T+1,0} \mid \bc_{(T+1-q):T,0}),
\ese
where $p_{0}$ and $f_{0}$ are generics for the true transition and emission distributions, respectively, with associated true parameters implicitly understood 
and $\bc_{0}=\bc_{1:T,0}$ denoting the true values of the latent sequence $\bc$.  
For continuous emission distributions, the integrated squared error (ISE) in estimating $f_{pred,T+r,0}(y_{})$ 
is estimated by $\sum_{i=1}^{N} \{f_{pred,T+r,0}(y_{i}^{\Delta})-\wh{f}_{pred,T+r}(y_{i}^{\Delta} \mid \by_{1:T})\}^{2}\Delta_{i}$, 
where $\{y_{i}^{\Delta}\}_{i=0}^{N}$ are a set of grid points on the range of $y$ and $\Delta_{i}=(y_{i}^{\Delta}-y_{i-1}^{\Delta})$ for all $i$. 
For Poisson emission distribution, the ISE is estimated as $\sum_{i=\max\{0,\min \by-1\}}^{\max\by+1} \{f_{pred,T+r,0}(i)-\wh{f}_{pred,T+r}(i\mid \by_{1:T})\}^{2}$.

We also evaluated the Hamming distance between the true and the estimated state sequences by the proposed HOHMM and the HDP-HMM. 
To calculate the Hamming distance, we used the Munkres algorithm \citep{munkres:1957}, 
mapping the indices of the estimated state sequence to the set of indices that maximize the overlap with the true sequence.

\begin{figure}[!ht]
\begin{center}
\includegraphics[height=10cm, width=15cm, trim=0cm 0cm 0cm 0cm, clip=true]{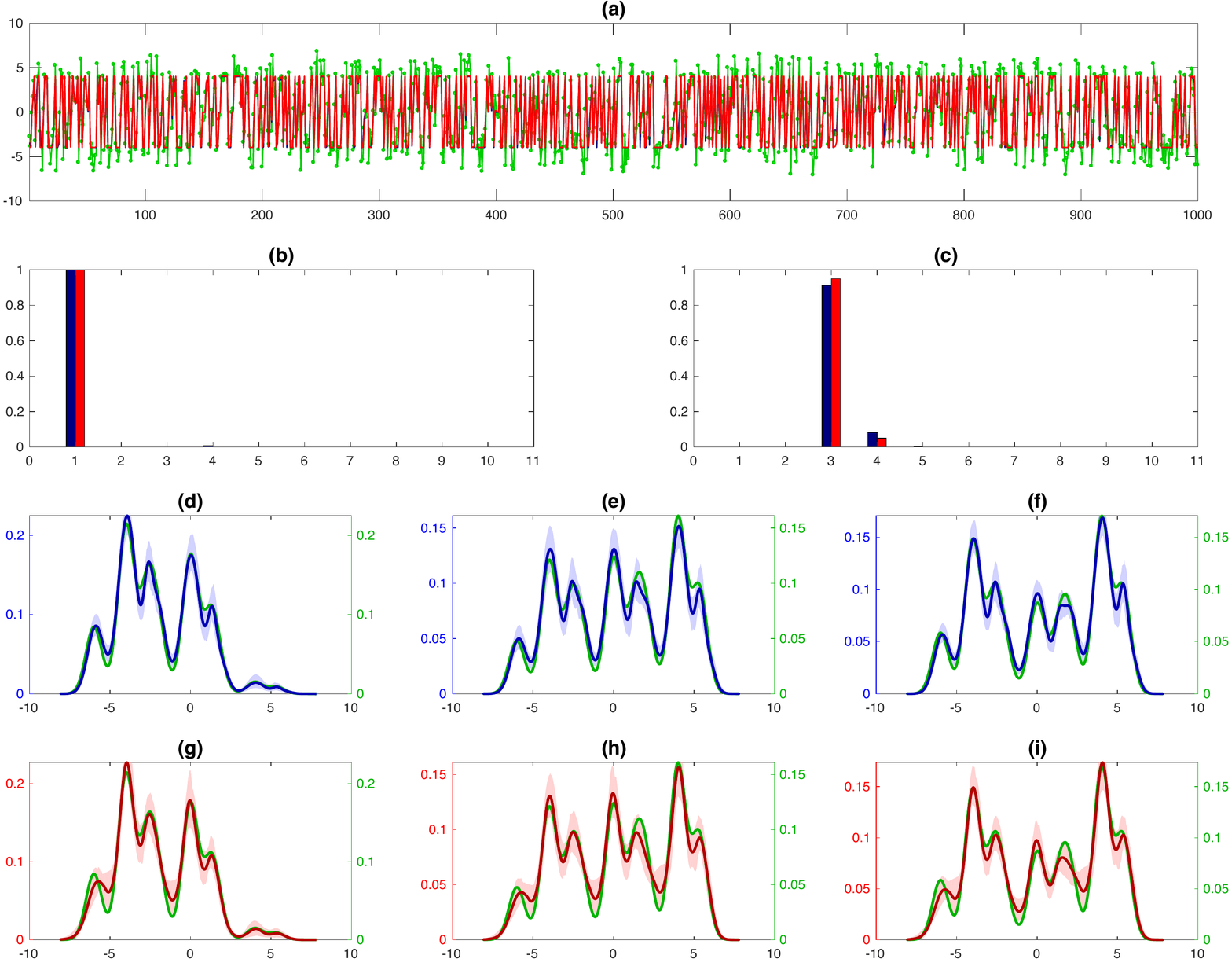}
\end{center}
\caption{\baselineskip=10pt 
Results for a synthetic dataset in the (A3) case described in Section \ref{sec: simulation experiments} (transition dynamics $3 \{1\}$ and translated Normal emission densities):    
CTF-HOHMM in blue and HDP-HMM in red 
(a) posterior means super-imposed over the observed time series in green; 
(b) the inclusion probabilities of different lags; 
(c) the distribution of different number of states; 
(d), (e), (f) estimated one, two and three steps ahead predictive densities, respectively, and their 90\% credible intervals by the CTF-HOHMM; 
(g), (h), (i) estimated one, two and three steps ahead predictive densities, respectively, and their 90\% credible intervals by the HDP-HMM. 
The true predictive densities are shown in green. 
}
\label{fig: simulation A3}
\end{figure}

The performances in estimating the one, two and three step ahead predictive densities and in clustering the observations $y_{t}$ 
are summarized in Tables \ref{tab: MISEs} and \ref{tab: Hamming Ds}, respectively. 
The reported results are based on $100$ simulated datasets in each case. 
The proposed approach vastly outperformed the HDP-HMM in the higher order cases and remarkably also in the first order parametric cases. 
Figure \ref{fig: simulation D1} summarizes the results for the data set corresponding to the median ISE in estimating the one-step ahead predictive density for the HOHMM 
in the (D1) case with $T=1000$. 
Panel (a) in Figure \ref{fig: simulation D1} suggests that the CTF-HOHMM provides a better fit to local variations in the dataset. 
The improvements in higher order cases are explained by the HDP-HMM's restrictive first order assumption. 
The proposed method, on the other hand, not just accommodates higher order lags, 
but also effectively eliminates the unnecessary ones, while also characterizing the dynamics using efficient sparse representations.  
As explained in Section \ref{sec: unknown order and state space}, 
the improvements in the first order cases can be attributed to 
this ability to effectively eliminate the unnecessary lags in correctly identifying the true first order dynamics 
and then sparsely characterizing the dynamics using soft allocation of the latent states, enabling better data compression than the hard allocation model implemented in HDP-HMM. 

The remarkable efficiency of the proposed HOHMM method even when the maximal lag is set at large conservative values 
is also seen from comparisons of the results when the maximal lag was set at $10$ 
with the results reported in parentheses in Tables \ref{tab: MISEs} and \ref{tab: Hamming Ds} 
that were produced by its first order restriction by prefixing the maximal lag at one.  
See Figure \ref{fig: simulation A3} that summarizes the results for the data set corresponding to the median ISE for the one-step ahead predictive density in the (A3) case with $T=1000$. 
%
%
Table \ref{tab: MISEs} shows that with increase in the prediction step 
the performance in estimating the predictive density improved for both the HDP-HMM and the proposed HOHMM. 
As the prediction step increases, 
the true and the estimated predictive densities approach the corresponding stationary distributions which are invariant to $c_{T-i_{1}+1},\dots,c_{T-i_{r}+1}$ 
and the error stabilizes. 
Improved estimation of the latent states, as evident from the estimated Hamming distances, 
can likewise be explained by the aforementioned novel aspects of the proposed HOHMM.

\section{Applications} \label{sec: applications}
In this section, we discuss results of the proposed CTF-HOHMM applied to a few real datasets. 
The datasets discussed here are all available publicly. 
In each case, we compare with HDP-HMM. 
Unless otherwise mentioned, each model is allowed a maximum of $C=10$ states; 
the HOHMM was allowed a maximal lag of $q=10$; 
and the model parameters were all initialized as in the simulation experiments.

\textbf{Old Faithful Geyser Data: }
We first consider the Geyser dataset, accompanying \cite{mcdonald_zuchhini:1997} and also available from the \texttt{MASS} package in \texttt{R}. 
The dataset comprises 299 sequential observations on eruption and waiting times (in minutes) of the Old Faithful geyser in Yellowstone National Park in the USA collected continually from August 1 to August 15, 1985. 
We focus here on modeling duration times using HMMs with Normal emission distributions. 
Empirical explorations of the dataset earlier in Section 3 in \cite{azzalini_bowman:1990} had suggested second order dynamics. 

\begin{figure}[!h]
\begin{center}
\includegraphics[height=10cm, width=15cm, trim=0cm 0.5cm 0cm 0.5cm, clip=true]{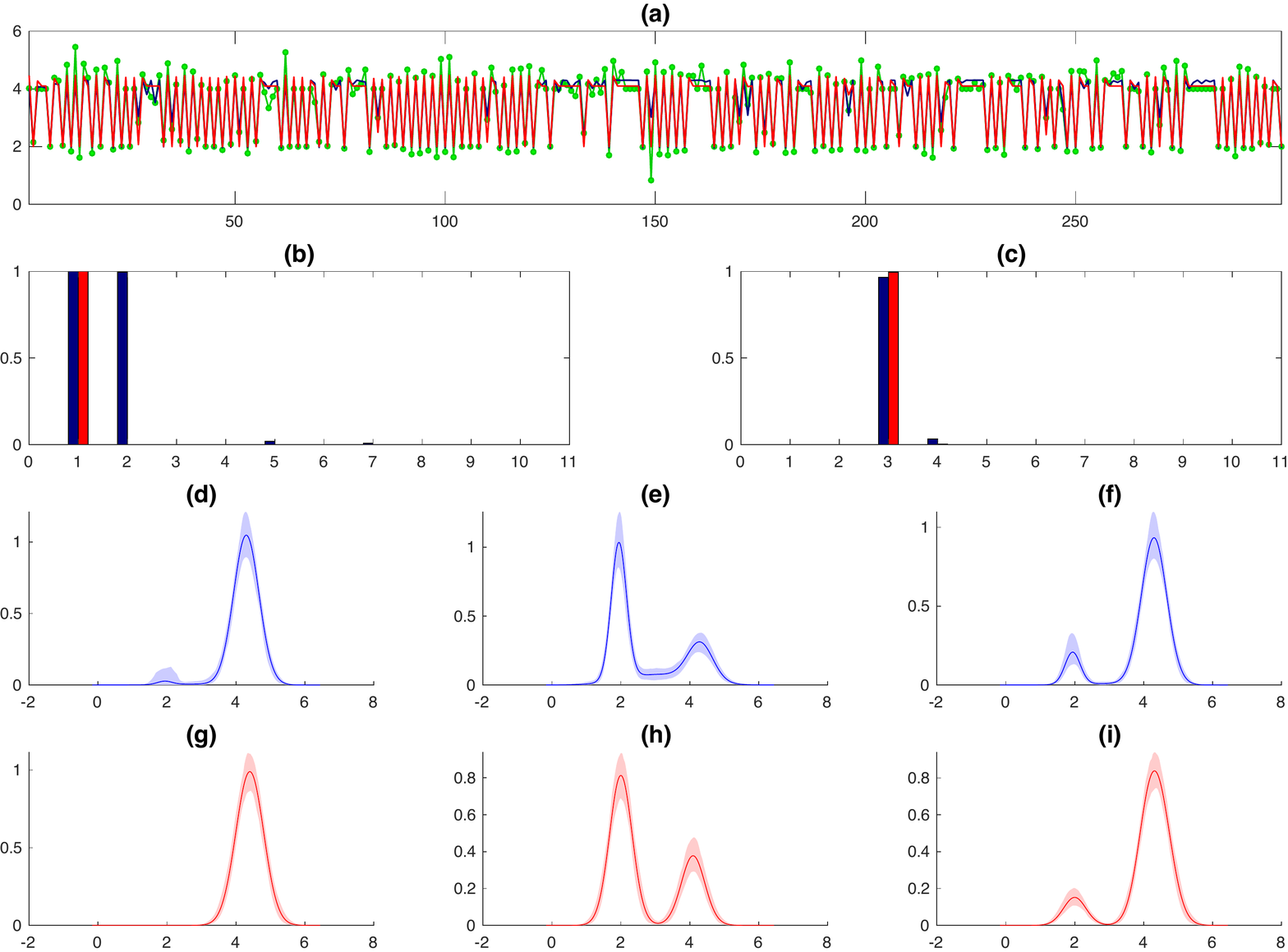}
\end{center}
\caption{\baselineskip=10pt 
Results for the Geyser dataset: CTF-HOHMM in blue and HDP-HMM in red 
(a) posterior means super-imposed over the observed time series in green; 
(b) the inclusion probabilities of different lags; 
(c) the distribution of different number of states; 
(d), (e), (f) estimated one, two and three steps ahead predictive densities, respectively, and their 90\% credible intervals by the CTF-HOHMM; 
(g), (h), (i) estimated one, two and three steps ahead predictive densities, respectively, and their 90\% credible intervals by the HDP-HMM. 
}
\label{fig: geyser duration times}
\end{figure}

Figure \ref{fig: geyser duration times} summarizes the results. 
Results produced by HDP-HMM and CTF-HOHMM are in general agreement, both models suggesting a three state dynamics. 
The results returned by HOHMM, however, suggest a second order HMM to provide the best fit, 
consistent with \cite{azzalini_bowman:1990}.

\begin{figure}[ht]
\begin{center}
\includegraphics[height=11cm, width=15cm, trim=3cm 2.25cm 3cm 1cm, clip=true]{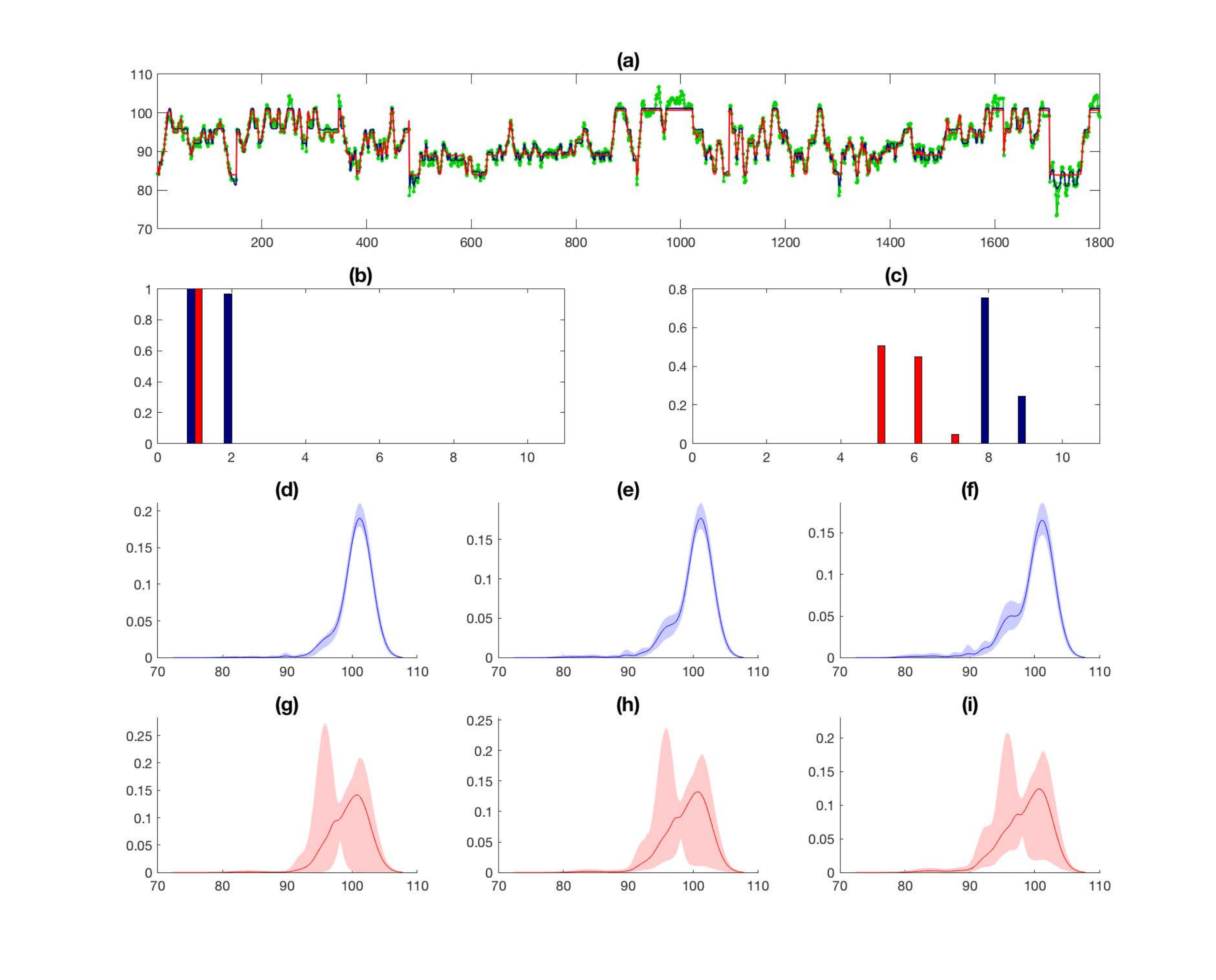}
\end{center}
\caption{\baselineskip=10pt 
Results for the MIT heart dataset: CTF-HOHMM in blue and HDP-HMM in red 
(a) posterior means super-imposed over the observed time series in green; 
(b) the inclusion probabilities of different lags; 
(c) the distribution of different number of states; 
(d), (e), (f) estimated one, two and three steps ahead predictive densities, respectively, and their 90\% credible intervals by the CTF-HOHMM; 
(g), (h), (i) estimated one, two and three steps ahead predictive densities, respectively, and their 90\% credible intervals by the HDP-HMM. 
}
\label{fig: heart1}
\end{figure}

\textbf{MIT Heart Data: } 
Next, we consider MIT heart data, a collection of 4 time series. 
The first two series contains 1800 evenly-spaced measurements of instantaneous heart rate from different subjects performing similar activities. 
The measurements (in units of beats per minute) occur at 0.5 second intervals over a 15 minute interval. 

Figure \ref{fig: heart1} summarizes the results for the series 1 dataset obtained by the CTF-HOHMM and the HDP-HMM with Normal emission distributions. 
The HDP-HMM results show uncertainty around the number of underlying latent states, suggesting a mixture of 5 and 6 latent states. 
The CTF-HOHMM results suggest second order dependencies. 
Like the HDP-HMM, CTF-HOHMM also accommodates uncertainty in the number of states, suggesting however a mixture of 8 and 9 states. 
Panel (a) in Figure \ref{fig: heart1} suggests that the CTF-HOHMM provides a better fit to local variations in the dataset. 
The predictive densities estimated by the two methods also look substantially different. 

The series 2 in the MIT heart dataset shows strong signs of irregular periodicity. 
HMMs are not suitable for modeling periodicity without additional modifications. 
We have thus not pursued modeling series 2. 
The series 3 and 4 were recorded in the same way but contain 950 measurements each, corresponding to 7 minutes and 55 seconds of data in each case.
CTF-HOHMM applied to these two datasets suggests first order dependencies in both cases. 
Results produced by HDP-HMM and CTF-HOHMM, not presented here, were very similar for these two series.

\textbf{E.coli Data: }
Next, we consider the E.coli data set available from \texttt{tscount} package in \texttt{R}. 
This dataset comprises weekly counts of reported disease cases caused by Escherichia coli in the state of North RhineWestphalia
(Germany) from January 2001 to May 2013.

\begin{figure}[!ht]
\begin{center}
\includegraphics[height=10cm, width=15cm, trim=0cm 0cm 0cm 0cm, clip=true]{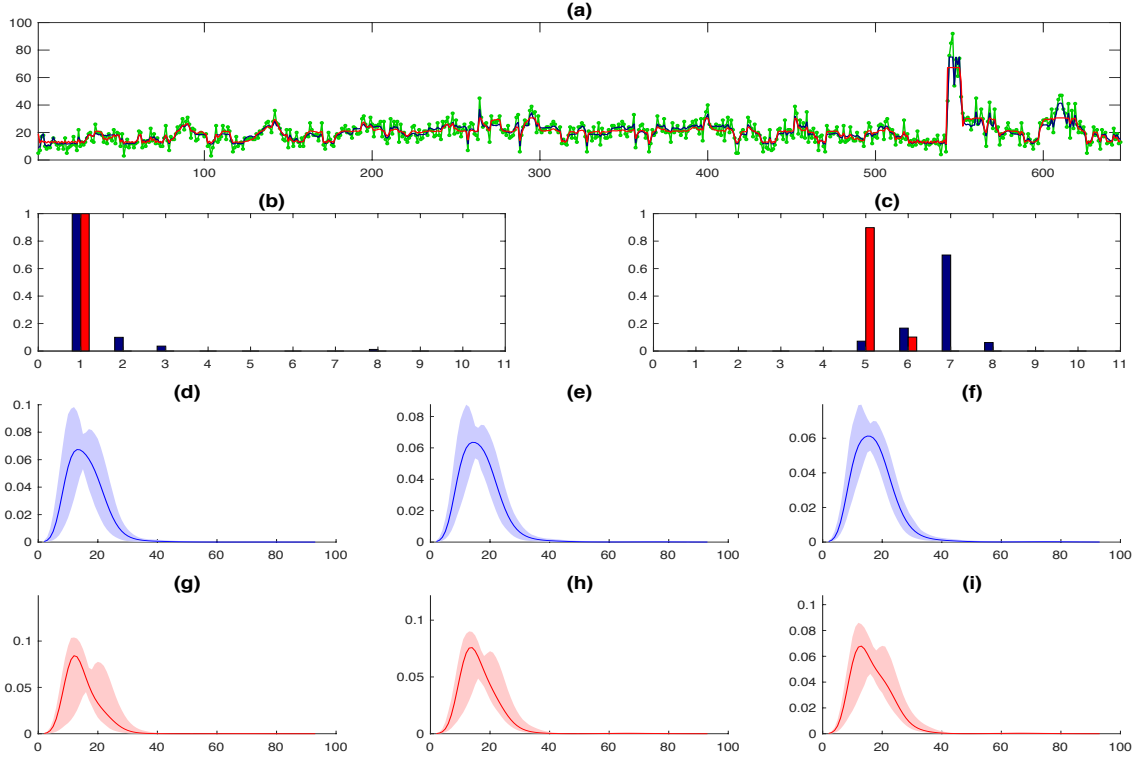}
\end{center}
\caption{\baselineskip=10pt 
Results for the E.coli dataset: CTF-HOHMM in blue and HDP-HMM in red 
(a) posterior means super-imposed over the observed time series in green; 
(b) the inclusion probabilities of different lags; 
(c) the distribution of different number of states; 
(d), (e), (f) estimated one, two and three steps ahead predictive densities, respectively, and their 90\% credible intervals by the CTF-HOHMM; 
(g), (h), (i) estimated one, two and three steps ahead predictive densities, respectively, and their 90\% credible intervals by the HDP-HMM. 
}
\label{fig: E.coli}
\end{figure}

Figure \ref{fig: E.coli} summarizes the results obtained by CTF-HOHMM and HDP-HMM with Poisson emission distributions. 
The HDP-HMM results suggests 5 latent states. 
The CTF-HOHMM results suggests a first order dynamics but 7 latent states. 
Panel (a) in Figure \ref{fig: E.coli} suggests that the CTF-HOHMM provides a better fit to local variations in the data. 
The one, two and three steps ahead predictive densities, however, look similar. 
 
\textbf{Coriell aCGH Data: }
Array comparative genomic hybridization (aCGH) studies are used to investigate the identification of DNA polymorphisms - 
deletions or sequences deviations, and duplications. 
The measurement at the $t\th$ location on the chromosome typically represents $\log_{2}$ ratio of copy numbers in the test genome to that in a reference genome. 
A zero value thus indicates the same copy number as that in the reference genome, 
positive values indicate gains or amplifications, whereas negative observations represent deletions.   

First and higher order HMMs have been used in the literature for modeling aCGH data. 
\cite{fridlyand2004hidden} used a first order HMM to segment aCGH data into sets with the same underlying copy number; 
\cite{guha2008bayesian} used a first order HMM with four latent copy number states with Gaussian emission densities; 
\cite{seifert2012parsimonious} used a tree based higher order HMM with three states with Gaussian emission densities; 
\cite{Yau_etal:2011} used a first order HMM with three states but a flexible infinite component translated-mixture of Normals as the emission distribution. 
The use of a flexible family of emission densities made this approach robust to the presence of outliers, skewness or heavy tails in the error process. 

We consider the Coriell aCGH dataset from the \texttt{DNAcopy} package in \texttt{Bioconductor}, 
originally presented in \cite{snijders2001assembly}. 
The data correspond to two array CGH studies of fibroblast cell strains. 
We chose the study GM05296 that comprised copy number ratios at 2271 consecutive genomic locations. 
We model the dataset using HDP-HMM and the proposed HOHMM with $C=3$ states. 
The unknown state specific means $\mu_{c}$'s are allowed to vary according to $\Normal(\mu_{c,0},\sigma_{c,0}^{2})$ hyper-priors 
with $\mu_{c,0}=-0.5,0.0,0.5$ and $\sigma_{c,0}=1/6,10^{-5},1/6$ for $c=1,2,3$, 
allowing $\mu_{c}$'s to vary over $[-1,1]$ essentially across disjoint intervals, thus ensuring identifiability of these states. 
%
As in \cite{Yau_etal:2011}, we also use a translated-mixture of Normals as our emission distribution with $S=5$ local components. 
Experiments with larger values of $S$ did not result in any practical difference in the results. 
 
\begin{figure}[!h]
\begin{center}
\includegraphics[height=10cm, width=15cm, trim=0cm 0.25cm 0cm 0.25cm, clip=true]{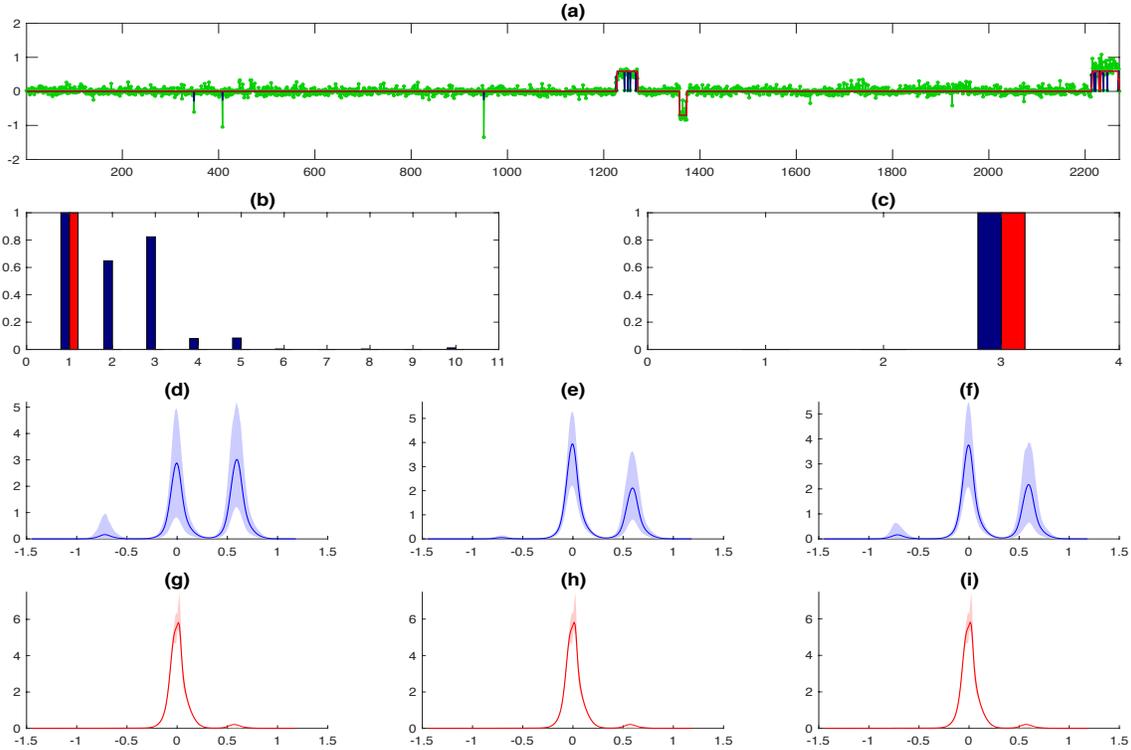}
\end{center}
\caption{\baselineskip=10pt 
Results for the aCGH dataset: CTF-HOHMM in blue and HDP-HMM in red 
(a) posterior means super-imposed over the observed time series in green; 
(b) the inclusion probabilities of different lags; 
(c) the distribution of different number of states; 
(d), (e), (f) estimated one, two and three steps ahead predictive densities, respectively, and their 90\% credible intervals by the CTF-HOHMM; 
(g), (h), (i) estimated one, two and three steps ahead predictive densities, respectively, and their 90\% credible intervals by the HDP-HMM. 
}
\label{fig: aCGH}
\end{figure}

Figure \ref{fig: aCGH} summarizes the results for the Coriell aCGH dataset obtained by CTF-HOHMM and HDP-HMM. 
The CTF-HOHMM results suggest higher order dependence with the first three lags being the important ones. 
This is reflective of the fact that copy number variations usually occur in clusters of adjacent locations. 
Panel (a) in Figure \ref{fig: aCGH} suggests that the CTF-HOHMM provides a better fit to local variations in the data, 
better capturing {focal aberrations} \citep{fridlyand2004hidden} due to alterations in very narrow regions. 
The predictive densities estimated by the two methods also look quite different. 
For example, for one step ahead prediction, the HDP-HMM basically predicts a Normal copy number state. 
The CTF-HOHMM, on the other hand,  
assigns equal probabilities to having either a Normal copy number or an increased copy number. 
It takes in account not just the immediately preceding location, which had a Normal copy number state, 
but also the variation in a few preceding locations which had amplified copy numbers.

\section{Discussion} \label{sec: discussion}
We proposed a flexible yet parsimonious nonparametric Bayesian approach to higher order hidden Markov models 
that allows automated identification of the important lags. 
The celebrated HDP-HMM is obtained as a special case 
when the order is restricted to one and the soft sharing feature of the model is turned off. 
In simulation experiments, our method vastly out-performed the HDP-HMM in higher order settings. 
Remarkably, the improvements were also substantial in the first order cases 
which may be attributed to greater levels of data compression achievable by the proposed model.

The focus of this paper has been on higher order homogeneous HMMs, 
but the proposed methodology can be easily extended to nonhomogeneous cases 
where the transition dynamics and the emission distributions are allowed to be influenced by exogenous predictors. 
We are also pursuing the development of faster algorithms for online inference in HOHMMs that scale better with larger datasets.
Additional important directions of ongoing research include extensions to other discrete state space dynamical systems, 
and models for spatial and spatio-temporal datasets.

\baselineskip=14pt
\bibliographystyle{natbib}
\bibliography{HMM,HOHMM,aCGH}

\clearpage\pagebreak\newpage
\newgeometry{tmargin=1.25in, textheight=8.5in, textwidth=6.5in,}
\pagestyle{fancy}
\fancyhf{}
\rhead{\bfseries\thepage}
\lhead{\bfseries SUPPLEMENTARY MATERIALS}

\baselineskip 20pt
\begin{center}
{\LARGE{Supplementary Materials} 
for\\ {\bf Bayesian Higher Order Hidden Markov Models}}
\end{center}

\setcounter{equation}{0}
\setcounter{page}{1}
\setcounter{table}{1}
\setcounter{figure}{0}
\setcounter{section}{0}
\numberwithin{table}{section}
\renewcommand{\theequation}{S.\arabic{equation}}
\renewcommand{\thesubsection}{S.\arabic{section}.\arabic{subsection}}
\renewcommand{\thesection}{S.\arabic{section}}
\renewcommand{\thepage}{S.\arabic{page}}
\renewcommand{\thetable}{S.\arabic{table}}
\renewcommand{\thefigure}{S.\arabic{figure}}
\baselineskip=15pt

\vskip 10mm

\begin{center}
Abhra Sarkar\\
Department of Statistics and Data Sciences,\\ University of Texas at Austin,\\ 2317 Speedway D9800, Austin, TX 78712-1823, USA\\
abhra.sarkar@utexas.edu\\ 
 and \\
David B. Dunson\\
Department of Statistical Science,\\ Duke University,\\ Box 90251, Durham NC 27708-0251\\
dunson@duke.edu\\
\end{center}

\vskip 15mm
\begin{center}
{\Large{\bf Summary}} 
\end{center}
The supplementary materials discuss details of the theoretical results on model identifiability and posterior convergence 
from Section \ref{sec: emission} and Section \ref{sec: posterior consistency} of the main paper 
and also provide some additional results. 
The supplementary document also presents details of the Chinese restaurant franchise process analog 
used to design the MCMC sampling algorithm described in Section \ref{sec: posterior computation} of the main paper. 
The document concludes with algorithms for sampling some of the prior hyper-parameters.

\newpage
\section{Proof of Lemma \ref{lem: identifiability}} \label{appendix: proof of identifiability}
We refer to an HOHMM with transition probability tensor $\bp$, initial and stationary distribution $\bpi$, and emission distributions $\b1f$ by $\HOHMM(\bpi,\bp,\b1f)$. 
The joint distribution of $\by_{1:T_{0}}$, $T_{0}=(2C^{q}+1)q$, under $\HOHMM(\bpi,\bp,\b1f)$ is then given by 
\bse
&& \hspace{-1cm} p_{\bp,\b1f}(\by_{1:T_{0}}) =  \sum_{c_{1},\ldots,c_{T_{0}}} \pi_{c_{1},\dots,c_{q}} p_{c_{q+1} \mid c_{1},\ldots,c_{q}} \dots p_{c_{T_{0}} \mid c_{T_{0}-q},\ldots,c_{T_{0}-1}}  f_{c_{1}} \dots f_{c_{T_{0}}} \\
&& = \sum_{c_{1},\ldots,c_{T_{0}}} \pi_{c_{1},\dots,c_{q}} p_{c_{q+1},\dots,c_{2q} \mid c_{1},\ldots,c_{q}} \dots p_{c_{T_{0}-q},\ldots,c_{T_{0}} \mid c_{T_{0}-2q},\ldots,c_{T_{0}-q-1}}  (f_{c_{1}} \dots f_{c_{q}}) \dots (f_{c_{T_{0}-q}} \dots f_{c_{T_{0}}}).
\ese

The elements of $\wt{P}^{q}$ are the probabilities $p_{c_{q+1},\dots,c_{2q} \mid c_{1},\ldots,c_{q}}=p\{(c_{q+1},\dots,c_{2q}) \mid (c_{1},\ldots,c_{q})\}$ 
of moving from $(c_{1},\ldots,c_{q})$ to $(c_{q+1},\dots,c_{2q})$ in $q$ steps by a Markov chain with transition probability matrix $\wt{P}$. 
Also, since $f_{c}, c\in\C,$ are all distinct in $\Y$, so are their ordered products $f_{c_{1}} f_{c_{2}} \cdots f_{c_{q}}$ in $\Y^{q}$. 
Additionally, since $\wt{P}$ is of full rank, so is $\wt{P}^{q}$. 
Straightforward application of Theorem 1 from \cite{alexandrovich2016nonparametric} then establishes 
separate nonparametric identifiability of $\wt{P}^{q}$ and $\{f_{c_{1}} f_{c_{2}} \cdots f_{c_{q}}: c_{j}\in\C, j=1,\dots,q\}$ up to label swapping of the states from the joint distribution of $(2C^{q}+1)q$ consecutive observations. 

Let $\HOHMM(\bpi^{\star},\bp^{\star},\b1f^{\star})$ be an HOHMM such that $p_{\bp,\b1f}(\by_{1:T_{0}}) = p_{\bp^{\star},\b1f^{\star}}(\by_{1:T_{0}})$.
There thus exists a permutation $\sigma$ such that 
\bse
& f_{c_{1}}^{\star}\dots f_{c_{q}}^{\star}= f_{\sigma_{1}(c_{1},\dots,c_{q})}\dots f_{\sigma_{q}(c_{1},\dots,c_{q})} ~~~~\text{and}\\ 
& p_{c_{q+1},\dots,c_{2q} \mid c_{1},\ldots,c_{q}}^{\star}  =  p_{\sigma_{1}(c_{q+1},\dots,c_{2q}),\dots,\sigma_{q}(c_{q+1},\dots,c_{2q}) \mid \sigma_{1}(c_{1},\dots,c_{q}),\ldots,\sigma_{q}(c_{1},\dots,c_{q})},
\ese 
where $\sigma_{j}(c_{1},\dots,c_{q})$ is the $j\th$ element of $\sigma(c_{1},\dots,c_{q})$. 

Since $f_{c}^{\star}$ and $f_{c}$ are probability distributions, for any pair $(j,c_{j})$
we have $f_{c_{j}}^{\star}= f_{\sigma_{j}(c_{1},\dots,c_{q})}$ for all $c_{\ell}\in\C, \ell\neq j$, which implies $\sigma_{j}(c_{1},\dots,c_{q})=\sigma(c_{j})$ for all $c_{j}\in \C, j=1,\dots,q$, and hence $f_{c}^{\star}= f_{\sigma(c)}$ for all $c\in\C$. 
This also implies that 
\bse
&& \hspace{-1cm} p_{c_{q+1},\dots,c_{2q} \mid c_{1},\ldots,c_{q}}^{\star}  =  p_{\sigma(c_{q+1}),\dots,\sigma(c_{2q}) \mid \sigma(c_{1}),\ldots,\sigma(c_{q})} ~\text{for all}~(c_{q+1},\ldots,c_{2q}).
\ese
Summing across $c_{q+2},\dots,c_{2q}$ then gives
\bse
&& \hspace{-1cm} p_{c_{q+1} \mid c_{1},\ldots,c_{q}}^{\star}  =  p_{\sigma(c_{q+1}) \mid \sigma(c_{1}),\ldots,\sigma(c_{q})} ~\text{for all}~(c_{1},\ldots,c_{q},c_{q+1}).
\ese
Finally, since the stationary distribution $\bpi$ is uniquely determined by $\wt{P}$, 
we have for all $(c_{1},\ldots,c_{q})$, $\pi^{\star}_{c_{1},\dots,c_{q}}=\pi_{\sigma(c_{1}),\dots,\sigma(c_{q})}$. 
This concludes the proof of Lemma \ref{lem: identifiability}.

\vskip 4mm
\section{Proof of Theorem \ref{thm: consistency} and Some Additional Results} \label{appendix: proof of consistency}

Using the first order representation of HOHMMs in blocks of size $q$ considered in Section \ref{sec: emission}, 
Theorem \ref{thm: consistency} can be proved with $\underline{p}$ replaced by $\underline{p}^{q}$ for values of $\ell$ that are multiples of $q$ 
by straightforwardly adapting the proofs of the results for first order HMMs in \cite{vernet2015posterior}.
For general values of $\ell \geq q$, we instead combine the first $q$ observations into a special $0\th$ initial state $(c_{1},\dots,c_{q})$ and then consider single step transitions as
\vspace{-4ex}\\
\bse
\begin{array}{c c c c c c c}
(c_{1},\dots,c_{q}) & \to & (c_{2},\dots,c_{q+1}) & \to & (c_{3},\dots,c_{q+2}) & \to & \cdots \\
\downarrow &  & \downarrow & & \downarrow &  &  \\
(\by_{1},\dots,\by_{q}) & & \by_{q+1} & &  \by_{q+2} & & \cdots.
\end{array} 
\ese 
\vspace{-2ex}\\
Using the above representation and adapting to \cite{vernet2015posterior}, the following additional results can also be easily established. 
Theorem \ref{thm: theorem2} below establishes consistency the $\ell\th$ order marginals in the weak topology (Ghosh and Ramamoorthi, 2003, page 12). 
Theorem \ref{thm: theorem3} establishes separate consistency of the transition probabilities and the emission distributions, the latter in the weak topology. 

\begin{Thm}	\label{thm: theorem2}
Under Assumptions \ref{assmp: 1}, 
for all weak neighborhood $U_{W}$ of $f_{1:\ell,0}^{\star}$, $\ell \geq q$, 
\bse
P_{\btheta_{0}} \left\{\lim_{T \to \infty} \Pi(U_{W} \mid \by_{1:T}) =1\right\} = 1. 
\ese
\end{Thm}

\begin{Thm}	\label{thm: theorem3}
Let $\btheta_{0}=(\bp_{0},\b1f_{0})$ characterize a stationary HOHMM such that $\wt{P}_{0}$, the first order representation of $\bp_{0}$, is full rank and the elements of $\b1f_{0}=(f_{1,0},\dots,f_{C,0})\trans$ are linearly independent.  
Also, let $P_{\btheta_{0}} \left[\lim_{T \to \infty}  \Pi\{\btheta: D_{\ell}(\btheta_{0},\btheta) < \epsilon \mid \by_{1:T}\} =1 \right] =1$ for $\ell \geq (2C^{q}+1)q$. 
Then, for all neighborhood $U_{\bp_{0}}$ of $\bp_{0}$ and all weak neighborhood $U_{f_{c,0}}$ of $f_{c,0}$, $c=1,\dots,C$, 
\bse
P_{\btheta_{0}} \left[\lim_{T \to \infty} \Pi\left\{\btheta=(\bp,\b1f): \exists\sigma, \sigma(\bp) \in U_{\bp_{0}},~ f_{\sigma(c)} \in U_{f_{c,0}}, c=1,\dots,C \mid \by_{1:T}\right\} =1\right] = 1, 
\ese
where $\sigma$ is a permutation with $\sigma(\bp) = \{p_{\sigma(c_{q+1}) \mid \sigma(c_{1}),\ldots,\sigma(c_{q})}, c_{j}=1,\dots,C, j=1,\dots,q+1\}$.  
\end{Thm}

Some details of the proofs, including coupling arguments, are provided below to show that the original calculations follow through in the HOHMM settings 
using the above first order representation with single step transitions. 
This also makes the article relatively self-contained.

\subsection{Background}

The following results are directly used in the proofs. 
In what follows, $\norm{\mu-\nu}_{TV}$ denotes the total variation distance between two probability measures $\mu$ and $\nu$.  
That is, $\norm{\mu-\nu}_{TV} = \sup_{A}\abs{\mu(A)-\nu(A)} = (1/2)\norm{\mu-\nu}_{L_{1}}$.
For a sequence $\{X_{n}\}_{n\geq 0}$, 
its $\phi$ mixing coefficients are defined as $\phi_{m} = \sup_{A,B: P(A)>0} \{P (X_{m} \in B) - P(X_{m} \in B \mid X_{0} \in A)\}$.

\begin{Thm} \label{thm: coupling}
\textbf{(Coupling Inequality, Doeblin, 1938, Lindvall, 1992 and Thorisson, 2000)} 
For any coupling $\Lambda$ of random variables $X \sim \mu$ and $Y \sim \nu$ we have
\bse
\norm{\mu-\nu}_{TV} 
\leq P_{\Lambda}(X \neq Y).
\ese
\end{Thm}

Consider a finite state Markov chain with state space $\C=\{1,\dots,C\}$ and transition probability matrix $P$.  
Let $P(x,y) \geq \epsilon_{0} \psi(y)$ for some probability measure $\psi(\cdot)$. 
We can `split' the transition kernel $P$ into two constituent parts, one of which does not depend on the current position in the chain, using the decomposition
\bse
P(x,\cdot) = \epsilon_{0} \psi(\cdot) + (1-\epsilon_{0})R(x,\cdot),
\ese
where $R(x,\cdot) = \{P(x,\cdot)-\epsilon_{0} \psi(\cdot)\}/(1-\epsilon_{0})$ is also a transition kernel. 
Generating the next sample in a Markov chain with kernel $P$ can therefore be done in two stages. 
First, draw a Bernoulli random variable with probability of success $\epsilon_{0}$, 
and then conditional on success draw from $\psi(\cdot)$, otherwise draw from $R(x,\cdot)$.

Consider the following coupling with the Markov chains $\{X_{t}\}_{t\geq 0}$ and $\{Y_{t}\}_{t\geq 0}$, 
both with state space $\C=\{1,\dots,C\}$, transition kernel $P$ and invariant distribution $\pi$.
\begin{enumerate}
\item
For the chain $\{X_{t}\}_{t\geq 0}$, set $X_{0} = x_{0}$, and for $\{Y_{t}\}_{t \geq 0}$, set $Y_{0} \sim \pi$.
\item
At each iteration $m$, draw $U_{m} \sim \Bern(\epsilon_{0})$. 
\item
If $U_{m} = 0$, then draw $X_{m} \sim R(x_{m-1},\cdot)$ and $Y_{m} \sim R(y_{m-1},\cdot)$ independently. 
\item
But if $U_{m} = 1$, draw $X_{m} \sim \psi$, and set $Y_{m} = X_{m}$.
In subsequent steps, draw values of $X_{t}$ as above but always set $Y_{t}=X_{t}$.
\end{enumerate}
We then have  
\bse
&& P( X_{1} \neq Y_{1} \mid X_{0},Y_{0}) = P( X_{1} \neq Y_{1} \mid U_{1}=0, X_{0},Y_{0}) (1-\epsilon_{0}) + P( X_{1} \neq Y_{1} \mid U_{1}=1, X_{0},Y_{0}) \epsilon_{0} \\
&& = P( X_{1} \neq Y_{1} \mid U_{1}=0, X_{0},Y_{0}) (1-\epsilon_{0}) \leq (1-\epsilon_{0}).
\ese
Under the coupling, once $X_{s} = Y_{s}$ we have $X_{t} = Y_{t}$ for all $t > s$. 
Therefore, 
\bse
&& P(X_{m} \neq Y_{m} \mid X_{0},Y_{0}) = P(X_{m} \neq Y_{m}, X_{m-1} \neq Y_{m-1}, \dots, X_{1} \neq Y_{1} \mid X_{0},Y_{0}) \\
&& = P(X_{m} \neq Y_{m} \mid X_{m-1} \neq Y_{m-1}) \cdots P(X_{2} \neq Y_{2}\mid X_{1} \neq Y_{1}) P( X_{1} \neq Y_{1} \mid X_{0},Y_{0}) \\
&& \leq (1-\epsilon_{0})^{m}. 
\ese
The bound is uniform since it does not depend on the starting value $X_{0}$. 
Therefore, we have
\bse
\phi_{m}[\{X_{t}\}_{t \geq 0}] \leq \sup_{x} \norm{P^{m}(x,\cdot)-\pi(\cdot)}_{TV} 
\leq P_{\Lambda}(X_{m} \neq Y_{m}) \leq (1-\epsilon_{0})^{m}.
\ese
If $P$ be such that $P(x,y) \geq \underline{p} >0$ for all $x,y \in \C$.
Choosing $\psi(y) = 1/C$ for all $y$, we can then set $\epsilon_{0}=C \underline{p}$. 
Therefore, in this case, we have 
\bse
\phi_{m}[\{X_{t}\}_{t \geq 0}] \leq \sup_{x} \norm{P^{m}(x,\cdot)-\pi(\cdot)}_{TV} 
\leq (1-C\underline{p})^{m}.
\ese

Consider now a stationary HMM with latent state sequence $X_{t}$ and observation sequence $Y_{t}$. 
Then, for the Markov sequence $\{(X_{t},Y_{t})\}_{t \geq 0}$, we have  
\bse
&& \hspace{-1cm} \phi_{m}[\{(X_{t},Y_{t})\}_{t \geq 0}] \leq  \sup_{x_{0},y_{0},B=B_{x} \times B_{y}} \abs{P\{(X_{m},Y_{m}) \in B\} - P\{(X_{m},Y_{m}) \in B \mid (X_{0},Y_{0})=(x_{0},y_{0})\}} \\
&& \hspace{-0.5cm} = \sup_{x_{0},y_{0},B=B_{x} \times B_{y}} \abs{\pi(X_{m} \in B_{x})  P(Y_{m} \in B_{y} \mid X_{m} \in B_{x}) - P(X_{m} \in B_{x} \mid X_{0}=x_{0}) P(Y_{m} \in B_{y} \mid X_{m} \in B_{x})} \\
&& \hspace{-0.5cm} = \sup_{x_{0},y_{0},B=B_{x} \times B_{y}} \abs{P(Y_{m} \in B_{y} \mid X_{m} \in B_{x}) \{\pi(X_{m} \in B_{x})   -  P(X_{m} \in B_{x} \mid X_{0}=x_{0})\}} \\
&& \hspace{-0.5cm} \leq \sup_{x_{0},B_{x}} \abs{\pi(B_{x})   -  P^{m}(B_{x} \mid X_{0}=x_{0})} \leq (1-C\underline{p})^{m}.
\ese
The mixing coefficients $\phi_{m}[\{(X_{t},Y_{t})\}_{t \geq 0}]$ of the Markov sequence $\{(X_{t},Y_{t})\}_{t \geq 0}$ thus admit the same bounds as those on the mixing coefficients of the original sequence $\{X_{t}\}_{t \geq 0}$.

Consider next the Markov sequence $\{Z_{t}\}_{t \geq 0}$ formed with blocks of the sequence $\{(X_{t},Y_{t})\}_{t \geq 0}$ of size $\ell$, 
that is, $Z_{t} = (X_{t\ell+1},\dots,X_{t\ell+\ell}, Y_{t\ell+1},\dots,Y_{t\ell+\ell})$, 
where $\{X_{t}\}_{t \geq 0}$ and $\{Y_{t}\}_{t \geq 0}$ are the latent and observed processes of an HMM as above. 
Then, proceeding as above, for the sequence $\{Z_{t}\}_{t \geq 0}$, we have  
\bse
&& \hspace{-1cm} \phi_{m}[\{Z_{t}\}_{t \geq 0}] \leq \sup_{x_{1},\dots,x_{\ell},B_{x,1},\dots,B_{x,\ell}} \vert P(X_{m\ell+1} \in B_{x,1},\dots,X_{m\ell+\ell} \in B_{x,\ell})  \\
&& \hspace{3cm}  -  P(X_{m\ell+1} \in B_{x,1},\dots,X_{m\ell+\ell} \in B_{x,\ell} \mid X_{1}=x_{1},\dots,X_{\ell}=x_{\ell}) \vert \\
&& \leq \sup_{x_{\ell},B_{x,1},B_{x,2},\dots,B_{x,\ell}} \vert P(X_{m\ell+2} \in B_{x,2},\dots,X_{m\ell+\ell} \in B_{x,\ell} \mid X_{m\ell+1} \in B_{x,1}) \\
&& \hspace{3cm}  \{P(X_{m\ell+1} \in B_{x,1})  -  P(X_{m\ell+1} \in B_{x,1} \mid X_{\ell}=x_{\ell})\} \vert \\
&& \leq \sup_{x_{\ell},B_{x,1}} \vert P(X_{m\ell+1} \in B_{x,1})  -  P(X_{m\ell+1} \in B_{x,1} \mid X_{\ell}=x_{\ell}) \vert \\
&& \leq (1-C\underline{p})^{(m-1)\ell+1} \leq (1-C\underline{p})^{m} .
\ese

\vspace*{-4ex}
Next, consider a $q\th$ order Markov chain $\{X_{t}\}_{t \geq 0}$ with transition probabilities $p(x_{q+1} \mid x_{1},\dots,x_{q})$ 
with $p(x_{q+1} \mid x_{1},\dots,x_{q}) \geq \underline{p}$ for all $x_{1},\dots,x_{q},x_{q+1} \in \C$. 
We can still `split' the transition kernel $p$ into two constituent parts using the decomposition
\bse
p\{(x_{1},\dots,x_{q}) \to \cdot\} = \epsilon_{0} \psi(\cdot) + (1-\epsilon_{0})R\{(x_{1},\dots,x_{q}) \to \cdot\}.
\ese
We now consider the following coupling for two higher order Markov chains $\{X_{t}\}_{t\geq 0}$ and $\{Y_{t}\}_{t\geq 0}$, 
both with transition kernel $p$, stationary distribution $\pi$, marginal one component stationary distribution $\pi_{1}$, 
and special initial states as described below. 
\begin{enumerate}
\item
For $\{X_{t}\}_{t\geq 0}$, set $X_{0} = (x_{-q+1},\dots,x_{0})$, and for $\{Y_{t}\}_{t \geq 0}$, set $Y_{0} = (y_{-q+1},\dots,y_{0}) \sim \pi$.
\item
At each iteration $m$, draw $U_{m} \sim \Bern(\epsilon_{0})$. 
\item
If $U_{m} = 0$, draw $X_{m} \sim R\{(x_{m-q},\dots,x_{m-1}) \to \cdot\}$ and $Y_{m} \sim R\{(y_{m-q},\dots,y_{m-1}) \to \cdot\}$ independently. 
\item
But if $U_{m} = 1$, draw $X_{m} \sim \psi(\cdot)$, and set $Y_{m} = X_{m}$.
In subsequent steps, draw values of $X_{t}$ as before but always set $Y_{t}=X_{t}$.
\end{enumerate}
Following an argument and calculations similar to the first order case, 
we then have 
\bse
\phi_{m}[\{X_{t}\}_{t \geq 0}] \leq \sup_{x_{1},\dots,x_{q}} \norm{P^{m}\{(x_{1},\dots,x_{q}),\cdot\}-\pi_{1}(\cdot)}_{TV} 
\leq (1-\epsilon_{0})^{m} = (1-C\underline{p})^{m}.
\ese
The sequence $\{Z_{j}\}_{j \geq 0}$, where $Z_{j} = (X_{j\ell+1}, . . . , X_{j\ell+\ell},Y_{j\ell+1}, \dots , Y_{j\ell+\ell})$ is also a Markov sequence whenever $\ell \geq q$.
Proceeding as above, under the assumption that $p(x_{q+1} \mid x_{1},\dots,x_{q}) \geq \underline{p}$ for all $x_{1},\dots,x_{q},x_{q+1} \in \C$, we have 
\bse
&& \hspace{-1cm} \phi_{m}[\{Z_{t}\}_{t \geq 1}] \leq \sup_{x_{1},\dots,x_{\ell},B_{x,1},\dots,B_{x,\ell}} \vert P(X_{m\ell+1} \in B_{x,1},\dots,X_{m\ell+\ell} \in B_{x,\ell})  \\
&& \hspace{3cm}  -  P(X_{m\ell+1} \in B_{x,1},\dots,X_{m\ell+\ell} \in B_{x,\ell} \mid X_{1}=x_{1},\dots,X_{\ell}=x_{\ell}) \vert \\
&& \leq \sup_{x_{\ell-q},\dots,x_{\ell},B_{x,1},B_{x,2},\dots,B_{x,\ell}} \vert P(X_{m\ell+2} \in B_{x,2},\dots,X_{m\ell+\ell} \in B_{x,\ell} \mid X_{m\ell+1} \in B_{x,1}) \\
&& \hspace{3cm}  \{P(X_{m\ell+1} \in B_{x,1})  -  P(X_{m\ell+1} \in B_{x,1} \mid X_{\ell-q}=x_{\ell-q},\dots,X_{\ell}=x_{\ell})\} \vert \\
&& \leq (1-C\underline{p})^{(m-1)\ell+1} \leq (1-C\underline{p})^{m}.
\ese

\begin{Thm} (\textbf{Theorem 5 of Barron, 1988})
\begin{enumerate}[label=\alph*.]
\item Let $p_{T,\btheta_{0}}(\by_{1:T})$ and $\int_{\Theta}p_{T,\btheta}(\by_{1:T}) \Pi(d\btheta)$ merge with probability one. That is, for every $\beta>0$
\bse
P_{\btheta_{0}}\left\{ \frac{\int_{\Theta} p_{T,\btheta}(\by_{1:T}) \Pi(d\btheta)}{p_{T,\btheta_{0}}(\by_{1:T})} \leq \exp(-T\beta)~\text{infintely often}\right\} = 0.
\ese 
\item The sets $A_{T}, B_{T}, C_{T}$ are such that 
	\begin{enumerate}
	\item $A_{T} \cup B_{T} \cup C_{T} = \Theta$,
	\item $\Pi(B_{T}) \leq \exp(-T \beta_{1})$ for some $\beta_{1}>0$, 
	\item there exists a uniformly exponentially consistent test $\psi_{T}(\by_{1},\dots,\by_{T})$ for testing $H_{0} : f=f(\btheta_{0})$ vs $H_{1} : f \in \{f(\btheta): \btheta \in C_{T}\}$. 
	That is, for some $\beta_{2} > 0$, 
	\bse
	\lim_{T\to \infty} \eE_{\btheta_{0}}(\psi_{T}) = 0 
	~~~\text{and}~~~ 
	\sup_{\btheta \in C_{T}} \eE_{\btheta}(1-\psi_{T}) \leq \exp(-T \beta_{2}).
	\ese 
	\end{enumerate}
\end{enumerate}
Then, $P_{\btheta_{0}}\{ \lim_{T \to \infty} \Pi(A_{T} \mid \by_{1:T}) = 1\} = 1$.  
\end{Thm}

\vskip 5mm
\begin{Thm}  (\textbf{Rio's inequality, Rio, 2000})\\
Let $\{X_{n}\}_{n\geq 0}$ be a sequence of random variables, each $X_{i}$ taking values in $E_{i}$, 
with mixing coefficients $\phi_{m} = \sup_{A,B: P(A)>0} \{P (X_{m} \in B) - P(X_{m} \in B \mid X_{0} \in A)\}$. 
Let $f$ be a function from $E^{n}=E_{1} \times \dots \times E_{n}$ to $\rR$ such that
\vspace{-2ex} 
\bse
\abs{f(x_{1},\dots,x_{n}) - f(y_{1},\dots,y_{n})} \leq \Delta_{1} 1_{x_{1} \neq y_{1}} + \dots + \Delta_{n} 1_{x_{n} \neq y_{n}}.
\ese
\vspace{-3ex}\\ 
Then, for any $x>0$
\vspace{-2ex} 
\bse
P\{f(X_{1},\dots,X_{n}) - \eE_{P}f(X_{1},\dots,X_{n}) \geq x\} \leq \exp\left\{ - \frac{2 x^{2} (\Delta_{1}^{2} + \dots + \Delta_{n}^{2})^{-1}}{(1 + 2\phi_{1} + \dots + 2\phi_{n-1})^{2}}\right\}.  \label{eq: Rio}
\ese
\vspace{-3ex}
\end{Thm}

For a stationary Markov sequence with transition probabilities $p(x_{2} \mid x_{1}) \geq \underline{p}$ for all $x_{1},x_{2} \in \C$, 
we have $\phi_{m} \leq \rho_{\btheta}^{-m}$ with  
$\rho_{\btheta} \geq (1-C\underline{p})^{-1}$. 
Therefore, $(1 + 2\phi_{1} + \dots + 2\phi_{n-1}) \leq (1 + 2 \rho_{\btheta}^{-1} + 2\rho_{\btheta}^{-2} + \dots ) = (2 + \rho_{\btheta}-1)/(\rho_{\btheta}-1) \leq (2-C \underline{p})/(C\underline{p})$.
Letting $f(x_{1},\dots,x_{n}) = \sum_{i=1}^{n} h(x_{i})$ with $0 \leq h(x) \leq 1$ so that $\Delta_{i} = 1$ for all $i$, we then have 
\bse
P\left[\sum_{i=1}^{n} \left\{h(X_{i}) - \eE_{P}h(X_{i})\right\} \geq x \right] \leq \exp\left\{ - \frac{2 x^{2}C^{2} \underline{p}^{2}}{n (2 - C\underline{p})^{2}}\right\} \leq \exp\left\{ - \frac{x^{2}C^{2} \underline{p}^{2}}{2 n}\right\}. 
\ese
Following the discussions above, the same inequality then also holds true for the sequence 
$\{Z_{t}\}_{t \geq 0}$ with $Z_{t} = (X_{t\ell+1},\dots,X_{t\ell+\ell}, Y_{t\ell+1},\dots,Y_{t\ell+\ell})$, 
where $\{X_{t}\}_{t \geq 0}$ and $\{Y_{t})\}_{t \geq 0}$ are the latent and observed processes of an HOHMM 
with transition kernel of the latent sequence $\{X_{t}\}_{t \geq 0}$ 
satisfying $p(x_{q+1} \mid x_{1},\dots,x_{q}) \geq \underline{p}$ for all $x_{1},\dots,x_{q},x_{q+1} \in \{1,\dots,C\}$.

\subsection{Proof of Theorem \ref{thm: theorem2}}
The likelihood function of an HOHMM is given by 
\bse
p_{T,\btheta} = p(\by_{1:T} \mid \btheta) = \sum_{c_{1},\dots,c_{T}} \pi(c_{1},\dots,c_{q}) \prod_{t=q+1}^{T} p(c_{t} \mid \bc_{(t-q):(t-1)}) \prod_{t=1}^{T} f(\by_{t} \mid c_{t}),
\ese
where $\pi_{c_{1},\dots,c_{q}}$ is the stationary as well as the starting distribution of the HOHMM. 
Following the steps similar to those the proof of Lemma 2.2 in \cite{vernet2015posterior}, we can show, under Assumptions \ref{assmp: 1}, that 
the Kullback-Leibler divergence between $p_{T,\btheta_{0}}$ and $p_{T,\btheta}$, denoted $KL(p_{T,\btheta_{0}}, p_{T,\btheta})$, satisfies 
\bse 
&& \hspace{-1cm} KL\left(p_{T,\btheta_{0}}, p_{T,\btheta} \right) = \eE_{\btheta_{0}} \log \left\{\frac{\sum_{c_{1},\dots,c_{T}} \pi_{c_{1},\dots,c_{q},0} \prod_{t=q+1}^{T} p_{c_{t} \mid \bc_{(t-q):(t-1),0}} \prod_{t=1}^{T} f_{c_{t},0}(\by_{t})}       {\sum_{c_{1},\dots,c_{T}}\pi_{c_{1},\dots,c_{q}} \prod_{t=q+1}^{T} p_{c_{t} \mid \bc_{(t-q):(t-1)}} \prod_{t=1}^{T} f_{c_{t}}(\by_{t})} \right\}\\
&& \leq \max_{\bc_{1:q}} \abs{\pi_{c_{1},\dots,c_{q},0} - \pi_{c_{1},\dots,c_{q}} } / \underline{p} + (T-q) \max_{\bc_{1:(q+1)}} \abs{p_{c_{q+1}\mid c_{1},\dots,c_{q},0} - p_{c_{q+1}\mid c_{1},\dots,c_{q}}} \\ 
&&~~~ + T \max_{c} \int f_{c,0}(\by) \max_{k} \log \frac{f_{c,0}(\by)}{f_{k}(\by)}\lambda(\by). 
\ese
This implies, for any $\epsilon>0$ and any $\btheta \in \Theta_{\epsilon}$, that
\bse
\frac{1}{T} KL\left(p_{T,\btheta_{0}}, p_{T,\btheta} \right) \leq  \frac{3}{\underline{p}} \epsilon.
\ese

Using the bound on the KL divergence, following the steps in the proof of Theorem 2.1 in \cite{vernet2015posterior}, we can show that for any $\beta>0$ 
\bse
P_{\btheta_{0}}\left\{ \frac{\int_{\Theta} p_{T,\btheta}(\by_{1:T}) \Pi(d\btheta)}{p_{T,\btheta_{0}}(\by_{1:T})} < \exp(-\beta T)~\text{infinitely often} \right\} = 0.  
\ese
Consistency on any weak neighborhood of $f_{1:\ell,0}^{\star}$ is equivalent to consistency in neighborhoods of the type $U_{h}=\{f_{1:\ell}^{\star}: \int h f_{1:\ell}^{\star} - \int h f_{1:\ell,0}^{\star}<\epsilon\}$ 
where $0 \leq h \leq 1$ is a bounded continuous function on $\Y^{\ell}$ (Ghosh and Ramamoorthi, 2003, page 131). 
Uniformly exponentially consistent tests for $H_{0} : f_{1:\ell}^{\star}=f_{1:\ell,0}^{\star}$ vs $H_{1} : f_{1:\ell}^{\star} \in U_{h}^{c}$ may then be constructed as $\psi_{T}(\by_{1:T}) =1_{\S_{T}}(\by_{1:T})$ where 
\bse
&& \S_{T} = \left\{\by_{1:T} : \frac{\ell}{T}\sum_{j=0}^{T/\ell -1}h(\by_{j\ell+1},\dots,\by_{j\ell+\ell}) > \frac{\alpha+\gamma}{2}\right\},\\
&& \alpha=\eE_{f_{1:\ell,0}^{\star}}\{h(\by_{1},\dots,\by_{\ell})\},~~~~~  \gamma=\inf_{f_{1:\ell}^{\star} \in U_{h}^{c}}\eE_{f_{1:\ell}^{\star}}\{h(\by_{1},\dots,\by_{\ell})\} > \alpha+\epsilon.  
\ese
Then, using Rio's inequality, we have 
\bse
&& \eE_{\btheta_{0}}(\psi_{T}) = P_{\btheta_{0}}(\S_{T}) \\
&& = P_{\btheta_{0}} \left[\sum_{j=0}^{T/\ell -1}\left\{h(\by_{j\ell+1},\dots,\by_{j\ell+\ell}) - \int h(\by_{j\ell+1},\dots,\by_{j\ell+\ell}) f_{1:\ell,0}^{\star}(\by_{j\ell+1},\dots,\by_{j\ell+\ell}) d\lambda^{\otimes\ell} \right\} > \frac{T(\gamma-\alpha)}{2\ell}\right] \\
&& 
\leq \exp\left\{ -\frac{T(\gamma-\alpha)^{2} C^{2} \underline{p}^{2} } {32\ell}\right\}.
\ese
Likewise, for all $\btheta \in \Theta(\underline{p})$ such that $f_{1:\ell}^{\star} \in U_{h}^{c}$ 
\bse
&& \eE_{\btheta}(1-\psi_{T}) = P_{\btheta}(\S_{T}^{c}) \\
&& = P_{\btheta} \left[\sum_{j=0}^{T/\ell -1}\left\{-h(\by_{(j\ell+1):(j\ell+\ell)}) + \eE_{f_{1:\ell}^{\star}} h(\by_{(j\ell+1):(j\ell+\ell)}) \right\} > \frac{T}{\ell}\eE_{f_{1:\ell}^{\star}} h(\by_{(j\ell+1):(j\ell+\ell)}) - \frac{T(\gamma+\alpha)}{2\ell}\right] \\
&& \leq P_{\btheta} \left[\sum_{j=0}^{T/\ell -1}\left\{-h(\by_{j\ell+1},\dots,\by_{j\ell+\ell}) + \int h(\by_{j\ell+1},\dots,\by_{j\ell+\ell}) f_{1:\ell}^{\star}(\by_{j\ell+1},\dots,\by_{j\ell+\ell}) d\lambda^{\otimes\ell} \right\} > \frac{T(\gamma-\alpha)}{2\ell}\right] \\
&& 
\leq \exp\left\{ -\frac{T(\gamma-\alpha)^{2} C^{2} \underline{p}^{2} } {32\ell}\right\}. 
\ese
Applying Barron's theorem with $A_{T}=U_{h}, B_{T}=\phi, C_{T}=U_{h}^{c}$, we have that 
$P_{\btheta_{0}}\{ \lim_{T \to \infty} \Pi(U_{h} \mid \by_{1:T}) = 1\} = 1$. 
This proves Theorem \ref{thm: theorem2}.

\subsection{Proof of Theorem \ref{thm: consistency}} 
Let $\norm{f_{1:\ell}^{\star}-f_{1:\ell,0}^{\star}}_{1}=\int \abs{f_{1:\ell}^{\star}-f_{1:\ell,0}^{\star}} \lambda^{\otimes \ell}$ and $\norm{\bp_{0} - \bp} = \max_{\bc_{(t-q):t}} \abs{p_{c_{t} \mid c_{t-q},\dots,c_{t-1},0} - p_{c_{t} \mid c_{t-q},\dots,c_{t-1}}}$. 
We now let $U=\left\{\btheta: D_{\ell}(\btheta_{0},\btheta)<\epsilon\right\}$ denote a $D_{\ell}$ neighborhood of $\btheta_{0}$. 
Since $\Pi\{\Theta(\underline{p})\}=1$, in what follows, complements are implicitly understood to be with respect the set $\Theta(\underline{p})$. 
That is, $U^{c} = \Theta(\underline{p}) \cap U^{c}$ and so on. 
To apply Barron's theorem, we then set 
\bse
A_{T}=U,~~~~B_{T}=\calP(\underline{p}) \times \F_{T},~~~~C_{T} = (A_{T} \cup B_{T})^{c}. 
\ese
Using Assumption 1C from the main paper, we have that 
\vspace{-3ex}
\bse
\Pi(B_{T}) = \Pi_{F}(\F_{T}^{c}) \leq \exp(-T\beta_{1}).
\ese 
We next have to show the existence of a uniformly consistent test $\psi_{T}$ for testing $H_{0}: f_{1:\ell}^{\star}=f_{1:\ell,0}^{\star}$ vs $H_{1}: f_{1:\ell}^{\star} \in C_{T}$. \\
Let $f_{1:\ell,j}^{\star} = f_{1:\ell}^{\star}(\by_{1},\dots,\by_{\ell} \mid \btheta_{j})$. 
Let $\btheta_{j}, j=1,\dots,N$, $N=\N(\delta,\F_{T}, D_{\ell})$ be a sequence such that for all $\btheta \in \calP(\underline{p}) \times \F_{T}$, 
$D_{\ell}(\btheta_{j},\btheta) \leq \delta$ with $\delta=\epsilon/12$. 
Define $\phi_{j}(\by_{1:T}) =1_{\S_{j}}(\by_{1:T})$, where
\bse
&& \S_{j} = \left\{\by_{1:T} : \sum_{j=0}^{T/\ell -1}\left[1\{(\by_{j\ell+1},\dots,\by_{j\ell+\ell})\in B_{j}\} - P_{\btheta_{0}}\{(\by_{1},\dots,\by_{\ell})\in B_{j}\}\right] > s_{j} \right\},\\
&& \text{with}~~~B_{j} = \{(\by_{1},\dots,\by_{\ell})\in \Y^{\ell} : f_{1:\ell}^{\star}(\by_{1},\dots,\by_{\ell} \mid \btheta_{0}) \leq f_{1:\ell}^{\star}(\by_{1},\dots,\by_{\ell} \mid \btheta_{j})\}\\
&& \text{and}~~~s_{j} = \frac{T\norm{f_{1:\ell,j}^{\star}-f_{1:\ell,0}^{\star}}_{1}}{4\ell}.
\ese
Then, we have 
\vspace{-2ex} 
\bse
P_{\btheta_{j}}\{(\by_{j\ell+1},\dots,\by_{j\ell+\ell})\in B_{j}\} - P_{\btheta_{0}}\{(\by_{1},\dots,\by_{\ell})\in B_{j}\} = \frac{1}{2}  \norm{f_{1:\ell,j}^{\star}-f_{1:\ell,0}^{\star}}_{1}.
\ese
Using Rio's inequality again, we have 
\bse
&& \eE_{\btheta_{0}}(\phi_{j})  =  P_{\btheta_{0}} (\S_{j}) = P_{\btheta_{0}} \left[\sum_{j=0}^{T/\ell -1}\{1(\by_{(j\ell+1):(j\ell+\ell)}\in B_{j}) - P_{\btheta_{0}}(\by_{1:\ell} \in B_{j})\} > s_{j}\right] \\
&& 
\leq \exp\left\{-\frac{T \norm{f_{1:\ell,j}^{\star}-f_{1:\ell,0}^{\star}}_{1}^{2} C^{2} \underline{p}^{2}}{32\ell}\right\}. 
\ese
We next define 
\vspace{-3ex}
\bse
\psi_{T} = \max_{1 \leq j \leq N: \btheta_{j} \in A_{T}^{c}} \phi_{j}. 
\ese
For $\btheta_{j} \in A_{T}^{c}$, $\norm{f_{1:\ell,j}^{\star}-f_{1:\ell,0}^{\star}}_{1} \geq \epsilon$. 
We then have 
\bse
&& \eE_{\btheta_{0}}(\psi_{T}) = \eE_{\btheta_{0}}\left(\max_{1 \leq j \leq N: \btheta_{j} \in A_{T}^{c}} \phi_{j} \right)  \leq \eE_{\btheta_{0}}\left(\sum_{1 \leq j \leq N: \btheta_{j} \in A_{T}^{c}} \phi_{j}\right)  =  \sum_{1 \leq j \leq N: \btheta_{j} \in A_{T}^{c}} \eE_{\btheta_{0}}\left(\phi_{j}\right)\\
&& \leq \N\left\{\frac{\epsilon}{12},\calP(\underline{p}) \times \F_{T},D_{\ell}\right\} \max_{1 \leq j \leq N: \btheta_{j} \in A_{T}^{c}} \eE_{\btheta_{0}} (\phi_{j}) \\ 
&& \leq \N\left\{\frac{\epsilon}{12},\calP(\underline{p}) \times \F_{T},D_{\ell}\right\} \exp\left(-\frac{T\epsilon^{2} C^{2} \underline{p}^{2} }{32\ell} \right).
\ese
Likewise, 
we have 
\bse
&& \eE_{\btheta}(1-\phi_{j}) = P_{\btheta}(\S_{j}^{c}) \\
&& = P_{\btheta} \left[\sum_{j=0}^{T/\ell -1}\{-1(\by_{(j\ell+1):(j\ell+\ell)}\in B_{j}) + P_{\btheta_{0}}(\by_{1:\ell} \in B_{j})\} > - s_{j}\right] \\
&& = P_{\btheta} \Bigg[\sum_{j=0}^{T/\ell -1}\{-1(\by_{(j\ell+1):(j\ell+\ell)}\in B_{j}) + P_{\btheta}(\by_{1:\ell} \in B_{j})\} \\
&& \hspace{3cm} > - s_{j} + \sum_{j=0}^{T/\ell -1}\{P_{\btheta}(\by_{(j\ell+1):(j\ell+\ell)}\in B_{j}) - P_{\btheta_{0}}(\by_{1:\ell} \in B_{j})\}\Bigg].
\ese
\vspace{-2ex}
Now, for all $\btheta \in A_{T}^{c}$, we have 
\bse
&& - s_{j} + \sum_{j=0}^{T/\ell -1}\{P_{\btheta}(\by_{(j\ell+1):(j\ell+\ell)}\in B_{j}) - P_{\btheta_{0}}(\by_{1:\ell} \in B_{j})\} \\
&& = - \frac{T\norm{f_{1:\ell,j}^{\star}-f_{1:\ell,0}^{\star}}_{1}}{4\ell} + \sum_{j=0}^{T/\ell -1}\{P_{\btheta_{j}}(\by_{(j\ell+1):(j\ell+\ell)}\in B_{j}) - P_{\btheta_{0}}(\by_{1:\ell} \in B_{j})\} \\
&& ~~~ + \sum_{j=0}^{T/\ell -1}\{P_{\btheta}(\by_{(j\ell+1):(j\ell+\ell)}\in B_{j}) - P_{\btheta_{j}}(\by_{1:\ell} \in B_{j})\} \\
&& = - \frac{T\norm{f_{1:\ell,j}^{\star}-f_{1:\ell,0}^{\star}}_{1}}{4\ell} + \frac{T\norm{f_{1:\ell,j}^{\star}-f_{1:\ell,0}^{\star}}_{1}}{2\ell} - \frac{T\norm{f_{1:\ell,j}^{\star}-f_{1:\ell}^{\star}}_{1}}{2\ell} \\
&& = \frac{T\norm{f_{1:\ell,j}^{\star}-f_{1:\ell,0}^{\star}}_{1}}{4\ell} - \frac{T\norm{f_{1:\ell,j}^{\star}-f_{1:\ell}^{\star}}_{1}}{2\ell} 
\geq \frac{T\epsilon}{4\ell} - \frac{T\epsilon}{24\ell} 
\geq \frac{T\epsilon}{8\ell}.
\ese
The last line follows since for all $\btheta \in A_{T}^{c}$, we have $\norm{f_{1:\ell}^{\star}-f_{1:\ell,0}^{\star}}_{1} > \epsilon$, 
and also by definition of $\btheta_{j}$, we have $\norm{f_{1:\ell,j}^{\star}-f_{1:\ell}^{\star}}_{1} < \epsilon/12$. 
Therefore, for any $\btheta \in A_{T}^{c}$, using Rio's inequality, we have
\vspace{-3ex}
\bse
&& \hspace{-1cm} \eE_{\btheta}(1-\phi_{j}) 
\leq P_{\btheta} \left[\sum_{j=0}^{T/\ell -1}\{-1(\by_{(j\ell+1):(j\ell+\ell)}\in B_{j}) + P_{\btheta}(\by_{1:\ell} \in B_{j})\}  > \frac{T\epsilon}{8\ell} \right] 
\leq \exp\left( - \frac{T\epsilon^{2} C^{2} \underline{p}^{2}}{32\ell} \right).
\ese
Therefore, we have 
\vspace{-3ex}
\bse
&& \sup_{\btheta \in A_{T}^{c} \cap \B_{T}^{c}} \eE_{\btheta}(1-\psi_{T}) \leq \exp\left(-\frac{T\epsilon^{2} C^{2} \underline{p}^{2}}{32\ell} \right) .
\ese
\vspace{-3ex}\\
For all $\btheta_{0},\btheta$, we have, using triangle inequality, that 
\vspace{-2ex}
\bse
D_{\ell}(\btheta_{0},\btheta) \leq \sum_{c_{1},\dots,c_{q}} \abs{\pi_{c_{1},\dots,c_{q},0} - \pi_{c_{1},\dots, c_{q}} } 
	+ C(\ell-q) \norm{\bp_{0}-\bp} + \ell \max_{c}\int \abs{f_{c,0}(\by_{1}) - f_{c}(\by_{1})}\lambda(d\by).
\ese
Therefore, $D_{\ell}(\btheta_{0},\btheta) \leq \epsilon/12$ if all three terms on the right hand side above are $\leq \epsilon/36$. 
The function $\bp \to \bpi(\bp)$ is uniformly continuous on $\calP(\underline{p})$. 
Hence, there exists $\epsilon_{1}>0$ such that for all $\bp_{0},\bp \in \calP(\underline{p})$ 
with $\norm{\bp_{0} - \bp} < \epsilon_{1}$, 
we have $\sum_{c_{1},\dots,c_{q}} \abs{\pi_{c_{1},\dots,c_{q},0} - \pi_{c_{1},\dots,c_{q}} } < \epsilon/36$. 
This implies
\bse
&& \hspace{-1cm} \N\left\{\frac{\epsilon}{12},\calP(\underline{p}) \times \F_{T},D_{\ell}\right\}   
\leq  \N\left[\min \left\{\frac{\epsilon}{36C(\ell-q)},\epsilon_{1}\right\},\Theta(\underline{p}),\norm{\cdot}\right]  ~ \N\left\{\frac{\epsilon}{36\ell},\F_{T}, d(\cdot,\cdot)\right\} \\
&& \leq  \left[\max \left\{\frac{36C(\ell-q)}{\epsilon},\frac{1}{\epsilon_{1}}\right\}\right]^{C^{q}(C^{q}-1)} ~ \N\left\{\frac{\epsilon}{36\ell},\F_{T}, d(\cdot,\cdot)\right\}.
\ese
Under Assumptions \ref{assmp: 1}, this implies 
\bse
&& \eE_{\btheta_{0}}(\psi_{T}) \leq \N\left\{\frac{\epsilon}{12},\calP(\underline{p}) \times \F_{T},D_{\ell}\right\} \exp\left(-\frac{T\epsilon^{2} C^{2} \underline{p}^{2} }{32\ell} \right) \to 0.
\ese
Applying Barron's theorem, 
we have that $P_{\btheta_{0}}\{ \lim_{T \to \infty} \Pi(U \mid \by_{1:T}) = 1\} = 1$. 
This concludes the proof of Theorem \ref{thm: consistency}.

\vskip 5mm
\subsection{Proof of Theorem \ref{thm: theorem3}}
It suffices to show that for any $U_{\bp_{0}}$ and $U_{f_{c,0}}$, 
there exists a $D_{\ell}$-neighborhood $U$ of $\btheta_{0}$ such that 
\bse
U\subset \left\{\btheta=(\bp,\b1f): \exists\sigma, \sigma(\bp) \in U_{\bp_{0}},~ f_{\sigma(c)} \in U_{f_{c,0}}, c=1,\dots,C \mid \by_{1:T}\right\}. 
\ese
This is equivalent to showing that for any $\btheta_{T}= (\bp_{T},\b1f_{T}) \in \Theta(\underline{p})$ with $D_{T_{0}}(\btheta_{0},\btheta_{T}) \to 0$, $T_{0}=(2C^{q}+1)q$,
there exists a subsequence $\wt\btheta_{T}=(\wt\bp_{T},\wt{\b1f}_{T})$ of $\btheta_{T}$ 
such that $\norm{\wt\bp_{T}-\bar\bp}\to 0$ and $\b1f_{c,T} \to \bar{f}_{c,0}$ in the weak topology 
for some $\bar\btheta \in \Theta(\underline{p})$ 
where $(\bar\bp,\bar{\b1f})$ and $(\bp_{0},\b1f_{0})$ (and hence $\bar\bpi$ and $\bpi_{0}$) are equivalent up to label swapping of the states.  
 
Let $\btheta_{T} = (\bp_{T},\b1f_{T}) \in \Theta(\underline{p})$ be such that $D_{T_{0}}(\btheta_{0},\btheta_{T}) \to 0$.  
Since $\calP(\underline{p})$ is compact, there exists a subsequence of $\bp_{T}$, denoted $\wt\bp_{T}$, such that $\wt\bp_{T} \to \bar\bp \in \calP(\underline{p})$. 
Let $\wt\bpi_{T}=\wt\bpi_{T}(\wt\bp_{T})$ denote the associated stationary distribution and $\wt{\b1f}_{T}$ the associated emission distributions.  
Using triangle inequality, we have 
\vspace{-2ex} 
\bse
&& \hspace{-1cm} D_{T_{0}}(\btheta_{0},\wt\btheta_{T}) = \int \bigg| \sum_{c_{1},\dots,c_{T_{0}}} \wt\pi_{c_{1},\dots,c_{q},T} \prod_{t=q+1}^{T_{0}} \wt{p}_{c_{t} \mid c_{t-q},\dots,c_{t-1},T} \prod_{t=1}^{T_{0}} \wt{f}_{c_{t},T}(\by_{t}) \\ 
&&~~~~~ - \sum_{c_{1},\dots,c_{T_{0}}} \pi_{c_{1},\dots,c_{q},0} \prod_{t=q+1}^{T_{0}} p_{c_{t} \mid c_{t-q},\dots,c_{t-1},0} \prod_{t=1}^{T_{0}} f_{c_{t},0}(\by_{t})   \bigg| \lambda(d\by_{1}) \dots \lambda(d\by_{T_{0}}) \\
&& \geq -  \sum_{c_{1},\dots,c_{T_{0}}} \bigg|  \pi_{c_{1},\dots,c_{q}} \prod_{t=q+1}^{T_{0}} p_{c_{t} \mid c_{t-q},\dots,c_{t-1}}  -  \bar\pi_{c_{1},\dots,c_{q}} \prod_{t=q+1}^{T_{0}} \bar{p}_{c_{t} \mid c_{t-q},\dots,c_{t-1}} \bigg|  \\
&& ~~~~~ + \int \bigg| \sum_{c_{1},\dots,c_{T_{0}}} \bar\pi_{c_{1},\dots,c_{q}} \prod_{t=q+1}^{T_{0}} \bar{p}_{c_{t} \mid c_{t-q},\dots,c_{t-1}} \prod_{t=1}^{T_{0}} \wt{f}_{c_{t},T}(\by_{t}) \\ 
&&~~~~~ - \sum_{c_{1},\dots,c_{T_{0}}} \pi_{c_{1},\dots,c_{q},0} \prod_{t=q+1}^{T_{0}} p_{c_{t} \mid c_{t-q},\dots,c_{t-1},0} \prod_{t=1}^{T_{0}} f_{c_{t},0}(\by_{t}) \bigg| \lambda(d\by_{1}) \dots \lambda(d\by_{T_{0}})
\ese
Since the left hand side and the first term on the right hand side both tend to zero, we have 
\vspace{-3ex} 
\bse
&& \int \bigg| \sum_{c_{1},\dots,c_{T_{0}}} \bar\pi_{c_{1},\dots,c_{q}} \prod_{t=q+1}^{T_{0}} \bar{p}_{c_{t} \mid c_{t-q},\dots,c_{t-1}} \prod_{t=1}^{T_{0}} \wt{f}_{c_{t},T}(\by_{t}) \\ 
&&~~~~~ - \sum_{c_{1},\dots,c_{T_{0}}} \pi_{c_{1},\dots,c_{q},0} \prod_{t=q+1}^{T_{0}} p_{c_{t} \mid c_{t-q},\dots,c_{t-1},0} \prod_{t=1}^{T_{0}} f_{c_{t},0}(\by_{t}) \bigg| \lambda(d\by_{1}) \dots \lambda(d\by_{T_{0}}) \to 0. 
\ese 
Since $ \sum_{c_{1},\dots,c_{T_{0}}} \bar\pi_{c_{1},\dots,c_{q}} \prod_{t=q+1}^{T_{0}} \bar{p}_{c_{t} \mid c_{t-q},\dots,c_{t-1}} \prod_{t=1}^{T_{0}} \wt{f}_{c_{t},T}(\by_{t})$ converges in total variation, it is tight, and hence so are the sequence of distributions $\wt{f}_{c,T}$. 
Using Prokhorov's theorem (Ghosh and Ramamoorthi, 2003, page 13), there exist subsequences of $\wt{f}_{c,T}$ that converge weakly to some $\bar{f}_{c_{t}}$.  
This implies 
\vspace{-2ex} 
\bse
\sum_{c_{1},\dots,c_{T_{0}}} \bar\pi_{c_{1},\dots,c_{q}} \prod_{t=q+1}^{T_{0}} \bar{p}_{c_{t} \mid c_{t-q},\dots,c_{t-1}} \prod_{t=1}^{T_{0}} \bar{f}_{c_{t},T}(\by_{t}) 
 = \sum_{c_{1},\dots,c_{T_{0}}} \pi_{c_{1},\dots,c_{q},0} \prod_{t=q+1}^{T_{0}} p_{c_{t} \mid c_{t-q},\dots,c_{t-1},0} \prod_{t=1}^{T_{0}} f_{c_{t},0}(\by_{t}).
\ese
Application of Lemma \ref{lem: identifiability} then concludes the proof of Theorem \ref{thm: theorem3}.

\vskip 7.5mm
As we discussed in 
Section \ref{sec: posterior consistency} in the main paper, 
for the proposed tensor decomposition based model (\ref{eq: TFM1}) for HOHMM transition probabilities, 
truncated Dirichlet priors on the parameters $\blambda_{h_{1},\dots,h_{q}}$, truncated below $\underline{p}$, 
satisfy the assumptions on the transition probabilities required in Theorem \ref{thm: consistency}. 
Results showing how the additional assumptions on the emission distributions and associated priors in Theorem \ref{thm: consistency} 
relate to the specific examples considered in the main paper 
can be derived along the lines of similar results in \cite{vernet2015posterior} and are omitted.

\section{Higher Order Chinese Restaurant Franchise (CRF)}
\subsection{The Original CRF} 
We first review the original CRF (Teh \etal 2006) before we describe how we adapted it to our HOHMM setting in the next subsection. 
Let there be $J$ groups, each with $N_{j}$ observations $\{y_{j,\ell}\}_{\ell=1}^{N_{j}}$ with a generative model as 
\bse
&& \blambda_{0} \mid \alpha_{0}   \sim  \Dir(\alpha_{0}/C,\dots,\alpha_{0}/C), \\
&& \blambda_{j} \mid \alpha,\blambda_{0}  \sim  \Dir(\alpha\blambda_{0}), ~~~~c_{j,\ell} \mid \blambda_{j} \sim \Mult\{\lambda_{j}(1),\dots,\lambda_{j}(C)\}\\
&& y_{j,\ell} \mid \{\theta_{c}\}_{c=1}^{C}, c_{j,\ell}=k \sim f(\theta_{k}), ~~~~\theta_{c} \sim p_{0}.
\ese
The model generating the labels $c_{j,\ell}$'s may be reformulated as  
\bse
&& \blambda_{0} \mid \alpha_{0}   \sim  \Dir(\alpha_{0}/C,\dots,\alpha_{0}/C), \\
&& G_{j} = \textstyle\sum_{k=1}^{C} \lambda_{j}(k) \delta_{k}, ~~~~ \blambda_{j} \mid \alpha,\blambda_{0}  \sim  \Dir(\alpha\blambda_{0}),~~~~ c_{j,\ell} \mid G_{j} \sim G_{j}.
\ese
Another representation is given by 
\bse
&& G_{0} = \textstyle\sum_{k=1}^{C} \lambda_{0}(k) \delta_{k}, ~~~~ \blambda_{0} \mid \alpha_{0}   \sim  \Dir(\alpha_{0}/C,\dots,\alpha_{0}/C), \\
&& G_{j} = \textstyle\sum_{\tau=1}^{\infty} \wt{\lambda}_{j}(\tau) \delta_{\psi_{j,\tau}}, ~~~~ \wt{\blambda}_{j} \sim \SB(\alpha),~~~~~ \psi_{j,\tau} \sim G_{0},~~~~c_{j,\ell} \mid G_{j} \sim G_{j}.
\ese

\begin{figure}[h!]
\begin{center}
\includegraphics[height=10cm, width=13.75cm, trim=2cm 1cm 1cm 1cm, clip=true]{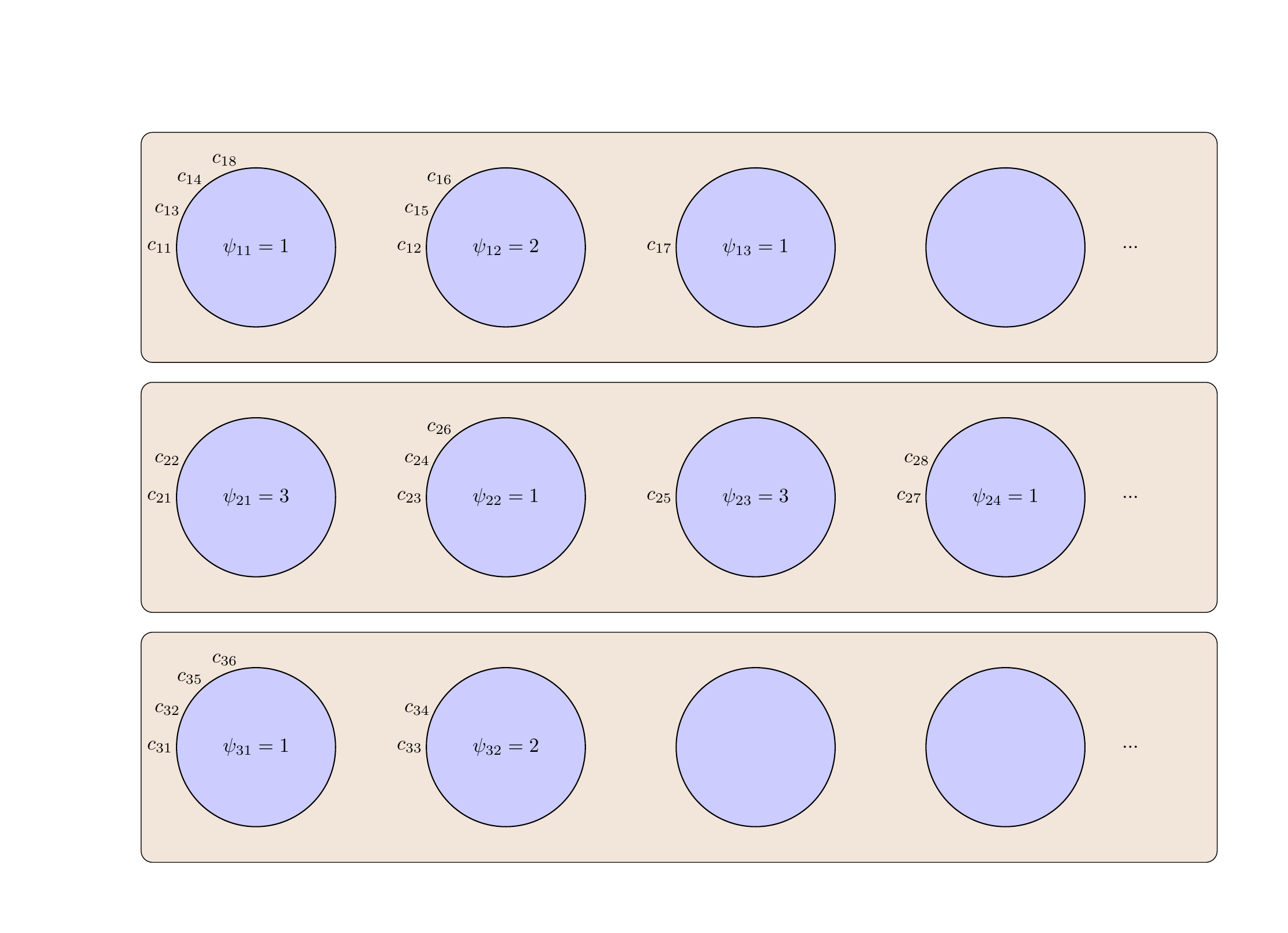}
\caption{The Chinese restaurant franchise. 
}
\label{fig: CRP}
\end{center}
\end{figure}

A CRF arising from this generative model is as follows. 
Corresponding to the $J$ groups, imagine $J$ restaurants, each with infinitely many tables but finitely many dishes $\C=\{1,\dots,C\}$ on their globally shared menu. 
The $\ell\th$ customer belonging to the $j\th$ group enters restaurant $j$, sits at a table $\tau_{j,\ell}$, and is served a dish $c_{j,\ell}$. 
While the restaurant assignments are predetermined by group memberships, 
the table assignment for the $\ell\th$ customer in restaurant $j$ is chosen as $\tau_{j,\ell} \sim \wt{\blambda}_{j}$, 
and each table $\tau$ is assigned a dish $\psi_{j,\tau} \sim \blambda_{0}$.
Customers sitting at the same table thus all eat the same dish. 
Multiple tables may, however, be served the same dish, 
allowing two customers enjoying the same dish to be seated at different tables. 
Given $c_{j,\ell}$ and the corresponding table assignment $\tau_{j,\ell}$, $\psi_{j,\tau_{j,\ell}}=c_{j,\ell}$. 
See Figure \ref{fig: CRP}.

Let $n_{j,\tau}$ denote the number of customers in restaurant $j$ at table $\tau$, 
$n_{j}(\psi)$ denote the number of customers in restaurant $j$ eating the dish $\psi$, 
and $n_{j}$ denote the total number of customers in restaurant $j$. 
Also, let $n_{j,\tau}(\psi)$ denote the number of customers in restaurant $j$ at table $\tau$ eating dish $\psi$. 
Clearly, $n_{j,\tau}(\psi)>0$ only when dish $\psi$ is served at an occupied table $\tau$. 
Finally, let $m_{j}(\psi)$ be the number of tables in restaurant $j$ serving dish $\psi$, 
and $m_{j}$ be the total number of occupied tables in restaurant $j$.

Given a posterior sample of the dish assignments, 
we can obtain a draw from the posterior of $\blambda_{j}$ by noting that a-priori 
$\blambda_{j} \sim  \Dir\{\alpha\lambda_{0}(1),\dots,\alpha\lambda_{0}(C)\}$
and that $c_{j,\ell}$ for each $j,\ell$ is a draw from $\blambda_{j}$. 
The number of different $c_{j,\ell}$'s that are associated with a specific dish $c$ 
thus equals the total number of customers in the restaurant $j$ eating the dish $c$, that is, $n_{j}(c)$. 
Using Dirichlet-Multinomial conjugacy, we then have
\bse
(\blambda_{j} \mid \alpha, \blambda_{0}, \bn, \bzeta) \sim \Dir\{\alpha\lambda_{0}(1)+n_{j}(1),\dots,\alpha\lambda_{0}(C)+n_{j}(C)\}.
\ese

Likewise, given a sample $(\btau,\bpsi)$ of the table and the dish assignments, 
we can obtain a draw from the posterior of $\blambda_{0}$ by noting that a-priori 
$\blambda_{0} \sim  \Dir(\alpha_{0}/C,\dots,\alpha_{0}/C)$
and that $\psi_{j,\tau}$ for each $\tau$ is a draw from $\blambda_{0}$. 
The number of different $\psi_{j,\tau}$'s that are associated with a specific dish $\psi$ 
is precisely the number of tables in the restaurant $j$ that served the dish $\psi$, that is, $m_{j}(\psi)$. 
The total number of tables serving dish $\psi$ across all restaurants is therefore $m_{0}(\psi)=\sum_{j}m_{j}(\psi)$. 
Using Dirichlet-Multinomial conjugacy, we then have
\bse
(\blambda_{0} \mid \bm, \bzeta) \sim \Dir\{\alpha_{0}/C+m_{0}(1),\dots,\alpha_{0}/C+m_{0}(C)\}.
\ese

The table assignments $\btau$ are also latent. 
To sample $\btau$ from the posterior, 
we first marginalize out their prior $\wt{\blambda}_{j} \sim \SB(\alpha)$ to obtain 
\bse
(\tau_{j,\ell} \mid \alpha, \btau_{j}^{-\ell}) \sim \textstyle \sum_{\tau \in \S_{j,\tau}^{-\ell}} \frac{n_{j,\tau}^{-\ell}}{n_{j}-1+\alpha} \delta_{\tau} + \frac{\alpha}{n_{j}-1+\alpha} \delta_{\tau_{new}},
\ese
where $n_{j,\tau}^{-\ell}$ denotes the number of customers sitting at table $\tau$ in restaurant $j$ excluding the $\ell\th$ customer, 
$\S_{j,\tau}^{-\ell}$ denotes the set of unique values in $\btau_{j}^{-\ell}=\{\tau_{j,s}: s=1,\dots,n_{j}, s\neq \ell \}$ 
and $\tau_{new}$ is a generic for any new value of $\tau$ not in $\S_{j,\tau}^{-\ell}$.
The distribution of the table assignment $\tau_{j,\ell}$ given $\btau_{j}^{-\ell}$ and the dish assignments $\bpsi$ may then be obtained as 
\bse
&& p(\tau_{j,\ell} = \tau \mid \psi_{j,\tau}=\psi,\alpha,\bpsi_{j}^{-\ell},\btau_{j}^{-\ell},\blambda_{0}) \propto \textstyle n_{j,\tau}^{-\ell}\delta_{\tau} ~~~~ \text{if}~\tau\in\S_{j,\tau}^{-\ell}, \\
&& p(\tau_{j,\ell} = \tau_{new} \mid \psi_{j,\tau_{new}}=\psi,\alpha,\bpsi_{j}^{-\ell},\btau_{j}^{-\ell},\blambda_{0}) \propto \textstyle \alpha\lambda_{0}(\psi) ~~~~ \text{if}~\tau_{new}\notin\S_{j,\tau}^{-\ell},
\ese
where $\bpsi_{j}^{-\ell}=\{\psi_{j,\tau_{j,s}}: s=1,\dots,n_{j}, s\neq \ell \}$. 
Since these assignments are restricted only to tables serving the dish $\psi$, 
the distribution reduces to 
\bse
(\tau_{j,\ell} \mid \psi_{j,\tau_{j,\ell}}=\psi,\alpha,\bpsi_{j}^{-\ell},\btau_{j}^{-\ell},\blambda_{0})     \sim    \textstyle \sum_{\tau \in \S_{j,\tau}^{-\ell}(\psi)} \frac{n_{j,\tau}^{-\ell}(\psi)}{n_{j}(\psi)-1+\alpha \lambda_{0}(\psi)} \delta_{\tau} + \frac{\alpha\lambda_{0}(\psi)}{n_{j}(\psi)-1+\alpha\lambda_{0}(\psi)} \delta_{\tau_{new}},
\ese
where $\S_{j,\tau}^{-\ell}(\psi)$ denotes the set of unique values in $\btau_{j}^{-\ell}(\psi)=\{\tau_{j,s}: s=1,\dots,n_{j}, s\neq \ell, \psi_{j,\tau_{j,s}}=\psi \}$, 
$n_{j,\tau}^{-\ell}(\psi)$ denotes the number of customers sitting at table $\tau$ in restaurant $j$ and enjoying the dish $\psi$ excluding the $\ell\th$ customer, 
and $\tau_{new}$ is a generic for any new value of $\tau$ not in $\S_{j,\tau}^{-\ell}(\psi)$.
This distribution can be identified with a marginalized conditional distribution of assignments of $n_{j}(\psi)$ observations to different components in a $\SB\{\alpha\lambda_{0}(\psi)\}$.
The full conditional for $\blambda_{0}$ given $(\bpsi,\btau)$ depends on the table assignments only via $m_{j}(\psi)$ 
which can be obtained from the table assignments $\btau_{j}$. 

Alternatively, for each of the $n_{j}(\psi)$ customers in restaurant $j$ enjoying the dish $\psi$, let
$m_{j,\ell}(\psi)=0$ if the $\ell\th$ customer sits at an already occupied table, and 
$m_{j,\ell}(\psi)=1$ if the $\ell\th$ customer goes to a new table.
Then, $m_{j}(\psi)=\sum_{\ell=1}^{n_{j}(\psi)}m_{j,\ell}(\psi)$. 
Using properties of a $\SB\{\alpha\lambda_{0}(\psi)\}$ distribution, 
we then have 
\bse
\{m_{j,\ell}(\psi) \mid \bm_{j}^{\ell-1}(\psi),\alpha,\blambda_{0}\} \sim \textstyle \frac{\ell-1}{\ell-1+\alpha\lambda_{0}(\psi)} \delta_{0} + \frac{\alpha\lambda_{0}(\psi)}{\ell-1+\alpha\lambda_{0}(\psi)} \delta_{1},
\ese
where $\bm_{j}^{\ell-1}(\psi)=\{m_{j,s}(\psi): s=1,\dots,\ell-1\}$. 
We can then sample the $m_{j,\ell}(\psi)$'s from the posterior by sequentially sampling them as 
\bse
\textstyle [\{m_{j,\ell}(\psi)\}_{\ell=1}^{n_{j}(\psi)} \mid \alpha,\blambda_{0}] \sim \prod_{\ell=1}^{n_{j}(\psi)} \Bern\left\{\frac{\alpha\lambda_{0}(\psi)}{\ell-1+\alpha\lambda_{0}(\psi)}\right\}.
\ese

\subsection{Higher Order CRF for CTF-HOHMM}
While customers in the CRF of the HDP are pre-partitioned into restaurants based on their fixed group assignments, 
in our HOHMM setting the restaurant assignments are latent and hence are also sampled.  
Specifically, they are determined by the labels $z_{j,t}$'s - when $(z_{1,t},\dots,z_{q,t})=(h_{1},\dots,h_{q})$, the customer enters the $(h_{1},\dots,h_{q})\th$ restaurant. 
There are thus a total of $\prod_{j=1}^{q}k_{j}$ restaurants. 

We recall that the $j\th$ lag $c_{t-j}$ is important in predicting the dynamics of $c_{t}$ only when $k_{j}>1$. 
In the culinary analogy, the $j\th$ lag is thus important if it has restaurants named (labeled) after it. 

The total number of customers entering the $(h_{1},\dots,h_{q})\th$ restaurant is now $n_{h_{1},\dots,h_{q}}=\sum_{t}1\{z_{1,t}=h_{1}, \dots, z_{q,t}=h_{q}\}$. 
Among them, the number of customers eating the dish $c$ is $n_{h_{1},\dots,h_{q}}(c)=\sum_{t}1\{z_{1,t}=h_{1}, \dots, z_{q,t}=h_{q}, c_{t}=c\}$.
Using Dirichlet-Multinomial conjugacy, we then have
\bse
(\blambda_{h_{1},\dots,h_{q}} \mid \bzeta) \sim \Dir\{\alpha\lambda_{0}(1)+n_{h_{1},\dots,h_{q}}(1),\dots,\alpha\lambda_{0}(C)+n_{h_{1},\dots,h_{q}}(C)\}.
\ese

We next define, for each $\ell=1,\dots,n_{h_{1},\dots,h_{q}}(c)$, $m_{\ell,h_{1},\dots,h_{q}}(c) = 0$ if the customer sits at an already occupied table
and $m_{\ell,h_{1},\dots,h_{q}}(c) = 1$ if the customer goes to a new table.
Then, we can sample $\{m_{\ell,h_{1},\dots,h_{q}}(c)\}_{\ell=1}^{n_{h_{1},\dots,h_{q}}(c)}$ from the posterior by sampling them sequentially from 
\bse
\textstyle \{m_{\ell,h_{1},\dots,h_{q}}(c)\}_{\ell=1}^{n_{h_{1},\dots,h_{q}}(c)} \mid \bzeta       \sim      \prod_{\ell=1}^{n_{h_{1},\dots,h_{q}}(c)} \Bern\left\{\frac{\alpha\lambda_{0}(c)}{\ell-1+\alpha\lambda_{0}(c)}\right\}.
\ese
Then, $m_{h_{1},\dots,h_{q}}(c)=\sum_{\ell=1}^{n_{h_{1},\dots,h_{q}}(c)}m_{\ell,h_{1},\dots,h_{q}}(c)$ 
gives the number of occupied tables serving the dish $c$ in the $(h_{1},\dots,h_{q})\th$ restaurant.

The table assignments in restaurants $(h_{1},\dots,h_{q})$ follow $\blambda_{0}$. 
Letting $m_{0}(c)=\sum_{h_{1},\dots,h_{q}}m_{h_{1},\dots,h_{q}}(c)$ 
denote the total number of tables serving dish $c$ across all such restaurants, 
we can update $\blambda_{0}$ using Dirichlet-Multinomial conjugacy as 
\bse
\{\lambda_{0}(1),\dots,\lambda_{0}(C)\} \mid \bzeta \sim \Dir\{\alpha_{0}/C+m_{0}(1),\dots,\alpha_{0}/C+m_{0}(C)\}. 
\ese

\newpage
\section{Sampling Prior Hyper-parameters}

The full conditional for the hyper-parameter $\alpha$ in the original CRF can be derived assuming a $\Ga(a,b)$ prior and adapting to West (1992). 
Following Antoniak (1974), integrating out $\blambda_{0}$, we have $p(m_{j} \mid \alpha,n_{j}) = \alpha^{m_{j}} s^{\star}(n_{j},m_{j})\Gamma(\alpha)  /  \Gamma(\alpha+n_{j})$, 
where $s^{\star}(n,v)$ are Stirling numbers of the first kind. 
Letting $\bn=\{n_{j}\}_{j=1}^{J}$, $\bm=\{m_{j}\}_{j=1}^{J}$ with $v=\sum_{j=1}^{J}m_{j}$, since the restaurants are conditionally independent, we have
\bse
&& \textstyle \hspace*{-1cm} p(\alpha \mid \bm, \bn, \bzeta) \propto p_{0}(\alpha \mid a,b) ~ p(\bm \mid \alpha,\bn) 
\textstyle \propto \exp(-\alpha b) (\alpha)^{a-1} ~\prod_{j=1}^{J}  \left\{ (\alpha)^{m_{j}} \frac{\Gamma(\alpha)}   {\Gamma(\alpha+n_{j})} \right\} \\
&& \textstyle \propto \exp(-\alpha b) (\alpha)^{a+v-1} \prod_{j=1}^{J}  \left\{\frac{(\alpha+n_{j}) ~ \Beta(\alpha+1,n_{j})}   {\alpha ~ \Gamma{(n_{j})}}\right\} \\
&& \textstyle \propto \exp(-\alpha b) (\alpha)^{a+v-1}   \prod_{j=1}^{J}  \left\{ \left(1+\frac{n_{j}}{\alpha}\right) \int r_{j}^{\alpha}(1-r_{j})^{n_{j}-1} dr_{j} \right\} \\
&& \textstyle \propto \exp(-\alpha b) (\alpha)^{a+v-1}  \prod_{j=1}^{J}  \left\{ \sum_{s_{j}=0}^{1} \left(\frac{n_{j}}{\alpha}\right)^{s_{j}}  \int r_{j}^{\alpha}(1-r_{j})^{n_{j}-1} dr_{j} \right\}.
\ese
Treating $\br=\{r_{j}\}_{j=1}^{J}$, $\bs=\{s_{j}\}_{j=1}^{J}$ as auxiliary variables, we have
\bse
\textstyle p(\alpha, \br, \bs \mid \bzeta) \propto \exp(-\alpha b) (\alpha)^{a+v-1} \prod_{j} \left\{ \left(\frac{n_{j}}{\alpha}\right)^{s_{j}}  r_{j}^{\alpha}(1-r_{j})^{n_{j}-1} \right\}.
\ese
The full conditionals for $\alpha$, $r_{j}$ and $s_{j}$ are then obtained in closed forms as 
\bse
\textstyle (\alpha \mid \bzeta) \sim \Ga(a+v-s, b-\log~r), ~~~(r_{j} \mid \bzeta) \sim \Beta(\alpha+1, n_{j}), ~~~(s_{j} \mid \bzeta) \sim \Bern\left(\frac{n_{j}}{n_{j}+\alpha}\right),
\ese
where $\log ~ r=\sum_{j=1}^{J} \log~r_{j}$, and 
$s=\sum_{j=1}^{J}s_{j}$.

To sample the hyper-parameter $\alpha$ in the HOHMM setting, we mimic the derivations in the CRF 
and introduce auxiliary variables $r_{h_{1},\dots,h_{q}}$ and $s_{h_{1},\dots,h_{q}}$ for each $h_{1},\dots,h_{q}$. 
Let $\bn_{0}=\{n_{h_{1},\dots,h_{q}}\}$; $\bm_{0},\br_{0},\bs_{0}$ are similarly defined. 
It can then follows that 
\bse
\alpha \mid \bzeta &\sim& \Ga (a_{0}+m_{0}-s_{0}, b_{0}-\log~r_{0}), \\ 
r_{h_{1},\dots,h_{q}} \mid \bzeta &\sim& \Beta(\alpha+1,n_{h_{1},\dots,h_{q}}), \\
s_{h_{1},\dots,h_{q}} \mid \bzeta &\sim& \Bern\left(\frac{n_{h_{1},\dots,h_{q}}}{n_{h_{1},\dots,h_{q}}+\alpha}\right), 
\ese
where $m_{0}=\sum_{y_{t}}\sum_{y_{t-1}}\sum_{h_{1},\dots,h_{q}} m_{h_{1},\dots,h_{q}}(c_{t})$,  
$\log ~ r_{0}=\sum_{h_{1},\dots,h_{q}} \log~r_{h_{1},\dots,h_{q}}$, and 
$s_{0}=\sum_{h_{1},\dots,h_{q}}s_{h_{1},\dots,h_{q}}$. 

Additionally, with an exponential prior $\varphi_{0}\exp(-\varphi_{0}\varphi)$ on $\varphi$, its full conditional is   
\bse
\textstyle \varphi \mid \bzeta \sim \exp\{-(\varphi_{0}+\sum_{j}jk_{j})\varphi\}.
\ese 

In simulation experiments and real data applications, we set the prior hyper-parameters at $a_{0}=b_{0}=1$ and $\varphi_{0}=2$. 
Our experiences with numerical experiments suggest the results to be highly robust to these choices.

\section*{References}

\refmark
Alexandrovich, G., Holzmann, H., and Leister, A. (2016).
Nonparametric identification and maximum likelihood estimation for hidden {M}arkov models.
{\em Biometrika\/}, {{103}}, 423-434.

\refmark
Allman, E., Matias, C., and Rhodes, J.~A. (2009). 
Identifiability of parameters in latent structure models with many observed variables.
\ANNALS\/, {{ 37}}, 3099-3132.

\refmark
Antoniak, C. E. (1974). Mixtures of Dirichlet processes with applications to Bayesian nonparametric problems. 
\ANNALS, {{2}}, 1152-1174.

\refmark
Barron, A. (1988). 
The exponential convergence of posterior probabilities with implications for Bayes estimators of density functions. 
Technical report.


\refmark
Doeblin, W. (1938). Expos\'{e} de la Th\'{e}orie des Cha\'{i}nes simples constants de Markoff \'{a} un nombre fini d'\'{E}tats.
{\em Revue Math. de l'Union Interbalkanique}, {{2}}, 77-105.

\refmark
Gassiat, E. and Rousseau, J. (2014).
About the posterior distribution in hidden Markov models with unknown number of states.
{\em Bernoulli}, {{20}}, 2039-2075.

\refmark
Ghosh, J. K. and Ramamoorthi, R. V. (2003). 
\emph{Bayesian nonparametrics}. 
Springer Verlag, Berlin.

\refmark
Lindvall, T. (1992). 
\emph{Lectures on the Coupling Method}. 
John Wiley \& Sons, New York.
Reprint: Dover paperback edition, 2002.

\refmark
Rio, E. (2000). In\'egalit\'es de Hoeffding pour les fonctions lipschitziennes de suites d\'ependantes. 
\emph{Comptes Rendus de l'Acad\'emie des Sciences-Series I-Mathematics}, {{330}}, 905-908.


\refmark
Teh, Y. W., Jordan, M. I., Beal, M. J., and Blei, D. M. (2006). 
Hierarchical Dirichlet processes. 
\emph{Journal of the American Statistical Association}, {{101}}, 1566-1581.

\refmark
Thorisson, H. (2000). 
\emph{Coupling, Stationarity and Regeneration}. 
Springer, New York.

 \refmark
Vernet, E. (2015b). 
Posterior consistency for nonparametric hidden Markov models with finite state space. 
\emph{Electronic Journal of Statistics}, {{9}}, 717-752.

\refmark 
West, M. (1992). 
Hyperparameter estimation in Dirichlet process mixture models. 
Institute of Statistics and Decision Sciences, Duke University, Durham, USA, Technical report.

\end{document}